\documentclass[12pt]{article}
\pdfoutput=1

\usepackage{graphicx,psfrag,epsf,xcolor}
\usepackage{amsmath,amssymb,amsfonts}
\usepackage{array}
\usepackage{cite}
\usepackage{rotating}
\usepackage{slashed, cancel}
\usepackage{scalerel,stackengine}
\usepackage{float}
\numberwithin{equation}{section}

\newcounter{MBQ}

\def\slash#1{#1 \hskip-0.45em /}

\newcommand{\be}{\begin{equation}}
\newcommand{\ee}{\end{equation}}
\newcommand{\bea}{\begin{eqnarray}}
\newcommand{\eea}{\end{eqnarray}}
\newcommand{\bi}{\begin{itemize}}
\newcommand{\ei}{\end{itemize}}
\newcommand{\ben}{\begin{enumerate}}
\newcommand{\een}{\end{enumerate}}
\newcommand{\bt}{\begin{tabular}}
\newcommand{\et}{\end{tabular}}

\newcommand{\nn}{\nonumber}

\newcommand{\nm}{n_-}
\newcommand{\np}{n_+}

\newcommand\sA{\ThisStyle{\ensurestackMath{%
			{%
\stackinset{r}{}{c}{}{\SavedStyle/}{\SavedStyle\mathcal{A}}}}}}

\definecolor{darkgreen}{rgb}{0.0,0.6,0.0}
\definecolor{cPurple}{RGB}{93,35,125}

\begin{document}
\allowdisplaybreaks

\begin{titlepage}
		
\begin{flushright}
{\small
TUM-HEP-1239/19\\
arXiv:1912.01585 [hep-ph]\\
March 06, 2020
}
\end{flushright}
		
\vskip0.7cm
\begin{center}
{\Large \bf\boldmath Threshold factorization of the   
Drell-Yan \\[0.2cm] 
process at next-to-leading power}
\end{center}
		
\vspace{0.5cm}
\begin{center}
{\sc Martin~Beneke},$^{a}$ 
{\sc Alessandro Broggio},$^{b,c}$\\ 
{\sc Sebastian Jaskiewicz},$^{a}$ 
and {\sc Leonardo Vernazza}$^{d}$\\[6mm]
{\it $^a$ Physik Department T31,\\
James-Franck-Stra\ss e~1, 
Technische Universit\"at M\"unchen,\\
D--85748 Garching, Germany\\[0.2cm]
}
{\it $^b$ Universit\`a degli Studi di Milano-Bicocca \\
Piazza della Scienza 3, I--20126 Milano, Italy\\[0.2cm]
}
{\it $^c$ INFN, Sezione di Milano-Bicocca \\
Piazza della Scienza 3, I--20126 Milano, Italy\\[0.2cm]}
{\it $^d$ Dipartimento di Fisica Teorica, 
Universit\`a di Torino \\
and INFN, Sezione di Torino, Via P. Giuria 1, 
I-10125 Torino, Italy 
}
\end{center}
		
\vspace{0.4cm}
\begin{abstract}
\vskip0.2cm\noindent
We present a factorization theorem valid near the kinematic threshold 
$z=Q^2/\hat{s}\to 1$ of the partonic Drell-Yan process 
$q\bar q\to\gamma^*+X$ for general subleading powers in the  
$(1-z)$ expansion. We then consider the specific case 
of next-to-leading power. We discuss the emergence of collinear 
functions, which are a key ingredient to factorization starting at 
next-to-leading power. We calculate the relevant collinear functions 
at $\mathcal{O}(\alpha_s)$ by employing an operator matching 
equation and we compare our results to the expansion-by-regions 
computation up to the next-to-next-to-leading order, finding 
agreement. 
Factorization holds only before the dimensional regulator is 
removed, due to a divergent convolution when the collinear and soft 
functions are first expanded around $d=4$ before the convolution is 
performed. This demonstrates an issue for threshold resummation 
beyond the leading-logarithmic accuracy at next-to-leading power.
\end{abstract}
\end{titlepage}
	
\section{Introduction}
\label{sec:introduction}
	
The study of soft emission in the threshold regime $z=Q^2/\hat{s} 
\to 1$ of the Drell-Yan (DY) process $A+B\to \gamma^*(Q)+X$ 
has a long history. The all-order summation of the leading-power 
(LP) logarithms in $(1-z)$ was pioneered in 
\cite{Sterman:1986aj,Catani:1989ne} and was later studied using 
soft-collinear effective theory (SCET) methods \cite{Idilbi:2005ky,Idilbi:2006dg,Becher:2007ty}.
Currently LP threshold logarithms can be resummed up to 
next-to-next-to-next-to-leading logarithmic accuracy 
\cite{Moch:2005ky,Becher:2007ty}. In comparison, the structure 
of factorization and resummation at the next-to-leading power (NLP), 
that is, the next order in the expansion in $(1-z)$, is not very 
well understood.
	
The DY process, given it is the simplest hadron-hadron collision 
process, has also been the target of several new calculations at 
subleading power. In this direction explicit computations of 
partonic cross sections at NLP up to 
next-to-next-to-leading order (NNLO) and partly beyond 
were performed in the coupling expansion 
by employing the expansion-by-regions method \cite{Bonocore:2014wua,Bahjat-Abbas:2018hpv} and diagrammatic factorization techniques \cite{Laenen:2008ux,Laenen:2008gt,Laenen:2010uz,Bonocore:2015esa,Bonocore:2016awd}.
The leading logarithmic (LL) resummation of the Drell-Yan processes  
$q\bar q\to\gamma^*+X$ and $gg\to H+X$ was first achieved using SCET methods \cite{Beneke:2018gvs,Beneke:2019mua} and soon after in the diagrammatic framework \cite{Bahjat-Abbas:2019fqa}.
Besides the threshold regime, the analysis of subleading power corrections for DY and single Higgs production has been investigated at fixed-order for resolution variables such as N-jettiness \cite{Boughezal:2016zws,Moult:2016fqy,Moult:2017jsg,Ebert:2018lzn,Boughezal:2018mvf,Boughezal:2019ggi} and the $q_T$ of the lepton pair or the Higgs boson \cite{Ebert:2018gsn,Cieri:2019tfv}. The resummation of NLP LLs for an event shape can be found in \cite{Moult:2018jjd}.

The resummation of NLP leading logarithms \cite{Beneke:2018gvs,Beneke:2019mua} relies on a factorization formula that was anticipated in 
these papers, and is also a prerequisite for taking the non-trivial 
step beyond LLs. In the present work we fill the theoretical 
gaps and provide the details of the derivation of the factorization 
formula beyond LP for $q\bar q\to\gamma^*+X$. 
The factorization formula, which achieves the 
separation of scales through operator definitions of the relevant 
functions, and its check against the existing NNLO NLP results from 
the expansion-by-regions approach, 
is the first main result of this paper. Nevertheless, 
it must be regarded as a formal result, because it applies to bare 
regularized quantities. As will be explained, when one attempts to 
renormalize these quantities by subtracting the divergent parts, 
the convolution of the various factors becomes itself divergent. 
This complicates the resummation of NLP logarithms beyond the 
LL accuracy with SCET renormalization group methods.
	
The second main result of this paper, already extensively used in 
\cite{Beneke:2018gvs,Beneke:2019mua}, is the identification of 
{\it{collinear functions}} or radiative jet functions 
at the amplitude level in the factorization formula at NLP.  We 
discuss their origin, why they do not appear in the well-known LP 
factorization formula, and provide their precise operator definition 
in SCET. We also calculate the collinear functions at 
$\mathcal{O}(\alpha_s)$, which 
illustrates the concept at the practical level and is required 
for the above mentioned NNLO comparison. 
	
The concept of a jet function radiating a soft gluon was originally 
introduced in \cite{DelDuca:1990gz} by way of extending the 
Low-Burnett-Kroll formula in QED. It has been extensively 
discussed in the diagrammatic factorization approach \cite{Bonocore:2015esa,Bonocore:2016awd,DelDuca:2017twk} for the production of a colourless final state in hadronic collisions. While these functions are 
closely related to the collinear functions above, since they 
describe the same physics, they are yet different. The collinear 
functions in SCET are defined at the operator level as the 
matrix elements of collinear fields. They are 
single scale objects, excluding soft contributions, and therefore 
appear suitable for the formulation of the NLP factorization 
formula. 
	
The paper is structured as follows. In Sec.~\ref{sec:collFuncns} we 
discuss the emergence of the collinear functions and provide their 
definition through an operator matching equation. The factorization 
formula valid at general subleading powers is derived in 
Sec.~\ref{sec:powerfact}. We then specialize this formula to NLP and 
identify the relevant soft and collinear functions which appear at 
this order in the power expansion. We present one of the main 
results of our paper in Sec.~\ref{sec:CollFuncsCalc}
where we extract the collinear functions at
$\mathcal{O}(\alpha_s)$ through a matching calculation. 
We compare the result that we obtain for the factorized cross
section, where we employ the newly computed collinear functions, to 
the expansion-by-regions results up to NNLO in fixed-order
perturbation theory \cite{Bonocore:2015esa,Bonocore:2016awd} 
in Sec.~\ref{sec:fixedresults}.
In particular, we find agreement with the collinear-soft NNLO 
contribution, if the convolution of collinear and soft function 
is performed in $d$ dimensions. In Sec.~\ref{sec:convolution} 
we demonstrate the appearance of 
a divergent convolution when we expand the collinear and soft 
function in $d-4$ before performing the convolution between 
the two. 
We conclude in Sec.~\ref{sec:summary}. The NLP SCET Lagrangian 
and supplementary results for the NLP one-loop soft emission 
amplitude are provided in Appendices~\ref{sec:appendixA} 
and~\ref{appendix:amplituderesults}, respectively.
	
\section{Threshold dynamics and collinear functions}
\label{sec:collFuncns}
	
The object of our investigation is the partonic DY process 
$q\bar q\to\gamma^* [\to\ell\bar{\ell}\,]+X$ in the kinematic region 
$z=Q^2/\hat{s}\to 1$, where $\hat{s}=x_ax_b \,s$ is the partonic 
centre-of-mass energy  squared, $x_a,x_b$ are the momentum 
fractions of the partons inside the incoming hadrons and $Q^2$ is 
the invariant mass squared of the lepton pair. The factorization 
of the Drell-Yan process near threshold at NLP 
will be conducted within the position-space formulation
\cite{Beneke:2002ph,Beneke:2002ni} of SCET 
\cite{Bauer:2000yr,Bauer:2001yt}. Four-momenta will often be  
decomposed using light-like vectors $n^{\mu}_+$ and $n^{\mu}_-$ 
satisfying $n_{+}\cdot n_{-} =2$ and $n_-^2=n_+^2 =0$, according to 
\bea
p^{\mu}=   (n_+p)
\frac{n^{\mu}_-}{2}+ (n_-p)\frac{n^{\mu}_+}{2}+p^{\mu}_{\perp}\,.
\eea 
At the partonic level, the threshold configuration constitutes a 
SCET$_{{\rm{I}}}$ problem, hence to capture the dynamics, collinear,
anticollinear, and soft fields are required. The scaling of the 
corresponding momenta, written in component notation
$(n_+p,n_-p,p_{\perp})$, is $Q(1,\lambda^2,\lambda)$, 
$Q(\lambda^2,1,\lambda)$, and  $Q(\lambda^2,\lambda^2,\lambda^2)$, 
respectively, where $Q$ is the hard scale of the process and 
$\lambda$ is the small power-counting parameter given by 
$\lambda = \sqrt{1-z}\,$. We note that threshold-collinear 
modes cannot be radiated into the final state $X$, since there 
is not enough energy available in threshold kinematics.
	
At the hadronic level, (anti)collinear-PDF modes with transverse 
momentum scaling $p_{\perp}\sim \Lambda$, where $\Lambda$ denotes 
the strong interaction scale, exist in addition 
to the above threshold-collinear modes. These  {\it{can}} be
radiated into the hadronic final state. The ordinary  
parton distribution functions are defined in terms of these modes. 
Concretely, the $c$-PDF modes have momentum scaling  
$(Q,\Lambda^2/Q,\Lambda)$, whereas it is $(\Lambda^2/Q,Q,\Lambda)$
for the $\bar{c}$-PDF modes. We assume that the scale $\Lambda$ of 
the strong interaction is  parametrically much smaller than the
threshold-collinear scale, $\Lambda \ll 
Q\lambda=Q(1-z)^{1/2}$. We consider power corrections in 
$\lambda$, but we always work at leading power in $\Lambda/Q$.
	
In the derivation of the 
LP factorization theorem the threshold-collinear fields are 
usually ignored, since they can be trivially integrated out. 
This can be traced to the soft-collinear decoupling transformation 
\cite{Bauer:2001yt}, which removes completely the soft-collinear 
interactions from the LP Lagrangian.
As is well known, the LP partonic cross section is then 
factorized at threshold into the convolution of a hard function, 
which is the square of a hard matching coefficient, and a soft
function, which is a vacuum matrix element of soft Wilson 
lines \cite{Korchemsky:1993uz}. At subleading power, soft-collinear 
interactions remain after the decoupling transformation, 
resulting in time-ordered product 
operators \cite{Beneke:2004in}. Threshold-collinear loops 
no longer vanish, and the threshold-collinear fields must now be 
matched to $c$-PDF collinear fields. The non-trivial 
matching coefficients constitute the {\it{amplitude-level collinear functions}}. In the following we will make these 
qualitative statements more precise. 
	
\subsection{Leading power and decoupling}
	

\begin{figure}
\begin{centering}
\includegraphics[width=0.28\textwidth]{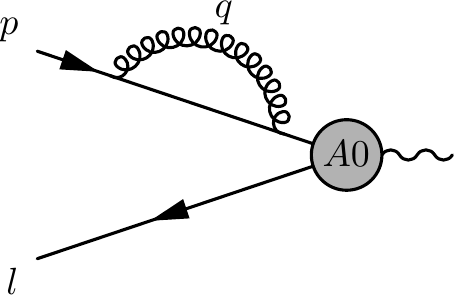}
\par\end{centering}
\caption{\label{fig:img3_c}   		
Example of a threshold-collinear loop attached to the external 
PDF-collinear line together with a LP (A0) current. This diagram, 
and the ones with more loops attaching to the 
(anti)collinear leg, yield scaleless integrals and therefore 
vanish in dimensional regularization.}
\end{figure}
	
We begin the discussion by considering the purely 
threshold-collinear\footnote{In the following, we will often 
refer to these as simply ``collinear''. ``Purely collinear'' means 
loops without soft attachments.} 
loop corrections to the DY process, as depicted in 
Figure~\ref{fig:img3_c}. The LP current, denoted $J^{A0,A0}$
following the notation of \cite{Beneke:2017ztn}, consists of 
a collinear quark field in the $n_-^{\mu}$ direction 
and an anticollinear antiquark field in the $n_+^{\mu}$ direction.
The first important observation is that on-shell such loops 
yield scaleless integrals, which vanish in dimensional 
regularization. 
	
\begin{figure}[t]
\begin{centering}
\includegraphics[width=0.28\textwidth]{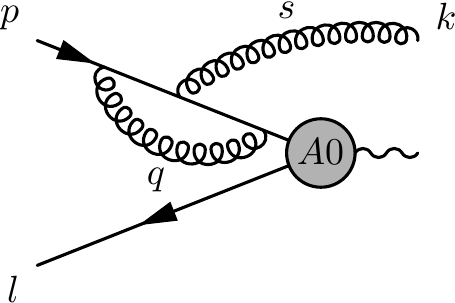}
\par\end{centering}
\caption{\label{fig:img3}   		
Example of a LP diagram with a collinear loop and a LP
soft emission. This diagram is non-vanishing.}
\end{figure}

In order to obtain non-vanishing corrections, the introduction 
of an additional scale is necessary, for example, through the 
injection of a soft momentum. This is possible in threshold
kinematics since the final state is composed of soft radiation. 
To this end, we consider the LP SCET Lagrangian written
in terms of standard SCET fields, 
\bea\label{eq:LagrangianLP}
\mathcal{L}^{(0)}_{}(z) =\bar{\xi}_c\left( 
\,in_-D\,+ i\slashed{D}_{c\perp}  \frac{1}{in_+D_c}
i\slashed{D}_{c\perp}   \right)\frac{\slashed{n}_{+}}{2}\,\xi_c 
+\mathcal{L}^{(0)}_{s}(z)+\mathcal{L}^{(0)}_{\rm{YM}}(z)  \,,
\eea
where the quark part is written explicitly as it will serve as 
an example. In this form, soft-collinear interactions are
present at LP since 
\bea
in_-D = in_-\partial +g_s\,n_-A_c(z)+g_s\,n_-A_s(z_-).
\eea
The $n_-$ component of the soft gluon field is unsuppressed with 
respect to the corresponding component of the collinear field, 
resulting in the well-known eikonal form of the soft-collinear 
interaction. This means diagrams of type shown in 
Figure \ref{fig:img3} exist.
	
The external soft line provides a scale to the collinear loop, 
and indeed, individually, such diagrams are non-vanishing.
Following the labeling in Figure \ref{fig:img3}, $k$, $p$, and $l$ are
soft, collinear, and anticollinear momenta, respectively. 
One can then form the collinear invariant 
$(n_-k)(n_+p) \sim \lambda^2$, resulting in dimensionally 
regulated results proportional to 
$[\mu^2/((n_-k)(n_+p))]^{\epsilon}$. It therefore appears that 
there should be collinear functions already at LP. 
		
However, at LP, the decoupling 
transformation \cite{Bauer:2001yt} 
$\xi_c(z) \to Y_+(z_-) \, \xi_c^{(0)}(z)$, 
$A^\mu_c(z) \to  Y_+(z_-) A^{(0) \mu}_{c}(z)Y^\dagger_+(z_-)$ 
can be applied, 
where the soft Wilson line is defined as 
\bea
	Y_{\pm}\left(x\right)&=&\mathbf{P}
	\exp\left[ig_s\int_{-\infty}^{0}ds\,n_{\mp}
	A_{s}\left(x+sn_{\mp}\right)\right].
	\eea
	Since 
	\be
	\bar{\xi}_c \, in_-D\, \frac{\slashed{n}_{+}}{2}\,\xi_c 
	= \bar{\xi}_c^{(0)} \, in_-D_c^{(0)}\, \frac{\slashed{n}_{+}}{2}\,
	\xi_c^{(0)}\,, 
\ee
this removes all soft-collinear interactions from the LP 
Lagrangian (\ref{eq:LagrangianLP}). It is 
often convenient to define the collinear gauge-invariant 
field $\chi_c = W_c^\dagger \xi_c$ involving the collinear Wilson 
line\footnote{Similar definitions apply to the collinear 
gluon field, and to anticollinear fields with $n_+\leftrightarrow 
n_-$.} 
\bea 
W_{c}\left(x\right)&=&\mathbf{P}
\exp\left[ig_s\int_{-\infty}^{0}ds\,n_+
A_{c}\left(x+sn_+\right)\right],
\eea 
in which case 
\be
\bar{\xi}_c \, in_-D\, \frac{\slashed{n}_{+}}{2}\,\xi_c 
= \bar{\chi}_c^{(0)}\left(in_-\partial + n_- \mathcal{A}_c^{(0)}
\right)\frac{\slashed{n}_{+}}{2}\,\chi_c^{(0)} 
\ee
with $\mathcal{A}_c^\mu = W_c^{\dagger}\left[
i D_c^\mu\,W_c\right]$, and the same conclusion applies. It is 
customary to drop the superscript $(0)$ on the fields after 
the decoupling transformation and we follow this convention below. 
	
Returning to the example diagram above, the decoupling 
transformation implies  that after summing together all 
collinear loop diagrams, they must cancel exactly. 
This is the reason why there are no collinear functions 
at LP.  The fact that the collinear scale shows up in intermediate 
steps of the calculation is a consequence of using Feynman rules 
derived from the SCET Lagrangian before the decoupling 
transformation. Indeed, if one employed the 
decoupled Lagrangian, diagrams like the one
in Figure \ref{fig:img3} would not be present from the beginning.
	
It follows from the absence of soft-collinear interactions at LP 
after the decoupling transformation that the matrix element 
relevant to DY production factorizes at LP into a product of 
(anti)collinear fields and the soft Wilson lines such that it can 
be written as 
\begin{eqnarray}
	\label{eq:2.2}
	\langle X|\bar{\psi}\, \gamma^\rho \psi(0)|A(p_A)B(p_B)\rangle
	&=&
	\int dt\, d\bar{t}\, \widetilde{C}^{A0,A0}(t, \bar{t}\,)\,
	\langle X^{{\rm PDF}}_{\bar{c}}|\bar{\chi}_{\bar{c}}
	\,(\bar{t} n_-)|B(p_B)\rangle  \,
	\gamma_{\perp}^\rho \nonumber\\ 
	&& \hspace*{-3cm}\times\, 
	\langle X^{{\rm PDF}}_{{c}}|
	\chi_{c}\left(tn_+\right)
	|A(p_A)\rangle\,\langle X_s|   {\mathbf{ T}}
	\left( \left[ Y_{-}^\dagger(0) Y_{+}(0) \right]_{}
	\right)|0 \rangle\,.
	\end{eqnarray}
Here $\widetilde{C}^{A0,A0}(t, \bar{t}\,)$ is the short-distance matching 
coefficient of the electromagnetic current to its leading-power 
SCET representation $J^{A0,A0}_{\rho}(t,\bar{t}\,)=\bar{\chi}_{\bar{c}}(\bar{t}n_-)\gamma^{}_{\perp\rho}\chi_c(tn_+)$. Due to threshold kinematics
the final state can only be composed of soft and
$(\bar{c})c$-PDF collinear modes, which are decoupled. Hence, in the
above equation the final state is factorized, 
$\langle X|=\langle X_s| \langle X^{{\rm PDF}}_c|\langle X^{{\rm PDF}}_{\bar{c}}|$.
	
Since soft-collinear interactions are absent and purely 
threshold-collinear loops are scaleless, the matching between 
the collinear field $\chi_{c}$ and the corresponding 
c-PDF field $\chi_c^{\rm{PDF}}$ is trivial: 
the threshold-collinear fields are 
simply identified with the $c$-PDF fields. Technically, 
the matching coefficient (collinear function) is  
a delta function  to all orders in perturbation theory converting
threshold-collinear fields to $c$-PDF fields, that is, 
\begin{eqnarray}
	\chi_c(tn_+) = \int\, du\,\tilde{J}(t,u)\, 
	\chi_c^{\rm{PDF}}(un_+)
\end{eqnarray}
with  $\tilde{J}(t,u) = \delta(t-u)$. Because of this trivial 
relation, LP collinear functions are not discussed in the 
context of LP factorization.
		
After this step, the computation proceeds in the usual manner. 
Squaring the amplitude and summing over the final state gives 
the usual PDFs $f_{a/A}(x_a)$ and $f_{b/B}(x_b)$ from the (anti)collinear matrix elements, and one arrives at
\begin{equation}
	\frac{d\sigma_{\rm DY}}{dQ^2} = 
	\frac{4\pi\alpha_{\rm em}^2}{3 N_c Q^4}
	\sum_{a,b} \int_0^1 dx_a dx_b\,f_{a/A}(x_a)f_{b/B}(x_b)\,
	\hat{\sigma}_{ab}(z)\,.
	\label{eq:dsigsq2}
\end{equation}
The partonic cross section $\hat{\sigma}_{ab}(z)$ factorizes 
into a hard function, originating from squaring the hard matching 
coefficient $\widetilde{C}^{A0,A0}(t, \bar{t}\,)$ in \eqref{eq:2.2}, and a
soft function:
\begin{equation}
	\hat{\sigma}(z) = H(Q^2) \,Q S_{\rm DY}(Q(1-z))\,.
	\label{eq:LPfact}
\end{equation}
The leading power DY soft function is given by 
\cite{Korchemsky:1993uz}
\begin{equation}
	S_{\rm DY}(\Omega) = \int \frac{dx^0}{4\pi}\,e^{i \Omega\, x^0 /2}\,
	\frac{1}{N_c}\,\mbox{Tr} \,
	\langle 0|\mathbf{\bar{T}}(Y^\dagger_+(x^0) Y_-(x^0)) 
	\,\textbf{T}(Y^\dagger_-(0) Y_+(0))
	|0\rangle\,.
	\label{eq:LPsoftfn}
\end{equation}
	
\subsection{Emergence of collinear functions}
	
The analysis becomes more involved when subleading-power
effects are studied. The framework employed here for 
the power-suppressed corrections in SCET was developed in 
\cite{Beneke:2004in,Beneke:2017ztn,Beneke:2018rbh,Beneke:2019kgv}. 
It makes use of collinear gauge-invariant building blocks,
which consist of collinear quark and gluon fields in a
particular collinear direction, and non-local operators with
insertions of terms from the power-suppressed SCET Lagrangian 
to systematically include subleading-power contributions in 
perturbative calculations. In what follows, we use this general framework
to derive power corrections to the LP factorization
formula for DY production at threshold. We find that
the new physical ingredients, the collinear functions, arise from
soft-collinear interactions present in the 
power-suppressed Lagrangian.
These technically appear as a consequence of Lagrangian insertions
in time-ordered product operators.
	
As an illustrative example, we consider the insertion of 
the NLP soft-collinear interaction Lagrangian 
\bea\label{eq:L2xi}
\mathcal{L}^{(2)}_{2\xi}(z) =\frac{1}{2}\, \bar{\chi}_c(z)\,
z^{\mu}_{\perp}\,z^{\nu}_{\perp}\,\Big[i\partial_{\nu}\,in_-\partial\,
\mathcal{B}^+_{\mu}(z_-) \Big]\frac{\slashed{n}_{+}}{2}\,\chi_c(z)
\eea
from \eqref{eq:quarkint}. The decoupling transformation 
has already been performed (and the superscript (0) on the 
collinear gauge-invariant quark field $\chi_c$ dropped), 
and the $\mathcal{B}_{\pm}$ field is a soft building block formed 
by a soft covariant derivative and soft Wilson lines (we also  
define the soft quark building block for completeness)   
\begin{eqnarray}\label{eq:softBB}
	\mathcal{B}_{\pm}^{\mu} &=& Y_{\pm}^{\dagger}
	\left[ i\,D^{\mu}_s\,Y_{\pm}\right] \,,
	\\ {q}^{\pm} &=& Y_{\pm}^{\dagger}\, q_s \,.
	\label{eq:softq} 
\end{eqnarray}  
	
\begin{figure}
\begin{centering}
\includegraphics[width=0.35\textwidth]{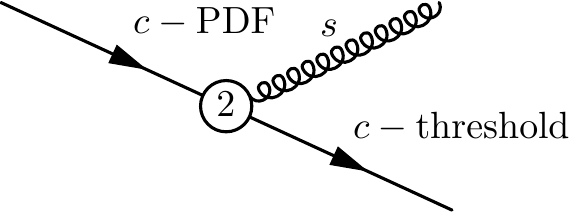}
\par\end{centering}
\caption{\label{fig:illustration}  
Insertion of the power-suppressed Lagrangian 
$\mathcal{L}^{(2)}_{2\xi}$ into a collinear quark line.}
\end{figure}
	
In contrast to LP, the decoupling transformation 
does {\it{not}} remove completely the soft-collinear interactions.
In fact, the insertions of Lagrangian terms appear in non-local
operators with an integral over the position of the insertion, 
\bea\label{eq:TprodExample}
J^{\,T 2}_{c}( t) = i \int d^4z \,\,\mathbf{T}\left[
\chi_c(tn_+) \, \mathcal{L}^{(2)}_{2\xi}(z)\right],
\eea 
where the field $\chi_c(tn_+)$ arises from the LP $J^{A0,A0}$ 
current. See Figure~\ref{fig:illustration} for illustration.
The collinear fields in \eqref{eq:L2xi} depend on all components 
of the $z$ coordinate. 
The soft $\mathcal{B}_{\pm}(z_-)$ field on the other hand 
has dependence only on  $z_-^\mu= (n_+z) \frac{n_-^\mu}{2}$ 
due to multipole expansion, but this dependence links the 
collinear and soft fields and leads to a collinear invariant 
for collinear loop integrals. Concretely, consider the 
DY matrix element with an insertion of the above Lagrangian, 
\begin{eqnarray}\label{1.19}
	\langle X|\bar{\psi} \gamma^\rho \psi(0)|A(p_A)B(p_B)\rangle
	&=&
	\int dt\, d\bar{t}\, \widetilde{C}^{A0,A0}(t, \bar{t}\,)\,
	\langle X^{{\rm PDF}}_{\bar{c}}|\bar{\chi}_{\bar{c},\alpha a}
	(\bar{t} n_-)|B(p_B)\rangle  
	\gamma_{\perp, \alpha \gamma}^\rho \nonumber\\ 
	&& \hspace*{-3.5cm} \times \,i\int d^4z \,\langle X^{{\rm PDF}}_{{c}}|
	\frac{1}{2}z_{\perp}^{\nu} z_{\perp}^{\mu} (in_-\partial_z)^2
	\,\mathbf{T}\left[ \chi_{c,\gamma f}\left(tn_+\right)
	\bar{\chi}_{c}\left(z\right)\textbf{T}^A \frac{\slashed n_+}{2} 
	\chi_{c}\left(z\right)\right]|A(p_A)\rangle \nonumber 
	\\ 
	&& \hspace*{-3.5cm}
	\times\,\langle X_s|   {\mathbf{ T}}
	\left( \left[ Y_{-}^\dagger(0) Y_{+}(0) \right]_{af}
	\frac{i\partial_{\perp}^{\mu}}{i\nm\partial}
	\mathcal{B}^{+A}_{{\perp}\nu}\left(z_{-}\right)
	\right)|0 \rangle\,.\quad
	\end{eqnarray}
Compared to the LP expression \eqref{eq:2.2},
there are extra collinear fields in the $c$-PDF matrix element 
and there is a convolution in $z_-$ between the soft and 
collinear matrix elements. It is precisely the presence of this 
extra convolution, injecting momentum with a soft scaling into 
the collinear matrix element, which induces a scale and leads 
to the emergence of collinear functions. 
The soft matrix element in the last line now contains an 
explicit gauge field insertion in addition to the Wilson lines, 
and will form a part of the {\it{generalized}} soft function. 
The anticollinear matrix element is the same as before, and will 
form part of a parton distribution function (PDF) upon squaring.
	
We now focus on the collinear matrix element, which appears in the 
second line. Due to threshold kinematics the threshold-collinear 
modes are forbidden from entering the final state. At leading power 
in the 
$\Lambda/Q$ expansion, the threshold-collinear fields in the collinear matrix element must be 
integrated out and matched to $c$-PDF mode operators consisting 
of a single quark (or gluon) field, which after squaring the 
amplitude will lead to the standard PDFs. 
A prototype for this matching step is the equation (refined later)
\begin{eqnarray}\label{eq:exampleMatch}
i\int d^4z \,
\,\mathbf{T}\left[ \,\{\psi_{c}(tn_+)\}
\,\mathcal{L}^{(2)}_{c}(z) \right] 
=2\pi \int du \int \,dz_{-}\, \tilde{J}(t,u;z_{-}) 
\,\chi^{\text{\scriptsize PDF}}_{c}(u n_{+})\,,
\end{eqnarray}
where $\mathcal{L}^{(2)}_c$ refers to only the collinear 
pieces of the Lagrangian insertion. 
The perturbative matching coefficient 
$\tilde{J}(t,u;z_{-})$ is the {\it{collinear function}}. 
It contains the collinear physics at the {\it{amplitude}} level. 
We stress once more that it appears first in power-suppressed 
corrections to the DY process.  The above equation 
provides an operator definition of the concept of 
the ``radiative jet 
amplitude''~\cite{DelDuca:1990gz,Bonocore:2015esa,Bonocore:2016awd}. 
The matching should be performed in the presence
of soft structures which, acting as projectors, define independent
collinear functions. We give formal definitions in 
Sec.~\ref{sec:formalDefinitions} below. 
	
For the above example, we calculate the tree-level contribution to 
the collinear terms in the second line of \eqref{1.19}.
To this purpose it is convenient to introduce the 
momentum-space operator
\begin{align}\label{eq:jmunu}
	\mathcal{J}^{\mu\nu,\,A}_{\gamma,f}
	\left(n_+p,\omega\right) \equiv &\int dt \, e^{i\, (n_+p)\, t}\,  i\int d^4 z\, e^{i \omega (n_+z)/2} \, \nonumber\\
	&\times \frac{1}{2}z_{\perp}^{\nu}
	z_{\perp}^{\mu} (in_-\partial_z)^2 \,\mathbf{T}\left[ \chi_{c,\gamma f}
	\left(tn_+\right)\bar{\chi}_{c}\left(z\right) 
	\textbf{T}^A   \frac{\slashed n_+}{2}
	\chi_{c}\left(z\right)\right],
\end{align}
which contains only collinear fields. To calculate the perturbative 
threshold-collinear matching coefficient, we consider the 
partonic analogue of the matrix element in  \eqref{1.19}, 
which amounts to replacing the incoming hadron by an incoming 
quark and the PDF-collinear final state by the vacuum. Hence, 
we calculate
\begin{align}
	\langle0| \mathcal{J}^{\mu\nu,\,A}_{\gamma,f}&
	\left(n_+q_a,\omega\right)|q(q)_e\rangle= \int dt e^{i(n_+q_a)\,t}\,i
	\int d^4z \left[(in_-\partial_z)^2 
	e^{i\omega \,({n_+z})/{2}}\right] \, \nonumber \\
	& \times \frac{1}{2}z_{\perp}^{\nu}
	z_{\perp}^{\mu} \,\langle 0| \mathbf{T}\left[ \chi_{c,\gamma f}
	\left(tn_+\right)\bar{\chi}_{c}\left(z\right) 
	\textbf{T}^A   \frac{\slashed n_+}{2}
	\chi_{c}\left(z\right)\right]|q(q)_e\rangle \,
	\nonumber\\=& -\frac{1}{2} i\omega^2\,(2\pi)
	\int \frac{d^4k}{(2\pi)^4} \,
	\delta\left(n_+q_a- n_+k\right) \int d^4z  
	\left[  \frac{\partial}{\partial k_{\perp \nu}}
	\frac{\partial}{\partial k_{\perp \mu}}
	\frac{i(n_+k)}{k^2}\right] 
	\nonumber\\ & \times  e^{i\omega \,({n_+z})/{2}}
	e^{ik\cdot z}
	\,\textbf{T}^A_{fe}\,
	{u}_{c,\gamma}(q)\,e^{-iz\cdot q} \,,
\end{align}
where we have contracted two of the collinear fields to form 
the collinear quark propagator, performed the 
$z$-derivatives and the integral over $t$, and used
\begin{equation}
	{\chi}_{c,\gamma d}(z)| q(q)_e\rangle =
	\delta_{de}\,{u}_{c,\gamma}(q)e^{-iz\cdot q}\,|0\rangle 
	\end{equation}
	for the incoming quark with fundamental colour index $e$.
	 The $z$-integral can next be performed,
	yielding delta functions which remove the remaining integral 
	over the momentum $k$. Then we find 
\begin{eqnarray}
\langle 0| \mathcal{J}^{\mu\nu,\,A}_{\gamma,f}
\left(n_+q_a,\omega\right)|q(q)_e\rangle&=&
(2\pi)\underbrace{ \delta\left(n_+q_a- n_+q \right) 
\,\frac{-g^{\nu \mu}_{\perp}}{(n_+q)}  \,
\textbf{T}^A_{fe}\, \delta_{\gamma \beta}}_{\equiv\,
J^{\mu\nu,\,A}_{2\xi,\gamma \beta,fe}
\left(n_+q_a,n_+q;\,\omega\right)}\,u_{c,\beta}(q) \,. \quad
\end{eqnarray}
	We have underbraced the matching coefficient which defines the 
	tree-level collinear function. The appearance of collinear 
	functions beyond LP is generic and constitutes a key concept in 
	NLP investigations.  In Sec.~\ref{sec:CollFuncsCalc} we will 
	calculate the one-loop corrections to these functions.

\subsection{Collinear matching: formal definitions}
\label{sec:formalDefinitions}
	
\begin{figure}[t]
\begin{centering}
\includegraphics[width=0.45\textwidth]{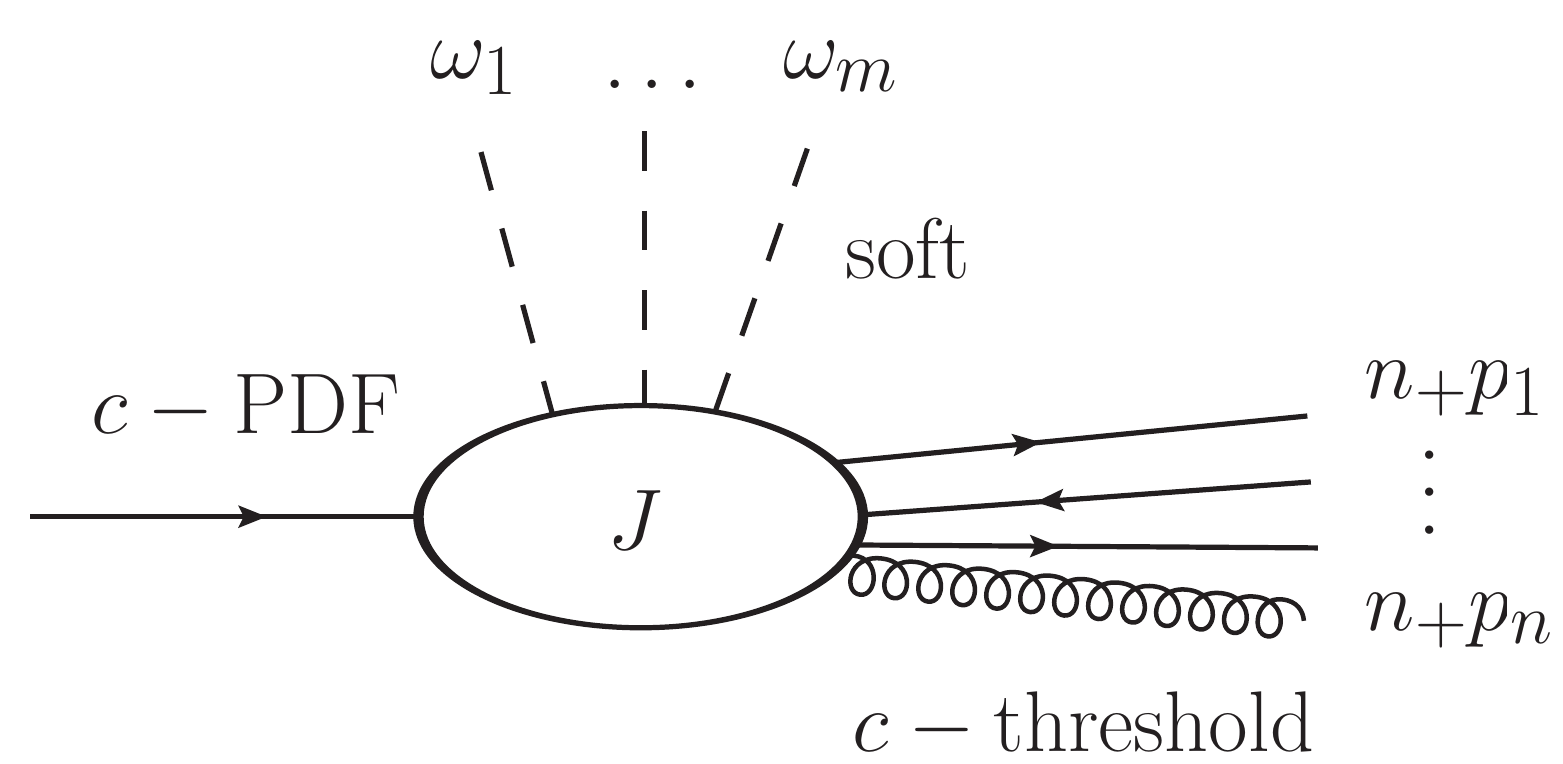}
\par\end{centering}
\caption{\label{fig:img4}   		
A momentum-space pictorial representation 
of the matching equation \eqref{eq:genMatch}. The oval labelled $J$ 
is a collinear function. The $\omega_i$ variables are conjugate
to the respective positions of the insertions of subleading-power 
Lagrangians. Many threshold-collinear fields 
may join to the (possibly power-suppressed) SCET currents of the 
$A,B,C...$ type \cite{Beneke:2017ztn}, but there is only a single $c$-PDF 
field at leading twist in the $\Lambda/Q$ expansion.}
\end{figure}

The general collinear matching equation, suppressing indices,
is given by 
\begin{eqnarray}
\label{eq:genMatch}
&& i^{\,m}\int \{d^dz_j\}\,\mathbf{T}\left[ \,\{\psi_{c}(t_kn_+)\}
\times\left\{\mathcal{L}^{(l)}_{}(z_j) \right\}\right] 
\nonumber \\ 
&&\hspace{1.0cm}
	= 2\pi \sum_i \int du \int \{dz_{j-}\}\,
	\tilde{J}_{i}\left(\{t_k\},u;\{z_{j-}\} \right)
	\,\chi^{\text{\scriptsize PDF}}_{c}(u n_{+})
	\,\mathfrak{s}_{i}(\{z_{j-}\})\,,\qquad
	\end{eqnarray}
where $\{d^dz_j\} =\prod^m_{j=0} \, d^dz_j$ and  
$\{dz_{j-}\}=  \prod^m_{j=0} \frac{dn_+z_j}{2}$.  
$\{z_{j-}\}$  denotes the set of $m$ positions
at which the soft building block insertions are located.
	$\left\{\mathcal{L}^{(l)}_{}(z_j) \right\}$ is a set of 
	$m$ $\mathcal{O}(\lambda^l)$-suppressed Lagrangian insertions. 
	$\{\psi_c(t_kn_+)\}$ denotes a set of
	$n$ fields chosen from the elementary 
	collinear-gauge-invariant
	collinear building blocks each dependent on  one variable
	from the $n$-sized set $\{t_k\}$. Here
	\begin{equation}
	\label{eq:Vccsc}
	\psi_{i}(t_in_{i+}) \in  
	\,\,\left\{ \begin{array}{ll}
	\displaystyle  \chi_i(t_in_{i+})\equiv W_i^{\dagger}
	\,\xi_i  & \quad{\rm{collinear\,quark} } \\[0.3cm]
	\displaystyle  \mathcal{A}^{\mu}_{i\perp}(t_in_{i+})
	\equiv W_i^{\dagger}\left[i\,D^{\mu}_{i\perp}\,W_i \right]
	& \quad{\rm{collinear\,gluon} }
	\end{array}\right.
	\end{equation}
for the collinear quark and gluon field in $i$-th direction, 
respectively. 
Furthermore, $\mathfrak{s}_{i}(\,\{z_{j-}\})$ is a soft 
operator and the sum over $i$ runs over a basis of soft structures, 
\begin{eqnarray}\label{eq:softSet}
	\mathfrak{s}_{i}(\{z_{j-}\})\hspace{+0.1cm} 
	&\in& \hspace{+0.15cm} \bigg\{
	\frac{i\partial_{\perp}^{\mu}}{in_-\partial}
	\mathcal{B}^+_{\mu_\perp}(z_{1-}) ,
	\frac{i\partial_{[\mu_{\perp}}}{in_-\partial}
	\mathcal{B}^+_{\nu_\perp]}(z_{1-}) , \\ \nn  & & \hspace{0cm}
	\frac{1}{(in_-\partial)^2}\left[ 
	\mathcal{B}^{+\,\mu_\perp}(z_{1-}),
	\left[in_-\partial\mathcal{B}^+_{\mu_\perp}(z_{1-})
	\right]  \right]\hspace{+0.15cm},
	\frac{1}{(in_-\partial)} \big[  \mathcal{B
	}^+_{\mu_\perp}(z_{1-}) ,\mathcal{B}^+_{
	\nu_\perp}(z_{1-})\,\big]
,\\ \nn  & & \hspace{0cm}  \mathcal{B}^+_{\mu_\perp}
(z_{1-})\mathcal{B	}^+_{\nu_\perp}(z_{2-}),
\frac{1}{(in_-\partial_{z_1})(in_-\partial_{z_2})} q_{+\sigma }(
z_{1-})\bar{q}_{+\lambda }(z_{2-}),\ldots\,
\bigg\}\,.
\end{eqnarray}
Here $[\mu,\nu]$ denotes antisymmetrization $\mu\nu-\nu\mu$,
and the ellipses indicate all possible {\it{independent}} soft 
structures\footnote{
The list (\ref{eq:softSet}) is still partially redundant. For 
later convenience, we have kept the two-gluon soft structures 
in the second line, although they can be considered as special 
cases of the bi-local structure $\mathcal{B}^+_{\mu_\perp}
(z_{1-})\mathcal{B}^+_{\nu_\perp}(z_{2-})$.}  
after utilizing the equation of motion
\begin{eqnarray}\label{eq:eom}
n_+\mathcal{B}^{+}(z_-) &=& -2
\frac{i\partial^{\mu}_{\perp}}{in_{-}\partial}  
\mathcal{B}_{\mu_\perp}^{+}(z_-)
-2 \frac{1}{(in_-\partial)^2}
\left[\mathcal{B}^{+\,\mu_\perp}_{}(z_-),\left[in_-\partial
\mathcal{B}^+_{\mu_\perp}(z_-) \right] \right]\nonumber 
\\ &&
-2 \frac{g_s^2}{(in_-\partial)^2}\textbf{T}^A\,
\bar{q}_+(z_-)\textbf{T}^A\slashed{n}_- q_+(z_-).
\end{eqnarray}
Eq.~(\ref{eq:genMatch}) is a formal all-order and all-power matching 
equation, and it will be used extensively in the following sections. 
A graphical illustration is given in Figure~\ref{fig:img4}. 

\section{Factorization near threshold}
\label{sec:powerfact}

We now turn to the formal derivation of the factorization 
formula beyond leading power. We recall that 
the SCET derivation of factorization at LP
\cite{Becher:2007ty} involves matching the coupling to the virtual 
photon to the LP SCET current,  
\bea\label{eq:LPmatching}
\bar{\psi} \gamma_\rho \psi(0) = \int dt\, 
d\bar{t}\,\widetilde{C}^{A0,A0}(t,\bar{t}\,) \,
J_\rho^{A0,A0}(t,\bar{t}\,)
\eea
where (prior to use of decoupling transformation
\cite{Bauer:2001yt})
\bea\label{eq:LPcurrent}
J_\rho^{A0,A0}(t,\bar{t}\,) =  \bar{\chi}_{\bar{c}} 
(\bar{t} n_-)\gamma_{\perp\rho} \chi_c(t n_+ )\,.
\eea
The matching coefficient is related to the corresponding 
momentum-space coefficient by
\bea\label{hardWilsonCoeff}
C^{A0,A0}(n_+p, n_-\bar{p}\,) = \int dt\, d\bar{t}\, 
e^{-i \, (n_+ p)\,t  -i \,(n_-\bar{p})\,\bar{t}\ } \,
\widetilde{C}^{A0,A0}(t,\bar{t}\,) \,.
\eea
The fields, denoted by $\chi_c$ 
(and $\mathcal{A}^{\mu}_{c\perp}$
further on), are the collinear-gauge-invariant collinear 
quark (and gluon) fields out of which building blocks
of general $N$-jet operators are formed  
\cite{Beneke:2017ztn}. To derive the factorization 
formula valid at subleading powers, the matching equation 
 (\ref{eq:LPmatching}) must be modified to include 
higher orders in the $(1-z)$ expansion. In general, this 
is accomplished through inclusion of all possible combinations
of power-suppressed currents and subleading Lagrangian 
insertions. We obtain such general factorization formula in 
the following, before specializing to the case of NLP in  
later sections where we provide explicit results for the
objects  appearing in the factorization formula. 

\subsection{Factorization at general subleading powers}
\label{sec:factGeneralPows}

Omitting the index structure for clarity, the general,
all power, hard matching of the vector current is given by 
\bea \label{eq:subleadingmatching}
\bar{\psi} \gamma_\rho \psi(0)
= \sum_{m_1, {m_2} }\, 
\int \{dt_k\} \, 
\{d\bar{t}_{\bar{k}}\}  \,\widetilde{C}^{\,m_1,m_2}
\left(\,\{t_k\} , \,\{\bar{t}_{\bar{k}}\}\right)\, J_s(0)
\,J_\rho^{\,m_1,m_2}\left(\,\{t_k\} , \,\{\bar{t}_{\bar{k}}\}
\right)\,.\quad
\eea
The sizes of the sets $\{dt_k\}$, $\{t_k\}$ 
(and the barred sets for the anticollinear direction)
for each term on the right-hand side of the matching equation 
depend on the type of current present in that term. Inclusion of all 
contributions  is accounted for by the sum over indices
$m_1$ and $m_2$, which label the basis  of SCET operators
(and their corresponding short-distance matching 
coefficients $\widetilde{C}^{\,m_1m_2} $)
depending   on the content of its building blocks
using the formalism and notation developed
in \cite{Beneke:2017ztn, Beneke:2018rbh}, for example 
$m_{1,2} = A0$ in the LP case  \eqref{eq:LPmatching}. 
Explicitly, the DY 
process consists of a collinear and an anticollinear direction
both of which can contain sources of power suppression, hence
the SCET currents are built as follows  
\bea \label{eq:DYcurrent}
J_\rho^{\,m_1,m_2}\left(\,\{t_k\} , \,\{\bar{t}_{\bar{k}}\}
\right) =  J^{\,m_1}_{\bar{c}}\left(\{\bar{t}_{\bar{k}}\}
\right)\,\Gamma_{\rho}^{m_1,m_2}\,J^{\,m_2}_c\left(\{t_k\}
\right)\,.
\eea 
As mentioned above, $J^{\,m_2}_c\left(\,\{t_k\}
\right)$ is constructed using collinear-gauge-invariant 
collinear building blocks given in \eqref{eq:Vccsc}.  
In this general construction, the letter $A$, $B$, $C$ etc.
used to label the operator denotes the  
number of fields in a particular collinear direction, 
and the number 0, 1, 2 etc. denotes the overall power of $\lambda$ 
of the current with respect to the LP, which is labelled
$0$. Hence, the $A$-type current consists of 
one field in one direction and derivatives of that field,
the $B$-type current contains two fields and their derivatives,
and so on.  $\Gamma_{\rho}^{m_1,m_2}$ in \eqref{eq:DYcurrent}
stands for the appropriate spinor and Lorentz structure 
of the operator. For 
instance, in (\ref{eq:LPcurrent}) $\Gamma^{\,A0,A0}_{\rho}=
\gamma_{\perp\rho}\,$. At $\mathcal{O}(\lambda)$, 
$\Gamma^{\,A0,A1}_{\rho}=   n_{+\rho}\,$ etc. 

In addition, there exist time-ordered 
products of currents  with subleading terms 
in the SCET Lagrangian. These are denoted by $Tn$, for example  
\bea\label{eq:Tprod}
J^{\,T2}_{c}( t\,) 
= i \int d^dz \,\,\mathbf{T}\left[  
J^{A0}_c(t) \, 
\mathcal{L}^{(2)}_{}(z)\right] \,
\eea
at $\mathcal{O}(\lambda^2)$,  where $\mathcal{L}^{(2)}= 
\mathcal{L}^{(2)}_{\xi} +\mathcal{L}^{(2)}_{\xi q}
+\mathcal{L}^{(2)}_{{\rm{YM}}}$ are the 
power-suppressed terms in the SCET Lagrangian 
 \cite{Beneke:2002ni}. As discussed in
Sec.~\ref{sec:collFuncns}, the 
decoupling transformation does not remove 
the soft-collinear interactions in the subleading 
SCET Lagrangian and the injection of soft momentum
into collinear loops is necessary to form non-vanishing 
collinear functions. For this reason the time-ordered
product terms are crucial ingredients of the 
factorization of the DY process at NLP.  
To yield a non-zero subleading power amplitude 
at least one leg must have such time-ordered product.
The other leg can then contribute to power suppression 
through power-suppressed currents of $A$, $B$, $C$ etc. 
type or another operator containing a time-ordered 
product. Starting from $\mathcal{O}\left(\lambda^3\right)$, 
in addition to collinear fields, the current operator can
contain purely soft building blocks \cite{Beneke:2017ztn},
denoted by $J_s(0)$ here. 

As discussed above, only the $(\bar{c})$c-PDF modes can be 
radiated into the final state. Eq.~\eqref{eq:DYcurrent}, however, 
contains threshold (anti)collinear modes, hence a second 
collinear matching onto a $(\bar{c})$c-PDF field 
must be performed using \eqref{eq:genMatch}.
The first line of \eqref{eq:genMatch} corresponds to the 
time-ordered product of the collinear part 
$J^{\,m_2}_c\left(\{t_k\}\right)$ 
of a general SCET operator~\eqref{eq:DYcurrent} 
with a subleading-power Lagrangian, 
hence applying \eqref{eq:genMatch} to the collinear and 
anticollinear sectors, we obtain the DY matrix 
element of \eqref{eq:subleadingmatching} in the form 
\bea
\label{eq:genrl_2}
\langle X|\bar{\psi} \gamma_\rho \psi(0)|A(p_A)B(p_B)\rangle
&=&\sum_{  	m_1, m_2}\,\sum_{i,\bar{i}}  \, \int\, \{dt_k\} \,
\{d\bar{t}_{\bar{k}}\}  \, \widetilde{C}^{ \, 	m_1, m_2}
\left(\,\{t_k\},\,\{\bar{t}_{\bar{k}}\} \right) \nn\\
&& \hspace*{-4cm}\times\,
2\pi \int d\bar{u} \int \{d\bar{z}_{\,\bar{j}+}\}\,
\bar{\tilde{J} }_{\,\bar{i}}^{\,m_1}\left(\,
\{\bar{t}_{\bar{k}}\},\bar{u};
\{\bar{z}_{\,\bar{j}+}\} \right) \, \langle 
X^{{\rm PDF}}_{\bar{c}}|\bar{\chi}^{{\rm PDF}}_{\bar{c}}
(\bar{u} n_-)|B(p_B)\rangle  \nn\\ 
&& \hspace*{-4cm}\times \,
2\pi  \int du   \int \{dz_{j-}\}\,
\tilde{J}^{\,m_2}_{i}\left(\,\{t_k\},u;\{z_{j-}\} \right)
\, \langle X^{{\rm PDF}}_{{c}}| \chi^{{\rm PDF}}_{c}(un_+)
|A(p_A)\rangle  \nonumber   \\ 
&& \hspace*{-4cm}\times  \,
\Gamma^{m_1,m_2}_{\rho}\,\langle X_s| \, {\mathbf{ T}}\left(
\bar{\mathfrak{s}}_{\,\bar{i}}\,(\,\{\bar{z}_{\,\bar{j}+}\})\,
\big[ Y_{-}^\dagger\, J_s \, Y_{+} \big](0)\,
\mathfrak{s}_{i}\,(\{z_{j-}\})
\right)|0 \rangle   \,.
\eea 
In this equation, the index $k$ $(\bar{k})$ counts the  
number of building block fields in the collinear (anticollinear)
direction within each current, and we sum over all currents. The 
index $j$ $(\,\bar{j}\,)$ refers to the number of Lagrangian 
insertions in the collinear (anticollinear) sector, where we 
also sum over all possibilities. 
Note that here, and throughout the text, the barred notation 
$(\,\bar{}\,)$ refers to the anticollinear direction, and the
tilde $(\,\,\widetilde{}\,\,)$ is used to denote the quantities with dependence
on the position arguments such as $t, \bar{t}$ and so on. This 
notation, also used for indices, is meant to facilitate keeping
track of the origin of the various contributions to the factorization
theorem.
 As discussed in 
Sec.~\ref{sec:collFuncns} at least one Lagrangian insertion is 
necessary to yield a non-vanishing subleading-power amplitude. 
Finally, the $\tilde{J}_{i}$ ($\bar{\tilde{J} }_{\,\bar{i}}$) 
are the (anti)collinear functions and 
$ \mathfrak{s}_{i}(\{z_{j-}\})$ 
($\bar{\mathfrak{s}}_{\,\bar{i}}\,(\,\{\bar{z}_{\,\bar{j}+}\}))$ 
are made up of explicit $\mathcal{B}^+$, $q_+$ 
($\mathcal{B}^-$, $q_-$) field products and their 
derivatives, as indicated in \eqref{eq:softSet}.\footnote{To be 
precise, at leading power and when only one time-ordered 
product is present, both or one of the collinear functions 
are trivial and the corresponding soft structure is unity.} 
The further derivation of the general factorization formula 
follows closely the steps presented in \cite{Beneke:2018gvs} 
for the derivation of the NLP leading logarithmic resummation.
Suppressing the $m_{1,2}$ labels, the hard 
matching coefficients and c-PDF fields are Fourier-transformed 
using  
\bea
\widetilde{C}\left(\{t_k\},\{ 
\bar{t}_{\bar{k}} \}\, \right) = 
\int \left\{\frac{dn_+p_{k} }{2\pi} \right\}
\left\{ \frac{dn_-\bar{p}_{\,\bar{k}}}{2\pi} \right\}  
\, e^{i  \,( n_+ p_k )\, t_k}
e^{ i  \,(n_- \bar{p}_{\,\bar{k}})\,\bar{t}_{\bar{k}}} 
\,C(\{ n_+p_k\}, \{ n_-\bar{p}_{\,\bar{k}}\})\quad
\eea
and 
\begin{eqnarray}
\label{eq:5.5c}
\chi^{{\rm PDF}}_{c}(un_+) = \int \frac{d(n_+p_a)}{2\pi}
e^{-i(n_+p_a)\,u}\,\hat{\chi}^{{\rm PDF}}_{c}(n_+p_a)\,,
\end{eqnarray}
respectively. For the collinear functions we define 
($z_{j-} = n_+ z_j/2$)
\bea\label{eq:colFT}
\int \{dt_k\}\int du \,  \tilde{J}^{\,m_2}_{\,{i}}
\left(\,\{t_k\},u;\{\, z_{j-}\}\right)
\, e^{i  \,( n_+ p_k )\,t_k}\,e^{-i(n_+p_a)\,u}
\nonumber \\= \int \left\{
\frac{d\omega_j}{2\pi}\right\}\,
\,  e^{-i \omega_j\, z_{j-}} \, 
{J}^{\,m_2}_{\,{i}}\left(\{ n_+p_k \} ,n_+p_a;
\{\omega_{j}\}\right)\,.
\eea 
The set $\{\omega_{j}\}$ is a set of variables with 
 soft scaling conjugate to $\{\, z_{j-}\}$, and 
$\left\{\frac{d\omega_j}{2\pi}\right\}= 
\frac{d\omega_1}{2\pi}\times...\times
\frac{d\omega_m}{2\pi}$. Einstein's summation 
convention is implied in the exponents. Equations analogous 
to \eqref{eq:5.5c} and \eqref{eq:colFT} are used for the 
anticollinear direction.  Implementing these in 
\eqref{eq:genrl_2} we arrive at  generalized version of
Eq.~(3.16) of \cite{Beneke:2018gvs}: 
\bea\label{eq:5.10c}
\langle X|\bar{\psi} \gamma_\rho \psi(0)|A(p_A)B(p_B)\rangle
&=& \sum_{	m_1, m_2} \,\sum_{i,\bar{i}}\,
\int \left\{\frac{dn_+p_{k} }{2\pi} \right\}
\, \left\{ \frac{dn_-\bar{p}_{\,\bar{k}}}{2\pi} \right\}
\nonumber \\ 
&& \hspace{-3cm} \times \int \,d(n_+p_a) \, 
\,d(n_-p_b)\, 
C^{\,m_1,m_2}(\{ n_+p_k\}, \,\{ n_-\bar{p}_{\bar{k}}\})
\nonumber \\ 
&& \hspace{-3cm}\times \, \int 
\left\{ \frac{d\bar{\omega}_{\,\bar{j}}}{2\pi}\right\} 
\, \bar{J}^{\,m_1}_{\,\bar{i}}
\left(\, \{ n_-\bar{p}_{\bar{k}}\},-n_-p_b;\,
\{ \bar{\omega}_{\,\bar{j}}\}\right) \,
\,\langle X^{{\rm PDF}}_{\bar{c}}|
\hat{\bar{\chi}}^{{\rm PDF}}_{\bar{c}}(n_-p_b)|B(p_B)\rangle 
\nonumber\\ 
&& \hspace{-3cm}\times \,  \int   
\left\{\frac{d\omega_j}{2\pi}\right\}
\, {J}^{\,m_2}_{\,{i}}\left(\,\{ n_+p_k \} ,n_+p_a;\,
\{\omega_{j}\}\right)  \,
\langle X^{{\rm PDF}}_{{c}}| \hat{\chi}^{{\rm PDF}}_{c}(n_+p_a)
|A(p_A)\rangle  \nonumber \\ 
&& \hspace{-2cm} \times  \,
\Gamma^{m_1,m_2}_{\rho} \int \{d\bar{z}_{\,\bar{j}+}\,\}
\int \{dz_{j-}\}\,e^{-i  \bar{\omega}_{\,\bar{j}}
	\,\bar{z}_{\,\bar{j}+}}\,
\, e^{-i \omega_j \,z_{j-}}\nonumber \\ 
&&  \hspace{-2cm} \times\,\langle X_s| \,  {\mathbf{ T}}\left(
\bar{\mathfrak{s}}_{\,\,\bar{i}}\,(\{
\bar{z}_{\,\bar{j}+}\})\left[ Y_{-}^\dagger\, J_s\, Y_{+} \right](0)\,
\mathfrak{s}_{i}(\{z_{j-}\})
\right)|0 \rangle   \,.
\eea  
For brevity, we define the coefficient function
\begin{eqnarray}
D^{\,m_1,m_2}_{\,{i}\,\bar{i}\,\,\rho}\,(n_+p_a, -n_-{p}_b;\{
\omega_j\}, \{\bar{\omega}_{\,\bar{j}}\}) &=& (2\pi)^2  
\int \left\{\frac{dn_+p_{k} }{2\pi} \right\}
\left\{ \frac{dn_-\bar{p}_{\,\bar{k}}}{2\pi} \right\} 
\nonumber \\ 
&& \hspace{-3cm} \times \,C^{\,m_1,m_2}(\{ n_+p_k\}
, \{ n_-\bar{p}_{\,\bar{k}}\})\,\bar{J}^{\,m_1}_{\,\bar{i}}
(\{ n_-\bar{p}_{\,\bar{k}}\},-\,n_-p_b;
\{ \bar{\omega}_{\,\bar{j}}\}) \nonumber 
\\[0.1cm] 
&& \hspace{-3cm} \times  \,\Gamma^{\,m_1,m_2}_{\rho} \,\,
{J}^{\,m_2}_{\,{i}}(\{ n_+p_k \} ,n_+p_a;\{\omega_{j}\}) 
\label{eq:Dcoeff}
\end{eqnarray} 
that contains both, the hard and collinear matching functions 
at the amplitude level. 

The next step in the derivation of the factorization formula is 
to square the amplitude, which gives the hadronic tensor
$W_{\mu\rho}$ defined below. Combined with the transverse lepton
tensor  belonging to the final-state lepton pair, for which the
phase-space integrals are computed  in $d$ 
dimensions, we obtain an expression for the cross section 
\begin{equation}\label{eq:comb_lept_2bc}
d\sigma = \frac{4\pi\alpha_{{\rm{em}}}^2}{3s\,q^2}
\frac{d^{d}q}{(2\pi)^d} \,\big(-g^{\mu\rho}W_{\mu\rho}\big)\,,
\end{equation}
where
\begin{eqnarray}\label{eq:ampTrans}
g^{\mu\rho} W_{\mu\rho} &=& \int d^dx \,e^{-iq\cdot x}\,
\langle A(p_A)B(p_B)|J^{\dagger\,\rho}(x)J_{\rho}(0)|A(p_A)B(p_B)
\rangle \nonumber \\ &=& \sum_X \, 
(2 \pi)^d \delta^{(d)}\left(p_A +p_B-q-p_{X_s}-p_{X^{\rm{PDF}}_{c}}
-p_{X^{\rm{PDF}}_{\bar{c}}}\right)\nonumber \\[-1ex]
&&\hspace{0cm}
\times\,\langle A(p_A)B(p_B)|
J^{\dagger}_{\rho}(0)| X\rangle 
\langle X|J^{\rho}(0)|A(p_A)B(p_B)\rangle \,.
\end{eqnarray}
At this point we transform the c-PDF 
fields back to coordinate space and use the standard 
definition 
\begin{eqnarray}
\langle A(p_A) |{\bar{\chi}}^{{\rm PDF}}_{c, \eta i}
(x+g' n_+ ){\chi}^{{\rm PDF}}_{c, \beta b}(g n_+ )
|A(p_A)\rangle && \nn \\&& \hspace{-5cm}= 
\frac{\delta_{bi}}{N_c}
\left(\frac{\slashed{n}_-}{4}\right)_{\beta\eta}(n_+p_A)
\int_0^1 dx_a\,e^{i(x+g'n_+-gn_+)\cdot p_A x_a}\,f_{a/A}(x_a)
\quad
\end{eqnarray}
for the PDF. 
After performing the integrations over $n_+p_a$, $n_-p_b$ 
and some further manipulations, we extract the convolution 
with the PDFs from the hadronic DY spectrum \eqref{eq:dsigsq2},
and obtain the expression 
\begin{eqnarray}
\hat{\sigma}&=& 
\sum_{ \substack{ 
		m'_1, {m}'_2\\ m_1, m_2}}
\, \sum_{\substack{i',\bar{i}'\\i,\bar{i}}}\,
\, \int 
\bigg\{\frac{d\bar{\omega}'_{\,\bar{j}'}}{2\pi}\bigg\}
\left\{\frac{d\omega'_{j'}}{2\pi}\right\}\,
\left\{ \frac{d\bar{\omega}_{\,\bar{j}}}{2\pi}\right\}
\left\{\frac{d\omega_j}{2\pi}\right\}  \nonumber \\ 
&& \hspace{0cm} \times\,(-Q^2) \,\bigg[ 
\left(\frac{\slashed{n}_-}{4}\right)
D^{*\,m'_1,m'_2\,\rho}_{\,{i}'\,\bar{i}'}\,(x_an_+p_A, x_bn_-{
	p}_B;\,\{ \omega'_{j'}\}, \,\{\bar{\omega}'_{\,\bar{j}'}\})  
\nonumber \\ 
&& \hspace{2cm} \times\,\left(\frac{\slashed{n}_+}{4}\right)  
D^{\,m_1,m_2}_{\,{i}\,\bar{i}\,\,\rho}\,(x_an_+p_A, x_b
n_-{p}_B; \{\omega_j\}, \,\{\bar{\omega}_{\,\bar{j}} \}) \bigg]
\nonumber\\ 
&& \hspace{0cm}  \times 
\int \frac{d^{d-1}\vec{q}}{(2\pi)^{d-1}\,2\sqrt{Q^2+\vec{q}^{\,\,2}}}
\,\frac{1}{2\pi}   
\int d^dx \,e^{i( x_ap_A+x_b\,p_B-q)\cdot x}\,
\nonumber\\
&&\hspace*{1cm}\times\,
\widetilde{S}_{\,{i}\,\bar{i}\,{i}'\,\bar{i}'}(x;
\{\omega_j\},\{\bar{\omega}_{\,\bar{j}}\},
\{{\omega}'_{{j}'}\},\{\bar{\omega}'_{\,\bar{j}'}\}) 
\label{eq:allpowerresult}
\end{eqnarray}
for the $q\bar{q}$-induced partonic cross section near threshold including 
power corrections in $(1-z)$ in the most general form. 
We recall that barred notation refers to the anticollinear 
direction, and the tilde denotes objects which depend on position-space arguments. 
Contributions to the factorization formula from the 
complex conjugate amplitude are marked here and throughout the text 
with a prime $(\,'\,)$ symbol.
This notation persists in the indices and is used in combination with each other,
such that $\bar{i}'$ refers to contribution from the anticollinear  part of the complex 
conjugate amplitude.
In the last line we introduced the generalized multi-local soft
function, 
$\widetilde{S}_{\,{i}\,\bar{i}\,{i}'\,\bar{i}'}(x;
\{\omega_j\},\{\bar{\omega}_{\,\bar{j}}\},
\,\{{\omega}'_{{j}'}\},\{\bar{\omega}'_{\,\bar{j}'}\})$, 
defined as
\begin{eqnarray}
&&\widetilde{S}_{\,{i}\,\bar{i}\,{i}'\,\bar{i}'}
(x; \{\omega_j\},\{\bar{\omega}_{\,\bar{j}}\},\{{\omega}'_{{j}'}\},
\{\bar{\omega}'_{\,\bar{j}'}\}) 
\nonumber \\[0.1cm] 
&& \hspace{1cm} = 
\int \{ d\bar{z}'_{\,\bar{j}'+}\} \int \{dz'_{j'-}\}
\int \{d\bar{z}_{\,\bar{j}+}\}
\int \{dz_{j-}\}\,\, 
e^{+i \bar{\omega}'_{\,\bar{j}'}
	\bar{z}'_{\,\bar{j}'+}}
\,e^{+i\omega'_{j'} z'_{j'-}}\,
\,e^{-i  \bar{\omega}_{\,\bar{j}} \bar{z}_{\,\bar{j}+}}\,
\, e^{-i\omega_j z_{j-}}
\nonumber   \\  && \hspace{2cm} 
\times \,\frac{ 1  }{N_c}\,{\rm{Tr}}\,
\langle 0|   \bar{\mathbf{ T}}\left(
\bar{\mathfrak{s}}'_{i'}\,(\,\{x+z'_{j'-}\})\,
\Big[ Y_{+}^\dagger\,J^{\dagger}_s\, Y_{-} \Big](x)\,
{\mathfrak{s}}'_{\,\bar{i}'}\,(\{x+ \bar{z}'_{\,\bar{j}'+}\})  \right)
\nonumber\\ 
&& \hspace{2cm}  \times \,{\mathbf{ T}}\left(
\bar{\mathfrak{s}}_{\,\bar{i}}\,(\{\bar{z}_{\,\bar{j}+}\})
\Big[ Y_{-}^\dagger\,J_s\,Y_{+} \Big](0)\,
\mathfrak{s}_{i}\,(\{z_{j-}\})
\right)|0 \rangle \,.
\label{eq:3.17}
\end{eqnarray}
This concludes the derivation of the general formula 
for the DY cross section near threshold including power corrections. Note that these
results were stated in Eqs.~(2.1) and (2.2) of
\cite{Beneke:2018gvs} without details, which are given here.

\subsection{Factorization at NLP}
\label{sec:NLPfact}

We next focus on the 
next-to-leading power effects where certain simplifications 
in the general formula (\ref{eq:allpowerresult}) can be made. 
We first note that since the $\omega$ variables are connected
to the soft emissions from collinear functions, and therefore 
come from insertions of subleading-power Lagrangians in 
a time-ordered product, at NLP their total number is 
highly constrained. On the one hand, there must be at least 
one $\omega$ present due to the fact that at
least one time-ordered product operator must appear
in the SCET amplitude in order to provide a threshold-collinear 
scale and not lead to a trivial null result,
as explained earlier in the text. On the other hand, the 
total power suppression at NLP is $\mathcal{O}(\lambda^2)$,
which means that  there can be at most two 
separate $\omega$ variables which correspond to two
$\mathcal{L}^{(1)}$ insertions, each contributing 
$\mathcal{O}(\lambda)$ suppression. 
The constraint on the number of subleading power interactions 
also limits the number of soft structures 
$\mathfrak{s}_i$ from the set \eqref{eq:softSet}, required 
at NLP.

In the position-space SCET framework, soft fields in the current 
operators appear only from $\mathcal{O}(\lambda^3)$ 
\cite{Beneke:2017ztn}. Hence, at NLP, the soft part 
$J_s(0)$ is not present, and the soft
structures come only from single insertions of the 
$\mathcal{O}(\lambda^2)$ SCET Lagrangian, 
$\mathcal{L}^{(2)}_{\xi}$ and $\mathcal{L}^{(2)}_{{\rm{YM}}}$, 
and double insertions of  the single power-suppressed terms, 
$\mathcal{L}^{(1)}_{\xi}$, 
$\mathcal{L}^{(1)}_{\xi q}$, and $\mathcal{L}^{(1)}_{{\rm{YM}}}$.

The next simplification is due to the fact that 
the kinematics of the process in the centre-of-mass frame does not support power suppression
created by a single operator with $\mathcal{O}(\lambda)$
scaling on a given leg. This is because the incoming 
collinear momentum can be chosen to carry only its large component,
$n_+p \sim Q$ ($n_- l\sim Q$ for the anticollinear leg), and all components of soft momentum scale
as $\mathcal{O}(\lambda^2)$. For the (anti)collinear direction
to carry $\mathcal{O}(\lambda)$ suppression, it would 
necessarily have to be proportional to the transverse 
component of the (anti)collinear vector, $p^{\mu}_{\perp}
(l^{\mu}_{\perp})\sim Q \lambda$, since no other momentum component
in the threshold kinematics carries $\mathcal{O}(\lambda)$ scaling, which, however, vanishes. 
This means that the $\mathcal{O}(\lambda^2)$ power suppression
cannot come from two insertions of $\mathcal{L}^{(1)}_{\xi}$ 
(or $\mathcal{L}^{(1)}_{\rm{YM}}$) on two  separate legs of a
diagram. Moreover, a  non-vanishing $\mathcal{O}(\lambda)$
{\em amplitude} also cannot exist in the $q\bar{q}$ 
channel.\footnote{Soft quark 
emission does yield a non-vanishing $\mathcal{O}(\lambda)$ 
amplitude, however this contributes to the (anti)quark-gluon 
($qg$, $\bar{q}g$) channel.}
In consequence, at cross section level at NLP, the 
$\mathcal{O}(\lambda^2)$ suppression must be generated in 
the amplitude which then interferes with the LP 
amplitude according to \eqref{eq:ampTrans}, yielding the 
$\mathcal{O}(\lambda^2)$ suppressed cross section. This still leaves
the possibility of $\mathcal{O}(\lambda^2)$ suppression to be 
generated by the $J^{\,T2}(t)$ operator formed by a  $\mathcal{L}^{(1)}$
insertion and a subleading current of $A1$ or $B1$-type.  
Due to chirality and helicity conservation in QCD, the possible
currents are 
\begin{eqnarray}\label{JA1.1}
J_{\rho}^{A0,A1}(t,\bar{t}\,) &=& { 
	\bar{\chi}_{{\bar{c}}}(\bar{t}n_-)\,n_{+\rho}\,
	i\slashed{\partial}_{\perp} \chi_{c}(t n_+) },
\\\label{JB1.1}
J_{\rho}^{A0,B1}(t_1,t_2,\bar{t}\,) &=& { 
	\bar{\chi}_{{\bar{c}}}(\bar{t} n_-)
	\,n_{\pm\rho} \,
	\slashed{\mathcal{A}}_{\perp c}(t_2 n_+) 
	\chi_{{c}}(t_1 n_+) },
\end{eqnarray}
and corresponding ones with power suppression in the 
anticollinear direction. The important detail to note is that 
both currents are each proportional to $n_{\pm\rho}$. However, 
the power-suppressed amplitude in which these 
currents could appear, is interfered with the LP 
amplitude, which is proportional to $\gamma_{\perp\rho}$, 
as can be seen in \eqref{eq:LPcurrent}. Contraction of 
these two structures makes such contribution vanish at
the cross section level to all orders in perturbation 
theory. This means that at NLP the sum over indices in $m_{1,2}$
in the formula derived in Sec.~\ref{sec:factGeneralPows} 
contains only the $A0$-type current, along with time-ordered
products of the LP current with Lagrangian insertions. 
Hence, only the hard matching 
coefficient $C^{\,A0,A0}$ of the LP current 
appears in the NLP factorization formula. 

These considerations lead to the conclusion that the soft 
structures relevant at NLP are in fact the terms already 
explicitly presented in \eqref{eq:softSet}
(dropping the ellipsis) after the use of the equation of motion 
to eliminate the redundant $n_+\mathcal{B}^+$ structure.       

The simplifications outlined above make it possible to write
down a NLP version of the general subleading-power factorization 
formula \eqref{eq:allpowerresult} in a more compact way. 
Namely, up to NLP \eqref{eq:allpowerresult} simplifies to 
\begin{eqnarray}
\label{eq:3.21}
\hat{\sigma}(z)&=&  \,\, \sum^5_{i,i'=0}\,\, \int 
\left\{\frac{d\omega_j}{2\pi}\right\} 
\left\{\frac{d\omega'_{j'}}{2\pi}\right\}
\,{\rm{Tr}}\,\bigg[ \left(\frac{\slashed{n}_-}{4}
\right) 
D_{\,i'}^{*\,\rho}(x_an_+p_A, x_bn_-{p}_B; \{ \omega'_{j'}\})
\nonumber \\ 
&& \hspace{1cm} \times\,
\left(\frac{\slashed{n}_+}{4}
\right) D_{\,{i}\rho}(x_an_+p_A, x_bn_-{p}_B; \{
\omega_j\}  ) \bigg] \nonumber\\ 
&& \hspace{0cm}  \times\,
(-Q^2) \int \frac{d^{d-1}\vec{q}}{(2\pi)^{d-1}
	\,2\sqrt{Q^2+\vec{q}^{\,2}}}\,
\frac{1}{2\pi}    
\int d^dx \,e^{i( x_ap_A+x_b\,p_B-q)\cdot x}
\widetilde{S}_{\,{i} i'}\,(x;\{\omega_j\},\{ \omega'_{j'}\}) 
\nonumber\\ 
&& \hspace{0cm}+ \mbox{ $\bar{c}$-terms}\,,
\end{eqnarray}
The set notation, with $\{\omega_j \}=\{\omega_1, \omega_2 \}$,
is only required for 
terms $i=4,5$ where the soft structures consist of insertions
of fields at different positions, as can be seen in the explicit
expressions below. All other terms require only a single 
$\omega$ variable, aside from the LP position-space 
soft function
\begin{eqnarray}\label{eq:LPsoft}
\widetilde{S}_{0}(x ) &=& 
\frac{1}{N_c}{\rm{Tr}}\, \langle 0| \bar{\mathbf{ T}} \left[ 
Y_{+}^\dagger(x)Y_{-}(x)  \right]  
{\mathbf{ T}}
\left[ Y_{-}^\dagger(0) Y_{+}(0) \right]
|0 \rangle\,.
\end{eqnarray}
In (\ref{eq:3.21}) the terms with power suppression placed
on the anticollinear leg, both in the amplitude and its 
conjugate, are indicated by ``$\bar{c}$-terms'' and 
not written explicitly, since eventually 
 they contribute a factor
of 2 to the power-suppressed terms in the above formula.

As explained above, the general structure $\Gamma^{\rho}$ defined
in \eqref{eq:DYcurrent} is simply $\gamma^{\,
\rho}_{\perp}$ at NLP, since only the $J^{\,A0,A0}$ current needs to be used 
in time-ordered products with Lagrangian insertions
in the matching to the DY current. Furthermore, the anticollinear 
functions $\bar{J}^{\,m_1}_{\,\bar{i}}
(\{ n_-\bar{p}_{\,\bar{k}}\},-\,n_-p_b;
\{ \bar{\omega}_{\,\bar{j}}\})$ in the general definition 
(\ref{eq:Dcoeff}) are delta functions in 
$D_{{i}\rho}(x_an_+p_A, x_bn_-{p}_B; \{ \omega_j\})$, 
which therefore simplify to\footnote{At LP,  collinear and anticollinear 
	functions are delta functions, and $D_{i\rho}$
	reduces to $\gamma_{\perp\rho}\,C^{A0,A0}$.}  
\begin{eqnarray}\label{eq:3.22}
D_{i\rho}(x_an_+p_A, x_bn_-{p}_B;\,\{ \omega_j\}\,) &=&  
\int d(n_+p) \, d(n_-\bar{p})  \, \,C^{A0,A0}(n_+p, n_-\bar{p})
\nonumber \\ && \hspace{-3.0cm} \times \,\delta(
n_- \bar{p} - x_b\,n_-p_B) \,\gamma_{\perp\rho}\,{J}_{i}
\left(n_+p,x_a\,n_+p_A; \{\omega_j\}\right) \,.
\end{eqnarray}
The index $i$, which is summed over in \eqref{eq:3.21},
stands in place of all indices---Dirac, Lorentz, and colour---required
by each term depending on the specific soft structure
appearing in the collinear  matching \eqref{eq:genMatch}.  
It is understood that one should perform the contraction 
of these indices prior to the spin trace in \eqref{eq:3.21},
because some soft functions, for example ${S}_5$ below, 
can  have open
spin indices which are connected to the collinear function. 
An expression similar to \eqref{eq:3.22} with $\{\omega'_{j'}\}$ 
variables holds for the conjugate amplitude.

Eq.~\eqref{eq:3.21} still contains the unexpanded final-state 
phase-space integral over the lepton-pair momentum $\vec{q}$. 
This means that in addition to the dynamical power corrections 
to the amplitude, there is a kinematic power correction 
from the phase-space integration over the LP amplitude, which 
will be discussed in more detail below. 

Next, we would like to draw attention to the collinear 
functions themselves. Since, as noted above, at NLP only the LP current 
$J^{A0,A0}$ is needed in time-ordered products with Lagrangian 
insertions, the set $\{\psi_{c}(t_kn_+)\}$ in the general 
collinear matching equation (\ref{eq:genMatch}) 
consists of a single quark (or antiquark) collinear field, 
and the set $\{\mathcal{L}^{(l)}_{}(z_j)\}$ of Lagrangian 
insertions is either $\{\mathcal{L}^{(2)}(z)\}$ or 
$\{ \mathcal{L}^{(1)}(z_1),\mathcal{L}^{(1)}(z_2)\}$. 
We also use momentum-space collinear functions as defined 
in (\ref{eq:colFT}), hence the collinear matching equation 
at NLP is either 
\bea\label{eqn:momentumMatch} 
i \int d^4z  \,\mathbf{T}\Big[ 
\chi_{c,\gamma f}\left(tn_+\right)
\, \mathcal{L}^{(2)}_{}(z) 
\Big] &&  = 2\pi \sum_i  
\int \frac{d\omega}{2\pi}
\int \frac{dn_+p}{2\pi}\,e^{-i\,(n_+p)\, t} 
\int 
\frac{dn_+p_a}{2\pi}
\nonumber \\
&&\hspace{-4cm}\times  \,{J}_{i;\gamma\beta,\mu,fbd}
\left(n_+p,n_+p_a;\omega \right)\, 
\hat{\chi}^{{\rm PDF}}_{c,\beta b}(n_+p_a)
\int dz_{-} 
\,e^{-i\,\omega\, z_-}
\,\mathfrak{s}_{i;\mu,d}(z_{-})\,,
\eea 
or the one with two $\mathcal{L}^{(1)}$ insertions, 
in which case the collinear function (soft function) has two 
arguments $\{\omega_1,\omega_2\}$ ($\{z_{1-},z_{2-}\}$) 
and the corresponding integrations must be added. 
The indices $\mu$ and $d$ carried by $\mathfrak{s}$ 
represent the collective Lorentz
and colour indices appropriate for the given soft structure.  
For each independent soft structure $\mathfrak{s}_{i}$ there exists a 
corresponding collinear function $J_i$ as shown on the right-hand
side of the above equation. 

Thus far we have focused on the derivation of the factorization 
formula for the bare partonic cross section $\hat{\sigma}$.
Indeed, $\hat{\sigma}$ still contains collinear 
singularities, which are usually subtracted by PDF renormalization. 
Therefore, care has to be taken when dealing with this 
$d$-dimensional quantity. 
For instance, the spin trace which appears at leading power  
gives a factor 
${\rm{Tr}}\left[ \left(\frac{\slashed{n}_-}{4}
\right) {\gamma}_{\perp\rho}  \left(\frac{\slashed{n}_+}{4}
\right) \gamma^{\rho}_{\perp} \right]=-(1-\epsilon)$.
In order to compare with with literature we find it convenient 
to consider the quantity $\Delta(z)$ defined through 
\bea 
\Delta(z) = \frac{1}{(1-\epsilon)} \frac{\hat{\sigma}(z)}{z}\,, 
\eea 
with the factor $(1-\epsilon)$ divided out 
compared to \cite{Matsuura:1987wt}.  

We next simplify the factorization formula in \eqref{eq:3.21} 
further by discussing separately the kinematic and dynamical 
NLP correction.

\subsubsection{NLP kinematic correction 
	$\Delta^{kin}_{{\rm{NLP}} }(z)$}
\label{subsec:kincorrs}

In the partonic centre-of-mass frame of the DY process, where
$x_a \,\vec{p}_A + x_b\,\vec{p}_B=0$, the three-momentum of the 
DY boson has to be balanced by the soft radiation, 
$\vec{q}=-\vec{p}_{X_s}$. The soft radiation energy is expanded 
in powers of $\lambda$ as follows:
\bea 
\big(x_ap_A + x_bp_B -q\big)^{0} = p^{0}_{X_s} = \sqrt{\hat{s}} 
-\sqrt{Q^2+\vec{q}^{\,\,2}} = 
\frac{\Omega_*}{2} - \frac{\vec{q}^{\,\,2}}{2Q}
+\mathcal{O}(\lambda^6)\,,
\label{eq:kinexp}
\eea 
where the first term has a further expansion in $(1-z)$,
\bea \label{eq:omStar}
\Omega_{*} = \frac{2\,Q(1-\sqrt{z})}{\sqrt{z}} = Q\,(1-z)+\frac{3}{4}
\,Q\,(1-z)^2 + \mathcal{O}(\lambda^6) \,.
\eea 
Starting with the LP soft function term in  \eqref{eq:3.21},
contributions to the NLP cross   section come from expanding the 
kinematic factors. Focusing on this LP soft function term
and recalling the simplification of the $D$ coefficients 
for this case noted after (\ref{eq:3.22}), we start from 
\begin{eqnarray}\label{eq:nlpKin}
{\Delta}^{kin}_{{\rm{NLP}} }(z)&=&  \,\,H(\hat{s})\, \frac{1}{z} \,
\frac{Q}{4\pi}  \int \frac{d^{d-1}\vec{q}}{(2\pi)^{d-1}\,}
\int d^dx \,e^{i (\Omega_*/2)x^0 
	- i(\vec{q}^{\,\,2}/(2Q))x^0-i\vec{q}\cdot \vec{x} }\,\,
\widetilde{S}_{0}\,(x) \,,
\end{eqnarray}
where $H(\hat{s})=|C^{A0,A0}(x_an_+p_A,x_bn_-p_B)|^2$. In the above equation 
a number of kinematic corrections can be identified. The first is 
due to power suppression provided by second term in the exponent. 
The second, originates in the expansion of $\Omega_*$ itself. 
The expansion of the $1/z$ factor gives the third kinematic 
correction, and a fourth kinematic correction comes from expansion of the argument of the hard function $H(\hat{s})$. After 
expanding out these terms, the integral over $\vec{q}$ 
can be performed, yielding a delta function, 
which sets $\vec{x}=0$ in the soft function. 
We write the four corrections in order as
\bea 
\Delta^{K1}_{{\rm{NLP}}}(\Omega) &=& H\big(Q^2\big) \label{3.29}
\frac{\partial}{\partial \Omega} \,\partial^2_{\vec{x}} 
\,S_{\rm DY}(\Omega,\vec{x})\lvert_{\vec{x}=0}\,,
\\[0ex] \Delta^{K2}_{{\rm{NLP}}}(\Omega) &=& H\big(Q^2\big)\,
\frac{3}{4}\,\Omega^2 \frac{\partial}{\partial \Omega}
S_{\rm DY}(\Omega,\vec{x})\lvert_{\vec{x}=0}\,, \label{3.30}
\\[1ex] \Delta^{K3}_{{\rm{NLP}}}(\Omega) &=& H\big(Q^2\big)\,
\,\Omega\, 
S_{\rm DY}(\Omega,\vec{x})\lvert_{\vec{x}=0}\,, \label{3.31}
\\[1ex]  \Delta^{K4}_{{\rm{NLP}}}(\Omega) &=& H'\big(Q^2\big)\,
Q^2\,\Omega\, S_{\rm DY}(\Omega,\vec{x})\lvert_{\vec{x}=0},\label{3.32}
\eea 
where $S_{\rm DY}(\Omega, \vec{x}\,)$ is the LP soft function 
defined in \eqref{eq:LPsoftfn}, but with argument $x^0$ generalized 
to non-zero $\vec{x}$. The full NLP kinematic correction, 
${\Delta}^{kin}_{{\rm{NLP}}}(z)$, is given by the sum of these  
four terms. In Sec.~\ref{sec:fixedresults} we present the result 
of evaluating these expressions up to NNLO. 

\subsubsection{Dynamical NLP power correction $\Delta^{dyn}_{{\rm{NLP}} }(z)$}
\label{sec:dynNLP}

Next we consider the contribution to the NLP cross section
due to insertions of subleading-power Lagrangians 
and LP kinematics. Thus we keep only the  
first term in the expansions (\ref{eq:kinexp}), \eqref{eq:omStar}. 
The $d^{d-1}\vec{q}$ integral then gives a delta function
for the spatial part of $x$, hence in the soft functions 
we can immediately set $\vec{x}=0$.

As opposed to the kinematic correction, 
the collinear functions appearing here are non-trivial. 
Note that the collinear functions will carry the same indices
as the corresponding soft function.  On top of the indices
connecting to the soft function, the collinear functions 
carry two Dirac and two colour indices,  $\gamma\beta$ and
$fb$, from the threshold-collinear and $c$-PDF fields in the
matching equation \eqref{eqn:momentumMatch}. 
It is understood that the collinear functions, $J_i$, in 
\eqref{eq:3.21} carry indices as prescribed by 
\eqref{eqn:momentumMatch}.
For instance,
the first soft structure in the set \eqref{eq:softSet} has one 
$\mathcal{B}^+$ field and therefore carries an adjoint 
index $A$ connecting to the collinear function.  This means that $J_1$
carries one additional 
adjoint index corresponding to the colour generator. Explicitly, 
${J}_{1}\left(n_+p,x_an_+p_A; \omega \right)$
in \eqref{eq:3.22} stands for
$ {J}^{A}_{1;\gamma\beta, fb}
\left(n_+p,x_an_+p_A; \omega \right)$.

In order to simplify the $\Delta^{dyn}_{{\rm{NLP}} }(z)$ part 
of the  factorization formula \eqref{eq:3.21} further, we 
decompose the collinear functions into all possible colour and 
spinor structures. Continuing with the example from above, this 
particular collinear function must be proportional to 
$\mathbf{T}^A_{fb}$ since this is the only structure 
which carries one adjoint, $A$, and two fundamental, $fb$, colour 
indices. At this point we can define a scalar collinear function
multiplied by $\mathbf{T}^A_{fb}$ and move the colour 
factor into the soft function where it forms part
of the trace over the colour indices. 
In a similar way, the colour factors in other collinear 
functions can be absorbed into their corresponding soft functions. 
The dynamical NLP part of \eqref{eq:3.21} can then be 
simplified to 
\begin{eqnarray}
\label{eq:3.24}
\Delta^{dyn }_{{\rm{NLP}} }(z)&=&- \frac{2}{(1-\epsilon)} \,  
Q \left[ \left(\frac{\slashed{n}_-}{4}
\right) {\gamma}_{\perp\rho}  \left(\frac{\slashed{n}_+}{4}
\right) \gamma^{\rho}_{\perp} \right]_{\beta\gamma}
\nonumber  \\ 
&&  \hspace{0cm} \times 
\, \int d(n_+p)\,C^{\,{A0,A0}\,}(n_+p, x_bn_-{p}_B  ) \,
C^{*A0,A0}\left(\,x_an_+p_A,\,x_b{n_-p_B}\right)
\nonumber \\ 
&&  \hspace{0cm} \times \,\sum^5_{i=1}
\,\int \left\{d\omega_j\right\}
{J}_{i,\gamma\beta}\left(n_+p,x_a n_+p_A; 
\left\{\omega_j\right\} \right) \, 
{S}_{i}(\Omega; \left\{\omega_j\right\} ) 
+\rm{h.c.}\,,\quad
\end{eqnarray}
where here $\Omega=Q(1-z)$. 
As in \eqref{eq:3.21}, the double-valued set 
$\{\omega_j \}=\{\omega_1, \omega_2 \}$,
is only required for terms $i=4,5$. For $i=5$,   
in addition to the Dirac indices $\beta\gamma$ written 
explicitly, $J_i$ and $S_i$ contain further indices, see the 
definition of $S_5$ below, which are contracted among them. 
As mentioned above, a factor of 2 in this formula comes from the 
$\bar{c}$-terms. Furthermore, one of the $D$ coefficients 
always reduces to the LP expression, since at NLP there is 
no $\mathcal{O}(\lambda)$ amplitude in the 
$q\bar{q}$-channel, as discussed above. 
We point out again the main difference to the LP factorization formula, 
namely the presence of the convolution of a jet function 
with multi-local, generalized soft functions.\footnote{This 
structure bears resemblance to the SCET treatment of $1/m_b$ 
suppressed power corrections to semi-leptonic $B$ decay 
in the so-called shape function 
region~\cite{Beneke:2004in,Bosch:2004cb,Lee:2004ja}.}

We define the multi-local, generalized soft functions in 
momentum space as the Fourier transforms 
\be
{S}_{i}(\Omega; \left\{\omega_j\right\} ) = 
\int \frac{dx^0}{4\pi} \,  e^{i \Omega\, x^0/2}  
\int \bigg\{\frac{dz_{j-}}{2\pi} \bigg\} \, e^{-i\omega_j{z_{j-}}}
{S}_{i}(x_0; \left\{z_{j-}\right\} ) \,.
\ee

The position-space soft functions appearing at NLP are given by 
\begin{eqnarray}
\label{eq:3.23}
&&{S}_{1}(x^0; z_- )\,=\,
\frac{1}{N_c}\, {\rm{Tr}} \langle 0| \bar{\mathbf{ T}} \left[ Y_{+}^\dagger(x^0)
Y_{-}(x^0)  \right] {\mathbf{ T}}\left(
\left[ Y_{-}^\dagger(0) Y_{+}(0) \right]
\frac{i\partial_{\perp}^{\nu}}{in_-\partial}
\mathcal{B}^{+}_{\nu_{\perp}}\left(z_{-}\right)
\right)|0 \rangle\,, \quad\qquad
\\[2ex] 
&&{S}^{}_{2;\mu\nu}(x^0; z_- )\,=\, \frac{1}{N_c}\,
{\rm{Tr}}\,\langle 0|  \bar{\mathbf{ T}} \left[ Y_{+}^\dagger(x^0) Y_{-}(x^0) 
\right]   \nn\\ 
&&\hspace{2.5cm} \times\, {\mathbf{ T}}\left(
\left[ Y_{-}^\dagger(0) Y_{+}(0) \right]
\frac{1}{(in_-\partial)}
\big[  \mathcal{B}^+_{\mu_\perp}(z_-)
,\mathcal{B}^+_{\nu_\perp}(z_-)\big] 
\right)|0 \rangle\,,
\\[2ex]
&&{S}^{}_{3}(x^0; z_-)\,=\,\frac{1}{N_c} \,{\rm{Tr}}\,\langle 0|  
\bar{\mathbf{ T}} \left[ Y_{+}^\dagger(x^0) Y_{-}(x^0) \right]
\nn \\&& \hspace{2.5cm}\times\, {\mathbf{ T}}\left(
\left[ Y_{-}^\dagger(0) Y_{+}(0) \right]
\frac{1}{(in_-\partial)^2}\left[ 
\mathcal{B}^{+\,\mu_\perp}(z_{-}),
\left[in_-\partial\mathcal{B}^+_{\mu_\perp}(z_{-})
\right]  \right] 
\right)|0 \rangle\,,
\label{eq:3.26}
\\[2ex]
&&{S}^{AB}_{4;\mu\nu,bf}(x^0; z_{1-},z_{2-} )\,=\,
\frac{1}{N_c} \,{\rm{Tr}}\, \langle 0| 
\bar{\mathbf{ T}} \left[ Y_{+}^\dagger(x^0) Y_{-}(x^0) \right]_{ba}
\nn\\&&\hspace{2.5cm} \times \,{\mathbf{ T}}\left(
\left[ Y_{-}^\dagger(0) Y_{+}(0) \right]_{af}
\mathcal{B}^{+A}_{\mu_\perp}(z_{1-})
\mathcal{B}^{+B}_{\nu_\perp}(z_{2-})
\right)|0 \rangle\,,
\\[2ex]
\label{eq:3.27}
&&{S}_{5;bfgh,\sigma\lambda}(x^0; z_{1-},z_{2-} )\,=\,
\frac{1}{N_c} \,\langle 0|  
\bar{\mathbf{ T}} \left[ Y_{+}^\dagger(x^0) Y_{-}(x^0) \right]_{ba}
\nn\\ &&\hspace{2.5cm}\times\,
{\mathbf{ T}}\left(
\left[ Y_{-}^\dagger(0) Y_{+}(0) \right]_{af}
\frac{g_s^2}{(in_-\partial_{z_1})(in_-\partial_{z_2})}
{q}_{+\sigma g}(z_{1-})
\bar{q}_{+\lambda h}(z_{2-})
\right)|0 \rangle\,.
\end{eqnarray}
We recall from the discussion of the list (\ref{eq:softSet}) that 
the soft functions $S_2$ and $S_3$ are redundant and could be 
eliminated by relating them to $S_4$.
There exists in principle another soft function,
\begin{eqnarray}
\widetilde{S}^A_{6;bf,\mu\nu}(x; \omega )&=&
\int {dz_-}\, e^{-i\omega\,z_-}
\frac{1}{N_c} \langle 0|  \bar{\mathbf{ T}} \left[
Y_{+}^\dagger(x)Y_{-}(x)  \right]_{ba} 
\nonumber \\ 
&& \times \, {\mathbf{ T}}\left(
\left[ Y_{-}^\dagger(0) Y_{+}(0) \right]_{af}
\frac{i\partial_{[\mu_{\perp}}}{in_-\partial}
\mathcal{B}^{+A}_{\nu_\perp]\,}(z_-)
\right)|0 \rangle\,,
\end{eqnarray}
with the soft structure given by the second
term in \eqref{eq:softSet}. This soft function is  
required  to obtain the NLP one-soft-gluon emission amplitude,  
see Appendix \ref{appendix:amplituderesults}, 
but does not contribute to the DY cross section at 
any order in perturbation theory. This is because
the soft functions are vacuum matrix elements of Wilson
lines and soft field insertions, hence, the only 
structure which can carry the Lorentz indices of the 
anti-symmetric structure $\frac{\partial_{[\mu_{\perp}
}}{in_-\partial} \mathcal{B}^{+A}_{\nu_\perp]\,}(z_-)$ 
in $\widetilde{S}^A_{6;bf,\mu\nu}(x; \omega )$ is the epsilon 
tensor. However, this is excluded in QCD by parity conservation. 
Therefore, only the five soft functions given in  
\eqref{eq:3.23} to \eqref{eq:3.27}	and their 
corresponding collinear functions appear in the factorization
formula. 
	
The above all-order formulation of NLP threshold factorization 
and the operator definition of the appearing jet and soft functions 
is one of the main results of this paper. 
	

\subsection{Expansion up to NNLO} 
\label{sec3.3}
	
In Sec.~\ref{sec:fixedresults} we will check the NLP factorization 
formula by comparing to existing fixed-order 
$\mathcal{O}(\alpha_s^2)$ results in 
the literature and to own expansion-by-region calculations. To 
prepare this discussion we consider here the terms that arise 
in the NNLO expansion of~\eqref{eq:3.24}.

Each of the objects in the formula, the hard matching 
coefficient $C^{A0,A0}(n_+p, n_-\bar{p})$, the collinear functions  
${J}_{i}\left(n_+p,x_a\,n_+p_A; \{\omega_j\} \right)$, 
the soft functions   
$\widetilde{S}_i(x;\{\omega_j\})$, has a perturbative 
expansion in the strong coupling.
Since at NLP the generalized soft functions 
contain explicit soft field insertions, as opposed
to simply being composed of Wilson lines as at LP, 
the lowest order at which they can contribute is $\alpha_s^1$. 
The  hard and the collinear functions can 
have tree-level contributions. This means that in 
order to reproduce NLO results, only one combination 
is needed, tree-level hard and collinear functions 
and a NLO soft function. 
Then, to reproduce the NNLO fixed order results, there are 
three contributions:  (1) Tree-level hard function together 
with one-loop collinear and soft functions, (2) one-loop 
hard function, tree-level collinear and one-loop soft function, 
and finally, (3) the soft functions at $\mathcal{O}(\alpha_s^2)$.
	
Before proceeding, it is important to note that
since the kinematic set-up allows only for soft
radiation, the large component of the incoming PDF-collinear 
momentum must be identical to the sum of the large components 
of the outgoing threshold momenta of the collinear function. 
Since for the A0 current there is only one outgoing momentum,
the collinear functions relevant to NLP will be
be proportional to $\delta(n_+p-x_an_+p_A)$. However,  
due to the presence of $n_- z$ in the soft-collinear 
interactions, which translates into a $n_+p$ derivative in momentum 
space, the momentum-space collinear 
functions can also contain derivatives 
of the momentum-conserving delta function. This occurs for 
$J_1(n_+p,x_a\,n_+p_A; \omega )$. Since it is 
also diagonal in the Dirac indices we write this collinear 
function in terms of two scalar components as follows:
\begin{eqnarray}\label{eq:collFuncDecomp}
	{J}^{}_{1;\gamma\beta}
	\left(n_+p,x_a\,n_+p_A; \omega \right) 
	&=& \delta_{\gamma\beta}\,
	\, \bigg[ {J}_{1,1}
	\left(x_an_+p_A; \omega \right) 
	\delta(n_+p-x_a n_+p_A)
	\nn \\ && \hspace{-2cm} + \, {J}_{1,2}
	\left(x_an_+p_A; \omega \right) 
	\frac{\partial}{\partial (n_+p)} 
	\delta(n_+p-x_a n_+p_A) \bigg]\,.
	\end{eqnarray}
In the factorization formula, we can integrate 
by parts the derivative such that it
acts on the amplitude hard-scattering coefficient. 
Hence the derivative collinear-function term contributes only 
when the hard matching coefficient is momentum-dependent, 
which happens only from the one-loop order on for 
$C^{A0,A0}$. Once the derivative on the coefficient 
function is taken, one can perform the remaining $ d(n_+p)$
integral using the extracted delta function. 
	
The only soft function from the list in 
\eqref{eq:3.23}--\eqref{eq:3.27} which begins at lowest,
next-to-leading order is $S_1$. The others contain at 
least two insertions of subleading soft fields, 
which implies that the leading contribution to the 
cross section is NNLO.
Therefore, expanded up to NLO, we have
\begin{eqnarray}\label{eq:3.32}
\Delta^{dyn\,(1)}_{{\rm{NLP}}}(z)&=& 4  Q \,  H^{(0)}(Q^2) 
\int d\omega\, {J}^{(0)}_{1,1}
\left(x_an_+p_A; \omega \right) {S}^{\,(1)}_{1}(\Omega;\omega)\,, 
\end{eqnarray}
where we have evaluated spin trace, 
${\rm{Tr}}\left[ \left(\frac{\slashed{n}_-}{4}
\right) {\gamma}_{\perp\rho}  \left(\frac{\slashed{n}_+}{4}
\right) \gamma^{\rho}_{\perp} \right]$, which 
gives a factor of $-(1-\epsilon)$. Also, here and below in 
this section, 
$\Omega$ is related to the threshold variable $1-z$ by 
$\Omega = Q(1-z)$. Eq.~(\ref{eq:3.32}) can be
simplified greatly by inserting the tree-level hard 
coefficient $H^{(0)}(Q^2)=1$, and the tree-level 
collinear function, which can be found in \eqref{eq:j011}:
\begin{eqnarray}\label{eq:3.41}
\Delta^{dyn\,(1)}_{{\rm{NLP}}}(z)&=& -4 \int d\omega
\,  {S}^{\,(1)}_{1}(\Omega; \omega )\,. 
\end{eqnarray}
	
Moving on to NNLO accuracy, the three contributions discussed 
above take the following expressions: 
\begin{itemize}
\item Collinear: one-loop collinear and NLO soft functions
	\begin{eqnarray}\label{eq:3.33}
	\Delta^{dyn\,(2)}_{{\rm{NLP-coll}}}(z)&=& 4  Q \, 
	H^{(0)}\big(Q^2\big) \int d\omega
	\,  {J}^{\,(1)}_{1,1}
	\left(x_an_+p_A; \omega \right) \, 
	{S}^{\,(1)}_{1}(\Omega; \omega ) \,.
	\end{eqnarray}
\item Hard: one-loop hard and NLO soft functions 
	\begin{eqnarray}\label{eq:3.44h}
	\Delta^{dyn\,(2)}_{{\rm{NLP-hard}}}(z) 
	&=& 2 Q \int d\omega  \,
	{S}^{(1)}_{\,{1}\,}\,(\Omega;\omega\,) 
	\, \, \bigg(\,H^{(1)}
	\big(Q^2\big)\,{J}^{(0)}_{1,1}
	\left(x_an_+p_A; \omega \right) 
	\nn \\ 
	&&-\,C^{*A0\,(0)}\left(x_an_+p_A,\, x_bn_-p_B\right) 
	{J}^{(0)}_{1,2}\left(x_an_+p_A; \omega \right) 
	\nonumber \\ 
	&&\times \,
	\frac{\partial}{\partial x_a(n_+p_A)}
	C^{A0\,(1)}(x_an_+p_A,x_bn_-{p}_B)\bigg)
	+\rm{h.c.}
	\end{eqnarray}
\item Soft: NNLO soft functions 
\begin{eqnarray}\label{eq:3.34s}
\Delta^{dyn\,(2)}_{{\rm{NLP-soft}}}(z) &=&- \frac{4}{(1-\epsilon)}   
	\,Q \left[ \left(\frac{\slashed{n}_-}{4}
	\right) {\gamma}_{\perp\rho}  \left(\frac{\slashed{n}_+}{4}
	\right)  \, \gamma^{\rho}_{\perp} \right]_{\beta\gamma}
	H^{(0)}\big(  Q^2 \big)
	\nonumber \\ 
	&& \times \, \sum^5_{i=1}
	\,\int \left\{d\omega_j\right\}
	\,  {J}^{(0)}_{i,\gamma\beta}
	\left(x_a\,n_+p_A; \left\{\omega_j\right\} \right) \, 
	{S}^{(2)}_{i}(\Omega; \left\{\omega_j\right\} ) \,.
	\end{eqnarray}
In $\Delta^{dyn\,(2)}_{{\rm{NLP-soft}}}(z)$ the derivative terms 
in the collinear functions do not contribute, since the hard 
function is taken at tree level. 
\end{itemize}
	
All of the above formulas can be simplified by using tree-level 
values for the relevant objects. In particular since $H^{(0)}(Q^2)=1$, 
we have
\begin{eqnarray}\label{eq:3.43}
\Delta^{dyn\,(2)}_{{\rm{NLP-coll}}}(z)&=& 4 Q \int d\omega
\,{J}^{(1)}_{1,1}(x_a\,n_+p_A; \omega)  
\,{S}^{(1)}_{1}(\Omega; \omega) 
\end{eqnarray}
for the collinear term. Next, the hard contribution in 
\eqref{eq:3.44h} can be simplified using tree-level values
for the collinear functions in \eqref{eq:j011}
and \eqref{eq:j012}.  Care has to be taken when 
dealing with this expression, since it refers to 
$d$-dimensional regularized objects. 
The one-loop 
$d$-dimensional hard matching coefficient depends on 
$Q^2 = x_a x_b n_+ p_A n_- p_B$ only through an overall
factor $(-Q^2/\mu^2)^{-\epsilon}$. 
Performing the derivative therefore gives back the 
hard matching coefficient multiplied by a factor of 
$-\epsilon/Q$.
Together with the hermitian conjugate term in 
\eqref{eq:3.44h}, we obtain $-\epsilon/Q\times 
\,(C^{*A0(0)}C^{A0(1)}+C^{*A0(1)}C^{A0(0)})=-\epsilon\,H^{(1)}/Q$ 
from the derivative term. Then we arrive at
\begin{eqnarray}\label{eq:3.44}
\Delta^{dyn\,(2)}_{{\rm{NLP-hard}}}(z)=  
- {4} \left(1-\epsilon\right) H^{(1)}(Q^2) \int \,d\omega  \,
{S}^{(1)}_{\,{1}\,}\,(\Omega;\,\omega\,) \,.
\end{eqnarray}
The tree-level collinear functions can also be utilized to simplify 
the soft term, but we do not present it here.
	
\section{Calculation of collinear functions }
\label{sec:CollFuncsCalc}

In this section  we present the computation of the collinear 
functions to one-loop accuracy. The presence of these functions 
at NLP is one of the main results of this paper, and we will 
need the one-loop calculation in the subsequent section to 
verify the NLP factorization formula to NNLO. 

The collinear functions are defined through the non-perturbative 
operator matching equation  \eqref{eqn:momentumMatch}.
The left-hand side includes the threshold-collinear fields 
originating from time-ordered products of the LP current with 
subleading-power Lagrangian terms. We introduce the  
abbreviation 
\begin{align}
\label{eqn:matching4.1} 
\tilde{\mathcal{T}}_{\gamma f}(t)&\equiv i \int d^4z  
\,\mathbf{T}\Big[ 
\chi_{c,\gamma f}\left(tn_+\right) \mathcal{L}^{(2)}_{}(z)\Big] \, ,
\end{align}
for the left-hand side of \eqref{eqn:momentumMatch}, and define 
its Fourier transform by
\begin{equation}
\label{eqn:matching4.1FT}
\mathcal{T}_{\gamma f}(n_+q) =  \int dt \, e^{i (n_+q)\,t}\,\tilde{\mathcal{T}}_{\gamma f}(t)\,  .
\end{equation}
The momentum-space matching equation reads
\bea \label{eqn:matching4.5}
\mathcal{T}_{\gamma f}(n_+q) & =& 2\pi \,
\sum_{i}  \, 
\int \frac{d n_+p_a}{2 \pi} \int du \, e^{i\, (n_+p_a)\, u}\,
\int \frac{d \omega}{2 \pi} 
\nonumber  \\ 
&&\times  \, {J}_{i;\gamma\beta,\mu,fbd}
\left(n_+q,n_+p_a;\omega \right)\, 
\chi^{{\rm PDF}}_{c,\beta b}(un_+)
\int d z_- 
\, e^{-i \omega z_-}\,\mathfrak{s}_{i;\mu,d}(z_{-})\,.
\eea
For soft structures with two soft gluon emissions the generalization 
explained below (\ref{eqn:momentumMatch}) applies. 

We recall that since the collinear scale $Q^2(1-z)\gg \Lambda^2$ 
by assumption, the collinear function is a perturbatively 
calculable  short-distance coefficient in the matching of 
\eqref{eqn:matching4.1} and \eqref{eqn:matching4.5}. 
We can therefore extract the collinear functions $J_i$ by 
taking an appropriate matrix element between partonic 
states. For example, in case of collinear functions with a 
single external soft gluon, the simplest choice is the matrix 
element $\langle g(k)|...|q(p)\rangle$ with a soft 
gluon and PDF-collinear quark. We then compute both sides 
of the matching equation with the LP collinear Lagrangian with 
soft fields decoupled, in which case the soft fields 
on both sides act only as external fields. Hence, 
the soft matrix element 
$\langle g(k)|\mathfrak{s}_{i;\mu,d}(z_{-})|0\rangle$ 
takes its tree-level expression (since only soft 
loops could contribute). The same is true for  
$\langle 0|\chi^{{\rm PDF}}_{c,\beta b}(un_+)|q(p)\rangle$, 
because loop corrections are scaleless in this case. 

\subsection{Collinear functions at $\mathcal{O}(\alpha_s^0)$}

For the $q\bar{q}$-induced DY process, only the insertions 
of the quark-gluon subleading SCET Lagrangian but not the 
Yang-Mills terms contribute at tree level to the collinear 
functions. Indeed, at least one collinear gluon loop would be 
needed to which a $\mathcal{L}^{(2)}_{\rm YM}$ insertion 
could be attached via a triple-gluon interaction.    

We use momentum-space Feynman rules for the 
soft-collinear interactions vertices from the power-suppressed 
SCET Lagrangian given in Appendix A of \cite{Beneke:2018rbh} to 
perform the computation. The collinear-quark soft-gluon 
interaction vertex is given by   
\begin{equation}
\label{eq:Vccs}
\raisebox{-13mm}{\includegraphics[width=0.2\textwidth]{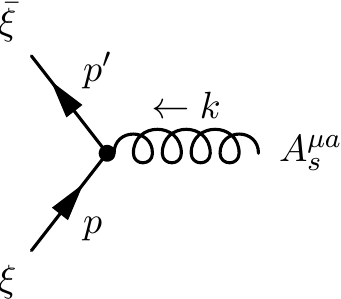}} 
\quad
i g_s \mathbf{T}^A \left\{ \begin{array}{ll}
\displaystyle 
\frac{\slashed{n}_{+}}{2} n_{-\mu}  & {\quad\mathcal O}(\lambda^0) \\[0.3cm]
\displaystyle 
\frac{\slashed{n}_{+}}{2} X_\perp^\rho n_-^{\nu} 
(k_\rho g_{\nu\mu}-k_\nu g_{\rho\mu}\,) & {\quad\mathcal O}(\lambda) \\
S^{\rho\nu}(k,p,p')  \displaystyle
\frac{\slashed{n}_{+}}{2} 
(k_\rho g_{\nu\mu}-k_\nu g_{\rho\mu})& {\quad\mathcal O}(\lambda^2)
\end{array}\right.\nonumber
\end{equation}
where 
\begin{equation}
\label{eq:Srnonu}
S^{\rho\nu}(k,p,p') \equiv \frac12 
\left[ { (n_- X) n_+^\rho n_-^\nu} + 
{(k X_\perp) X_\perp^\rho n_-^\nu}
+{X_\perp^\rho \left( \frac{\slashed{p}'_\perp}{n_+ p'} 
	\gamma_\perp^\nu   	+ \gamma_\perp^\nu
	\frac{\slashed{p}_\perp}{n_+ p} \right)} \right]
\end{equation}
and 
\bea
\label{eq:Xdef}
X^{\sigma}_{}=-\frac{\partial}{\partial p'_{\sigma}}
\big((2\pi)^d \delta^{(d)}(p - p'+k_+)\big)\,\,.
\eea
The momentum $k_+$, which appears in the argument of the delta 
function above, is defined as $k^{\mu}_+= (n_-k)\frac{n^{\mu}_+}{2}$. 
The three terms in the
${\mathcal O}(\lambda^2)$ vertex \eqref{eq:Srnonu} correspond 
directly to the three terms in the power-suppressed
SCET Lagrangian given in Eq.~(28) of \cite{Beneke:2002ni}. 
This has been rewritten in terms of gauge-invariant 
building blocks in \eqref{eq:quarkint}
such that the first term in \eqref{eq:Srnonu} corresponds to 
$\mathcal{L}^{(2)}_{1\xi}$, the second term to 
$\mathcal{L}^{(2)}_{2\xi}$, and the third term to 
$\mathcal{L}^{(2)}_{4\xi}$. In Appendix~\ref{sec:appendixA} 
we also provide the soft-quark and Yang-Mills SCET 
Lagrangian in this notation, which are needed for the 
one-loop calculation and for soft structures with soft quarks.

An important feature of the NLP Feynman rules is that they
contain derivatives of momentum-conservation delta functions 
at the subleading-power vertices. This is due to the 
appearance of explicit position-space arguments, $x^{\mu}$, in
the SCET Lagrangian terms owing to multipole expansion
\cite{Beneke:2002ph}. These derivatives must first be 
integrated by parts to act on the rest of the amplitude
{\it{before}} imposing momentum conservation.

\subsubsection{Single soft gluon structures}

Inspection of the subleading-power SCET 
Lagrangian \eqref{eq:quarkint} shows that only the two soft 
gluon structures 
\begin{equation}
\mathfrak{s}^{A}_{1}(z_{-})= 
\frac{i\partial_{\perp}^{\mu}}{in_-\partial}
\mathcal{B}^{+A}_{\mu_\perp}(z_{-}) \qquad\mbox{and} \qquad
\mathfrak{s}^{A}_{6}(z_{-})= 
\frac{i\partial_{[\mu_{\perp}}}{in_-\partial}
\mathcal{B}^{+A}_{\nu_\perp]}(z_{-})
\end{equation}
can have tree-level single-gluon matrix elements at 
$\mathcal{O}(\lambda^2)$. Hence the sum over $i$ in 
\eqref{eqn:matching4.5} reduces to $i=1,6$. Explicitly, 
\eqref{eqn:matching4.5} turns into 
\bea 
\label{eq:4.7}
\langle g(k)^K|\mathcal{T}^{1g}_{\gamma f}(n_+q)|q(p)_e
\rangle  &=& 2\pi  \int \frac{dn_+p_a}{2\pi}
du\,e^{i \,(n_+p_a)\, u} \int 
\frac{d\omega}{2\pi} \int  dz_{-} 
\, e^{-i\,\omega\, z_-}  
\nonumber \\
&& \hspace{-3.5cm}\times\,
\Big( \,{J}^A_{1;\gamma\beta,fb}
\left(n_+q,n_+p_a;\omega \right)\,\langle 0| 
\chi^{{\rm PDF}}_{c,\beta b}(un_+)|q(p)_e\rangle
\,\langle g(k)^K|\mathfrak{s}^{A}_{1}(z_{-})\,
|0\rangle 
\nn \\
&& \hspace{-2.8cm}
+\,{J}^{\mu\nu,A}_{6;\gamma\beta,fb}
\left(n_+q,n_+p_a;\omega \right)\,\langle 0| 
\chi^{{\rm PDF}}_{c,\beta b}(un_+)|q(p)_e\rangle
\,\langle g(k)^K|\mathfrak{s}^{A}_{6;\mu\nu}(z_{-})\,
|0\rangle   \Big) \,,\quad
\eea 
where $K,e$ refer to the colour of the external state and 
the superscript $1g$ reminds us that we consider the 
collinear functions for single soft gluon emission.

The $c$-PDF collinear matrix element on the right-hand side 
equals 
\begin{eqnarray}\label{eq:incQuark}
\langle 0|{\chi}^{\text{\scriptsize PDF}}_{c,\beta b}(un_+)|
q(p)_e\rangle =   \delta_{be} \sqrt{Z_{q,\text{PDF}}}\, 
{u}_{c,\beta}(p)\,e^{-i (n_+ p)\, u} \,,
\end{eqnarray}
where $\sqrt{Z_{q,\text{PDF}}}$ is the on-shell wave 
renormalization factor of the $c$-PDF field.
The soft matrix elements are found to give 
\begin{eqnarray}
\label{Feynm1.2}
\langle g(k)^K|  
\frac{i\partial_{\perp}^{\nu}}{in_-\partial}
\mathcal{B}^{+\,A}_{\nu_{\perp}\,}\left(z_{-}\right)
|0 \rangle &=& \delta^{AK}\,
\frac{g_s}{(n_-k)}  \,\left[ k^{\eta}_{\perp}
-\frac{k^2_{\perp}}{(n_-k)}n_-^{\eta} \right] 
\epsilon^*_{\eta\,}(k)
\, e^{iz_-k}\,,\quad
\\ [1ex]
\label{Feynm2.2}
\langle g(k)^K| 
\frac{i\partial_{[\mu_{\perp}}}{in_-\partial}
\mathcal{B}^{+\,A}_{\nu_\perp]\,}(z_-)
|0 \rangle &=&\delta^{AK}\,
\frac{g_s}{(n_-k)} \,\Big[ k^{\mu}_{\perp}
\,g^{\nu \eta}_{\perp} 
-k^{\nu}_{\perp}
\,g^{\mu \eta}_{\perp} \Big] 
\epsilon^*_{\eta}(k)
\, e^{iz_-k}\,.
\end{eqnarray}
Inserting these results into \eqref{eq:4.7}, we obtain 
\bea \label{eq:4.10} 
\langle g(k)^K|\mathcal{T}^{1g}_{\gamma f}(n_+q)|q(p)_e
\rangle  &=& 2\pi\,\frac{g_s}{(n_-k)}  \, \nn \bigg(
{J}^K_{1;\gamma\beta,fe}
\left(n_+q,n_+p;n_-k \right) \left[ k^{\eta}_{\perp}
-\frac{k^2_{\perp}}{(n_-k)}n_-^{\eta} \right]
\\
&& \hspace{-4.0cm}+ \,
{J}^{\mu\nu,K}_{6;\gamma\beta,fe}
\left(n_+q,n_+p;n_-k \right) 
\,\Big[ k^{\mu}_{\perp}
\,g^{\nu \eta}_{\perp} 
-k^{\nu}_{\perp}
\,g^{\mu \eta}_{\perp}  \Big]\bigg)
\,\sqrt{Z_{q,\text{PDF}}}\, {u}_{c,\beta}(p)\epsilon^*_{\eta\,}(k)
\,.
\eea
This is the final expression for the right-hand 
side of the matching equation \eqref{eqn:matching4.5} for single soft 
gluon structures for the chosen partonic state. 
We note that this expression is exact to all orders in 
perturbation theory, since, as mentioned above, there are 
no loop corrections to the above matrix elements.

We next turn our attention to the computation of the left-hand 
side of the matching equation \eqref{eqn:matching4.5}. The 
relevant terms in $\mathcal{L}^{(2)}$ are 
$\mathcal{L}^{(2)}_{1\xi, 2\xi, 4\xi}$, which give rise 
to the NLP soft-gluon vertex \eqref{eq:Srnonu}. A 
straightforward  tree-level calculation gives 
\begin{eqnarray}
\label{eq:treeMatch_bef_eom}
\langle g(k)^K|\mathcal{T}^{1g}_{\gamma f}(n_+q)|q(p)_e\rangle
&=& 
2\pi \, \frac{g_s}{(n_-k)} \mathbf{T}^{K}_{fe} \,\Bigg\{
\nonumber\\
&&\hspace{-2cm} 
- \left[ k^{\eta}_{\perp}
-\frac{k^2_{\perp}}{(n_-k)}n_-^{\eta} \right]
\frac{1}{n_+p}\delta(n_+q -n_+p)\delta_{\gamma\beta}
\nonumber \\[1ex] &&\hspace{-2cm}
 - \left[(n_-k)n^{\eta}_{+} - (n_+k)n^{\eta}_{-}  
\right]
\frac{\partial}{\partial n_+q}
\delta(n_+q -n_+p)\delta_{\gamma\beta}
\nonumber \\[1ex]
&&\hspace{-2cm}
-\,\Big[ k^{\mu}_{\perp}g_{\perp}^{\nu\eta}
-k^{\nu}_{\perp}g^{\mu\eta}_{\perp}\Big]
\nonumber 
\,\frac{1}{2} \frac{1}{n_+p}\delta(n_+q -n_+p) 
\big[\gamma^{\mu}_{\perp}\gamma^{\nu}_{\perp} \big]_{\gamma\beta}
\Bigg\}
\nonumber \\[1ex]
&&\hspace{-4cm}\times \,\epsilon^{*}(k)_{\eta}\, 
\sqrt{Z_{q,c}}|_{\text{tree}} \,u_{c,\beta}(p)
+ \mathcal{O}(\alpha_s) \,,
\end{eqnarray}
where $\sqrt{Z_{q,c}}|_{\text{tree}}=1$ is the tree-level value 
of the on-shell wave function renormalization factor of the 
quark field in the effective theory including the 
threshold-collinear mode. 
Calculating the contribution directly using the Feynman rule 
\eqref{eq:Srnonu} gives three contributions
proportional to different soft structures.  However, they are not
independent, as they are connected via the equation-of-motion  
identity \eqref{eq:eom}. We can use the transversality and on-shell 
conditions  $k\cdot \epsilon^*=0$ 
and $k^2=0$, respectively, for the emitted gluon, which have  
not yet been exploited in obtaining~\eqref{eq:treeMatch_bef_eom}. 
The relation $k\cdot \epsilon^*=0$  can be written in 
light-cone components as 
\begin{eqnarray}
\label{eq:onshelltransversality}
(n_+k)(n_-\epsilon^{*}\,)= 2 \left(-\frac{(n_-k)
(n_+\epsilon^{*}\,)}{2}-k_{\perp}\cdot\epsilon^{*}_{\perp}\, 
\right),
\end{eqnarray}
at which point we see that indeed we can express the second soft 
structure in the curly bracket of \eqref{eq:treeMatch_bef_eom} 
in terms of the first, 
\bea\label{eq:softrelation}
\left[(n_-k) n^{\nu}_{+}-
({n}_{+}k) {n}^{\nu}_{-}\right] \epsilon^*_{\nu }(k) =
-2 \left[k^{\nu}_{\perp}-\frac{k^2_{\perp}}{(n_-k)}
{n}^{\nu}_{-}\right]
\epsilon^*_{\nu }(k)\,. \eea 
This is expected as we know that the insertions of 
$\mathcal{L}^{(2)}_{1\xi}$ and $\mathcal{L}^{(2)}_{2\xi}$ 
contribute to the same collinear function $J_{1}$, since the 
soft structures are connected via \eqref{eq:eom}. 
Using this relation, we arrive at
\begin{eqnarray}
\label{eq:4.5}
\langle g(k)^K|\mathcal{T}^{1g}_{\gamma f}(n_+q)|q(p)_e\rangle
&=& 
2\pi \, \frac{g_s}{(n_-k)} \mathbf{T}^{K}_{fe} \,\Bigg\{
\nonumber\\
&&\hspace{-2cm} 
\left[ k^{\eta}_{\perp}
-\frac{k^2_{\perp}}{(n_-k)}n_-^{\eta} \right]
\left(-\frac{1}{n_+p}\delta(n_+q -n_+p) 
+2 \,\frac{\partial}{\partial n_+q}
\delta(n_+q -n_+p)\right)\delta_{\gamma\beta}
\nonumber \\[1ex]
&&\hspace{-2cm}-\,
\Big[ k^{\mu}_{\perp}g_{\perp}^{\nu\eta}
-k^{\nu}_{\perp}g^{\mu\eta}_{\perp}\Big]
\nonumber 
\,\frac{1}{2} \frac{1}{n_+p}\delta(n_+q -n_+p) 
\big[\gamma^{\mu}_{\perp}\gamma^{\nu}_{\perp} \big]_{\gamma\beta}
\Bigg\}
\nonumber \\[1ex]
&&\hspace{-4cm}\times \,\epsilon^{*}(k)_{\eta}\, 
\sqrt{Z_{q,c}}|_{\text{tree}} \,u_{c,\beta}(p)
+ \mathcal{O}(\alpha_s) \,.
\end{eqnarray}
Through comparison of (\ref{eq:4.5}) to 
\eqref{eq:4.10}, we find the tree-level 
collinear functions
\bea
{J}^{K(0)}_{1;\gamma\beta,fe}(n_+q,n_+p;\omega)
&=&   \mathbf{T}^{K}_{fe}  \delta_{\beta\gamma } 
\left(-\frac{1}{n_+p} \delta(n_+q -n_+p)  
+2 \,\frac{\partial}{\partial n_+q}
\delta(n_+q -n_+p) \right),\qquad\,\,
\label{eq:J1fn}
\\[1ex]
{J}^{\mu\nu,K(0)}_{6;\gamma\beta,fe}(n_+q,n_+p;\omega) &=&
-\frac{1}{2} \frac{1}{n_+p} \mathbf{T}^{K}_{fe} \,
\big[\gamma^{\mu}_{\perp}\gamma^{\nu}_{\perp} \big]_{\gamma\beta}
\,\delta(n_+q -n_+p)\,.
\quad
\eea 
We would like to draw attention to 
the factor of $-2$ in the second term of \eqref{eq:J1fn}
relative to the first that was not present in 
(\ref{eq:treeMatch_bef_eom}). Its origin can be traced 
back to the fact that the soft fields in the two terms 
giving rise to this contribution are connected by the 
equation-of-motion relation~\eqref{eq:eom}
precisely with this weight.
For the decomposition of the scalar collinear 
function $J_1$ introduced in \eqref{eq:collFuncDecomp}, 
Eq.~(\ref{eq:J1fn}) implies
\bea
{J}^{(0)}_{1,1}\left(n_+p;\omega\right) &=& 
-\frac{1}{n_+p}\,,\label{eq:j011}\\
{J}^{(0)}_{1,2}\left(n_+p;\omega \right) &=& 2\,.
\label{eq:j012}   
\eea
We recall from Sec.~\ref{sec:dynNLP} that the collinear function 
$J_6$ does not contribute to the DY cross section to 
any order in perturbation theory.

\subsubsection{Double soft parton structures}

\begin{figure}[t]
\begin{centering}
\includegraphics[width=0.85\textwidth]{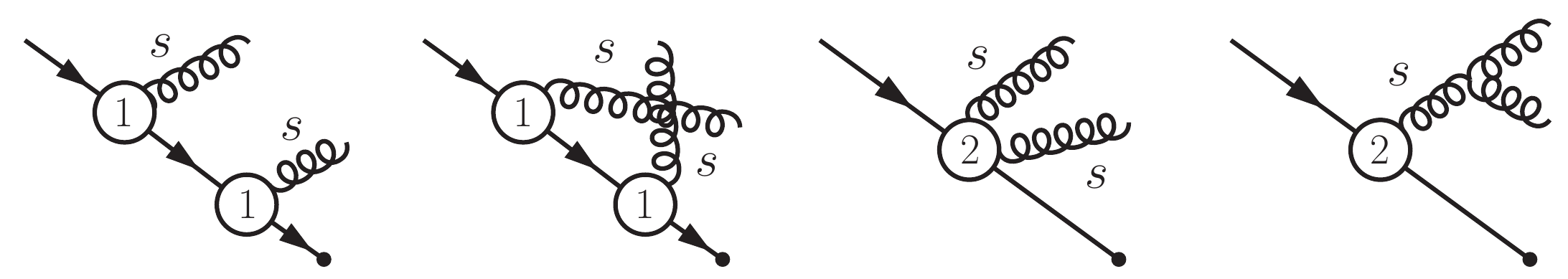}
\vskip0.2cm
\end{centering}
\begin{centering}
\includegraphics[width=0.4\textwidth]{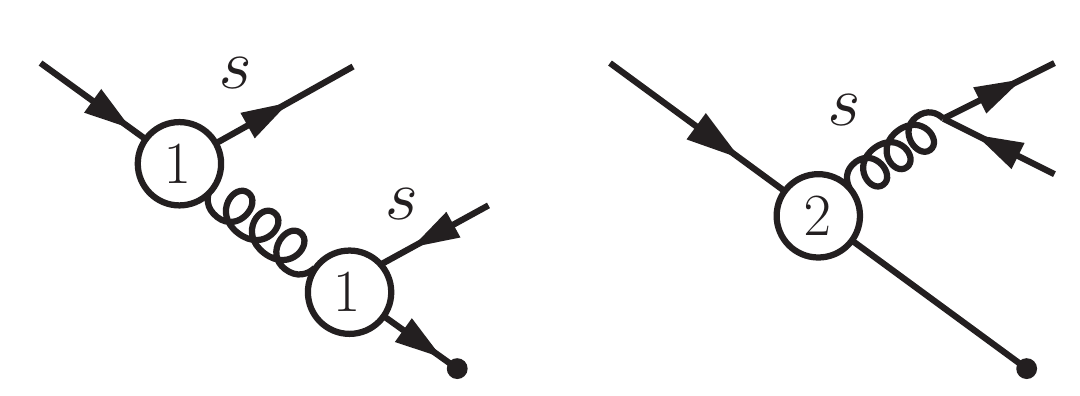}
\vskip0.2cm
\par\end{centering}
\caption{\label{fig:oneprdiagrams} 
Diagrams contributing to the matching of the two soft parton 
collinear functions.  Soft lines are labelled with an ``s''. 
The contributions from the one-soft-particle 
reducible diagrams, when the internal gluon originates from
$n_+\mathcal{B}^+$ term  in $\mathcal{L}^{(2)}$, 
are reproduced by the two parton terms in the 
equation of motion relation \eqref{eq:eom} applied to 
\eqref{eq:treeMatch_bef_eom}.}
\end{figure}

We now consider the collinear functions multiplying soft 
structures with at least two soft fields. In the graphical 
representation of Figure~\ref{fig:img4}, these correspond 
to diagrams with one external quark to the left and right, 
and two external soft gluons or a soft quark-antiquark 
pair attaching to $J$. 
The diagrams relevant to the tree-level 
matching computation are shown in Figure~\ref{fig:oneprdiagrams}.
Specifically, we require the single insertions 
of the $\mathcal{L}^{(2)}_{3\xi}$ and $\mathcal{L}^{(2)}_{5\xi}$
Lagrangians, and the double insertions of 
$\mathcal{L}^{(1)}_{\xi}$ and $\mathcal{L}^{(1)}_{\xi q}$, see
Appendix~\ref{sec:appendixA} for the definition of 
these terms. In addition, there exist 
one-soft-particle-reducible diagrams with an insertion of 
$\mathcal{L}^{(2)}_{1\xi}$, see the last diagram in 
each row in Figure~\ref{fig:oneprdiagrams}, since we 
eliminated  $n_+\mathcal{B}^+$ from the list of soft structures 
by the equation-of-motion relation \eqref{eq:eom}.

We start with collinear functions 
associated with two soft gluon emission at 
the same position $z_-$. The collinear 
functions due to insertions of $\mathcal{L}^{(2)}_{3\xi}$ and 
$\mathcal{L}^{(2)}_{5\xi}$ are calculated 
as for the single gluon emission, with a generalization 
of \eqref{eq:4.7} to the two-parton case and the 
$\mathfrak{s}_i$ structures given by third and fourth 
terms in \eqref{eq:softSet}. Since both terms involve  
$\mathcal{B}^+_{\mu_\perp}$ only, we choose the external 
soft gluon polarizations to be $\perp$ to extract the 
collinear function. The left-hand side of the 
matching equation is obtained by calculation of the third diagram
in Figure~\ref{fig:oneprdiagrams} with the appropriate 
Lagrangian insertions. The collinear function $J_3$, 
as defined by~\eqref{eq:3.24} with soft function 
\eqref{eq:3.26}, is given by  
\begin{eqnarray}\label{eq:J3collFuncDecomp}
	{J}^{}_{3;\gamma\beta}
	\left(n_+p,x_a\,n_+p_A; \omega \right) 
	&=& \delta_{\gamma\beta}\,
	\, \bigg[ {J}_{3,1}
	\left(x_an_+p_A; \omega \right) 
	\delta(n_+p-x_a n_+p_A)
	\nn \\ && \hspace{-2cm} 
    + \, {J}_{3,2}
	\left(x_an_+p_A; \omega \right) 
	\frac{\partial}{\partial (n_+p)} 
	\delta(n_+p-x_a n_+p_A) \bigg]\,
	\end{eqnarray}
with ${J}_{3,1}$ and ${J}_{3,2}$ to be determined.
A closer inspection of the subleading-power
SCET Lagrangian \eqref{eq:quarkint} shows that after the 
soft fields are stripped off, the remaining collinear 
parts of $\mathcal{L}^{(2)}_{3\xi}$ and 
$\mathcal{L}^{(2)}_{5\xi}$ are identical to those of 
$\mathcal{L}^{(2)}_{2\xi}$ and $\mathcal{L}^{(2)}_{4\xi}$, 
respectively.
This means that the collinear functions are the same, 
that is $J_{3,1}$ is equal to $J_{1,1}$, and 
$J_{2}$ to $J_{6}$. 
The one-soft-particle reducible diagram is 
only partly reproduced by the already determined single-gluon 
emission collinear function, since the $n_+\mathcal{B}^{+}$ soft 
field was eliminated from the basis of soft structures. 
The unaccounted piece in this diagram can be determined 
by explicit matching, or by making use of the single-gluon 
matrix element (\ref{eq:treeMatch_bef_eom}) before the 
on-shell and transversality of the external soft gluon 
was enforced. Replacing $n_+^\eta$ by the operator 
$n_+\mathcal{B}^+$, and then employing the operator 
equation-of-motion identity (\ref{eq:eom}) results in a term 
proportional to the two-soft gluon structure 
$\mathfrak{s}_3$. In this way, we deduce that $J_{3,2}$ in 
\eqref{eq:J3collFuncDecomp} is equal to $J_{1,2}$.
Alternatively, we could use  \eqref{eq:eom} directly in 
$\mathcal{L}^{(2)}_{1\xi}$, which then contains the same 
soft-gluon structure as  $\mathcal{L}^{(2)}_{3\xi}$, and 
derive  $J_{3,2}$ from the newly generated $q\bar{q} gg$ 
vertex.


It remains to consider the contribution from the 
double $\mathcal{L}^{(1)}$ insertions. 
The collinear matching equation for double Lagrangian insertions 
is
\bea\label{eqn:momentumMatch5} 
i^{\,2} \int d^4z_1\, d^4z_2  \,
\mathbf{T}\Big[ \chi_{c,\gamma f}
\left(tn_+\right)
\, \mathcal{L}^{(1)}_{}(z_1) 
\mathcal{L}^{(1)}_{}(z_2) 
\Big] && = 2\pi \sum_i \int 
\frac{dn_+p_a}{2\pi} du \,e^{i \,(n_+p_a) u} 
\nn \\&& \hspace{-6.5cm}\times 
\int \frac{d\omega_1}{2\pi}\, dz_{1-}
\,  e^{-i \omega_1\, z_{1-}}
\int \frac{d\omega_2}{2\pi}\, dz_{2-} 
\, e^{-i \omega_2\, z_{2-}}
\int \frac{dn_+p}{2\pi}e^{-i\,(n_+p) t}
\nn\\[0.1cm]
&& \hspace{-6.5cm}\times\,
{J}_{i;\gamma\beta,\mu,fbd}
\left(n_+p,n_+p_a;\omega_{1},\omega_{2} \right)
\chi^{{\rm PDF}}_{c,\beta b}(un_+)
\,\mathfrak{s}_{i;\mu,d}(z_{1-},z_{2-})\,.
\eea  
The partonic matrix elements to be calculated
here is $\langle g(k_1)g(k_2)| ...|q(p)\rangle$. 
The right-hand side of the matching equation is then
obtained as
\bea \label{eq:4.7.2}
\langle g(k_1)^{K_1}g(k_2)^{K_2}|
\mathcal{T}^{2g}_{\gamma f}(n_+q)|q(p)_e
\rangle  &=&  2\pi \int \frac{dn_+p_a}{2\pi}
du\,e^{i \,(n_+p_a)u}  \int\frac{d\omega_1}{2\pi} \, dz_{1-} 
\, e^{-i \omega_{1} z_{1-}}
\nn \\
&&\nn \hspace{-4.0cm}\times 
\int \frac{d\omega_2}{2\pi} \, dz_{2-} 
\, e^{-i \omega_{2}z_{2-}}\,
\Big( \,{J}^{\mu\nu,AB}_{4;\gamma\beta,fb}
\left(n_+q,n_+p_a;\omega_1,\omega_2 \right)\,
\\
&& \hspace{-4cm}\times \,\langle 0| 
\chi^{{\rm PDF}}_{c,\beta b}(un_+)|q(p)_e\rangle
\,\langle g(k_1)^{K_1}g(k_2)^{K_2}|
\mathfrak{s}^{AB}_{4;\mu\nu}(z_{1-},z_{2-})\,
|0\rangle \Big) \,,
\eea
The left-hand side is calculated as for the single soft gluon 
case, with $ \mathcal{L}^{(1)}_{\xi}$ 
in (\ref{eqn:momentumMatch5}). The relevant diagrams 
are the first two in the first line of 
Figure~\ref{fig:oneprdiagrams}.
After matching both sides of the equation we find
\begin{eqnarray}
J^{\mu\nu,AB\,(0)}_{4;\gamma\beta  ,fb }
\left(n_+q,n_+p;\omega_1,\omega_2\right) &=&\frac{2g^{\mu\nu}_{\perp}}{n_+p\,(\omega_1+\omega_2   )^2}
\left(  \, \omega_1   
\, \mathbf{T}^{A}\mathbf{T}^{B}  +\omega_2 \, \mathbf{T}^{B}_{ }\mathbf{T}^{A} 
\, \right)_{fb}\nonumber\\
&&\times\, \delta_{\gamma\beta} \,\delta(n_+q -n_+p)\,.
\end{eqnarray}
The calculation of the tree-level 
soft quark-anti-quark collinear function proceeds in the 
same way. The double $ \mathcal{L}^{(1)}_{\xi q}$ 
Lagrangian insertion contribution to the partonic matrix element
$\langle q(k_1)_{k_1}\bar{q}(k_2)_{k_2}| ...|q(p)\rangle$
can be written as 
\bea
\label{eq:4.7.3}
\langle q(k_1)_{k_1}\bar{q}(k_2)_{k_2}|
\mathcal{T}^{2q}_{\gamma f}(n_+q)|q(p)_e
\rangle  &=&  2\pi \int \frac{dn_+p_a}{2\pi}
du\,e^{i \,(n_+p_a)u} \int \frac{d\omega_1}{2\pi} \, dz_{1-} 
\, e^{-i\omega_{1}z_{1-}} 
\nn \\
&& \hspace{-4.0cm}\times
\nn \int 
\frac{d\omega_2}{2\pi} \, dz_{2-} 
\, e^{-i\omega_{2}z_{2-}}
\Big( \,\,{J}^{f g h b}_{5;\gamma\sigma\lambda\beta }
\left(n_+q,n_+p_a;\omega_1,\omega_2 \right)
\\&& \hspace{-4cm} \times\,\langle 0| 
\chi^{{\rm PDF}}_{c,\beta b}(un_+)|q(p)_e\rangle
\, \langle q(k_1)_{k_1}\bar{q}(k_2)_{k_2}|
\mathfrak{s}_{5;\sigma\lambda,gh}(z_{1-},z_{2-})\,
|0\rangle   \Big).
\eea 
The left-hand side corresponds to the first diagram in the 
second line of Figure~\ref{fig:oneprdiagrams}. Since for the 
quark-antiquark case we employed a non-redundant soft basis 
with the single bi-local soft structure $\mathfrak{s}_{5}$, 
the one-soft-particle reducible diagram in the same figure 
also contributes to $J_5$. The piece not already accounted for 
by the single-soft emission followed by a purely soft intercation 
can be obtained as for the 
two-gluon case from the quark-antiquark term in the operator 
equation-of-motion identity (\ref{eq:eom}). Adding both 
contributions, we obtain
\begin{eqnarray}
\label{eq:twoquarkcolfunc}
{J}^{f k_1 k_2 e\,(0)}_{5;\gamma\sigma\lambda\beta }
(n_+q,n_+p; \omega_1, \omega_2)&=&
-\mathbf{T}^A_{f k_2}\mathbf{T}^A_{k_1 e}\,
\frac{1}{n_+p}
\frac{  \omega_2}{(\omega_1+\omega_2)}
\frac{\slashed{n}_{-\gamma\eta}}{2}
\gamma^{\mu}_{\perp,\eta\sigma}
\gamma^{}_{\perp\mu,\lambda\beta} \,\delta(n_+q -n_+p)
\nonumber \\ && \hspace{-2cm}
+\,2 \,\mathbf{T}^{K}_{fe} \textbf{T}^{K}_{k_1k_2}
		\frac{ \omega_1 \omega_2}{(\omega_1+\omega_2)^2}    
	\slashed{n}_{-\lambda\sigma}
\delta_{\gamma\beta}  \,\frac{\partial}{\partial n_+q}
\delta(n_+q -n_+p)\,.\qquad
\end{eqnarray}

\subsection{Collinear functions at $\mathcal{O}(\alpha_s)$}
\label{subsec:cfunc1loop}

In this section we focus on demonstrating the consistency of 
the concept of collinear functions by calculating $J_1$ and $J_6$ at 
the one-loop level. $J_1$ is also the only collinear function  
which is needed at the one-loop order 
to verify the NLP factorization formula at NNLO accuracy, 
see \eqref{eq:3.43}. We do not calculate the loop 
correction to the collinear functions of the 
two soft-parton structures, since it is a next-to-next-to-next-to-leading 
order (NNNLO) effect.

\begin{figure}[t]
\begin{centering}
\includegraphics[width=0.65\textwidth]{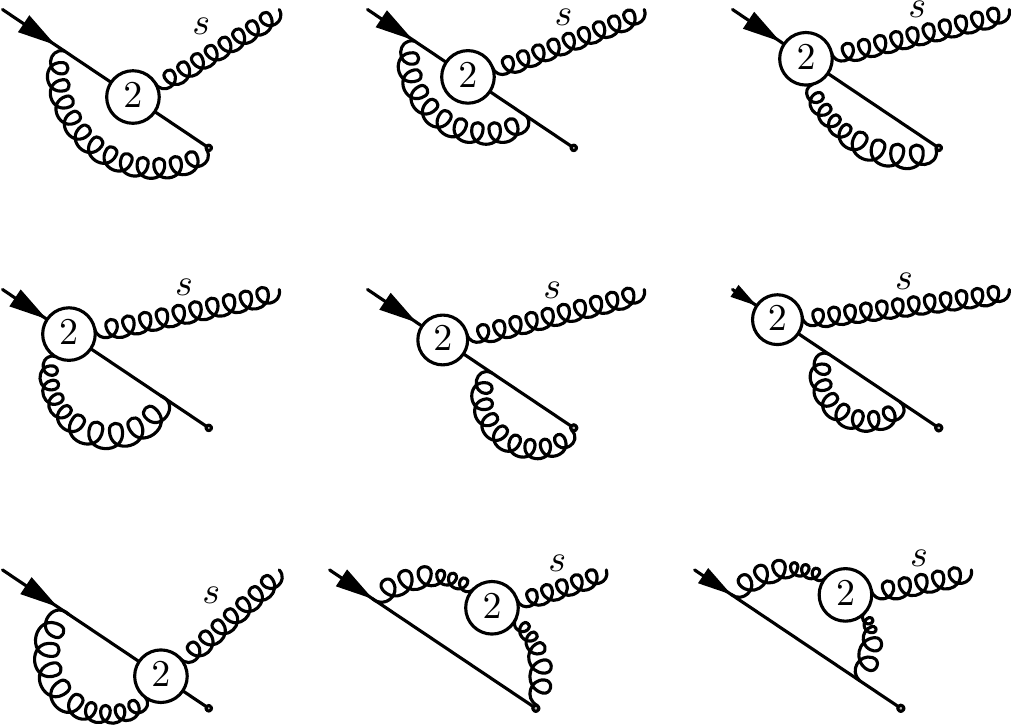}
\vskip0.2cm
\par\end{centering}
\caption{\label{fig:CollinearFunctionAmplitudes} 
One-loop collinear diagrams with one external soft gluon (labelled 
``s''). The dot at the right end of the solid quark line denotes the 
$\chi_c$ field from the LP current.  
The collinear gluon in the loop attaches either 
to the collinear quark or to the collinear Wilson 
line in the definition of the $\chi_c$ field.}
\end{figure}

The right-hand side of the matching equation has already been 
obtained in \eqref{eq:4.10}, which is valid to all orders 
in $\alpha_s$. The on-shell 
wave function renormalization factor should now be evaluated 
with one-loop accuracy. However, when dimensional regularization 
is used for ultraviolet and infrared divergences, 
$\sqrt{Z_{q,\text{PDF}}}=1$ to all orders, because the 
loops are scaleless.\footnote{The same statement applies to 
$\sqrt{Z_{q,c}}$ on the left-hand side of the matching equation, 
which will be used below.} 
The coupling renormalization is also the same on both sides of 
the matching equation, and drops out at the one-loop order. 

We therefore focus on the calculation of
$\langle g(k)^K|\mathcal{T}^{1g}_{\gamma f}(n_+q)|q(p)_e\rangle$ 
on the left-hand side of \eqref{eqn:matching4.5}, which requires the 
calculation of the Feynman diagrams with one collinear loop and 
a single soft emission, generated by insertions of the 
power-suppressed Lagrangian. The relevant SCET diagrams are 
shown in Figure \ref{fig:CollinearFunctionAmplitudes}.  
The circled vertex denotes the subleading-power Lagrangian 
insertion, while all other vertices are LP interactions. 

\subsubsection{Detailed computation}

\begin{figure}
\begin{centering}
\hskip-1cm
\includegraphics[width=0.27\textwidth]{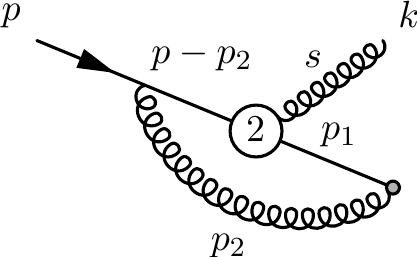}
\par\end{centering}
\caption{\label{fig:dy2_colour2} 
One of the diagrams contributing to the one-loop collinear 
functions. Through calculation of this diagram using Feynman 
rules from \cite{Beneke:2018rbh} we can obtain the $J_{1}$ and 
$J_6$ collinear functions, corresponding to insertions of 
$\mathcal{L}^{(2)}_{1\xi}$ and $\mathcal{L}^{(2)}_{2\xi}$, and 
$\mathcal{L}^{(2)}_{4\xi}$, respectively.}
\end{figure}

We illustrate the computation by considering as an example 
the top-left diagram in 
Figure~\ref{fig:CollinearFunctionAmplitudes}, which we 
draw again with momentum labels in Figure \ref{fig:dy2_colour2}.
All necessary Feynman rules were provided in Appendix A of 
\cite{Beneke:2018rbh} and (\ref{eq:Srnonu}). Applying them 
to the diagram under consideration leads to
\begin{eqnarray}
\label{eq:8.41}
\langle g(k)^K|\mathcal{T}^{1g}_{\gamma f}(n_+q)|q(p)_e
\rangle_{{\rm{fig}}\ref{fig:dy2_colour2}} &=& - 2 (2\pi) i g^3_s
\left(C_F -\frac{1}{2}C_A  \right) \mathbf{T}^K_{fe}
 \int \frac{d^dp_1}{(2\pi)^d}\int \frac{d^dp_2}{(2\pi)^d} 
\nn \\ 
&& \hspace{-2cm}
\times  \,\delta(n_+q-n_+p_1-n_+p_2)  \, \frac{1}{(n_+p_2)}
\frac{n_{+}p_1}{p_1^2} \,\frac{n_{+}(p-p_2)}{(p-p_2)^2} \, 
\frac{1}{p_2^2}
\nn  \\ 
&& \hspace{-2cm}\times 
\left[S^{\sigma\delta}(-k,p-p_2,p_1)u_{c}(p)\right]_{\gamma}
 \big(-k_{\sigma}g_{\delta\nu}+k_{\delta}g_{\sigma\nu}\big) \, \epsilon^{*\nu}(k).\quad
\end{eqnarray}
After substituting the expression for 
$S^{\sigma\delta}(-k,p-p_2,p_1)$ from~\eqref{eq:Srnonu} 
and performing an integration by parts of the derivative with 
respect to $p_1$ contained in $S$, the derivative 
acts on the integrand 
including the delta function in the second line. At this point, the momentum 
conservation delta function at the subleading-power interaction 
vertex can be imposed by performing the integral over $p_1$. 
This identifies $p^{\mu}_1= p^{\mu}-p_2^{\mu} -k^{\mu}_+$ and 
results in   
\begin{eqnarray}
\label{eq:8.44}
\langle g(k)^K|\mathcal{T}^{1g}_{\gamma f}(n_+q)|q(p)_e
\rangle_{{\rm{fig}}\ref{fig:dy2_colour2}} &=& -  (2\pi) i g^3_s
\left(C_F -\frac{1}{2}C_A  \right) \mathbf{T}^K_{fe}
 \,\int \frac{d^dp_2}{(2\pi)^d} \, 
\nn \\
&& \hspace{-2.0cm}\times\,\frac{1}{(n_+p_2)}\,
\frac{n_{+}(p-p_2)}{(p-p_2)^2} \, \frac{1}{p_2^2}\,   
\big(-k_{\sigma}g_{\delta\nu}+k_{\delta}g_{\sigma\nu}\big) 
\, \epsilon^{*\nu}(k)\nn \\ \nn
&&\hspace{-2.0cm} \times \,
\Bigg\{ n_{-}\cdot\frac{\partial}{\partial p_{1}}
\bigg(\delta(n_+q-n_+p_1-n_+p_2)  \,
\frac{n_{+}p_1}{p_1^2}\bigg) \,n_+^{\sigma} n_-^{\delta}\, 
u_{c,\gamma}(p)\, \nn \\
&&\hspace{-2.0cm}  - \, \bigg(k_\perp\cdot \frac{\partial}{\partial p_{1 \perp}}\bigg) \frac{\partial}{\partial p^{}_{1 \perp \sigma}}\bigg(\delta(n_+q-n_+p_1-n_+p_2)  \,
\frac{n_{+}p_1}{p_1^2}\bigg)\,n_-^{\delta} \, u_{c,\gamma}(p)\,\nn \\ \nn 
&& \hspace{-2.0cm}+\, \frac{\partial}{\partial p^{}_{1 \perp \sigma}}\Bigg[\bigg(\delta(n_+q-n_+p_1-n_+p_2)  \,
\frac{n_{+}p_1}{p_1^2}\bigg)\\  
&& \hspace{-2.0cm}\times \bigg(\frac{\slashed{p}_{1\,\perp}}{n_+p_1} \gamma^{\delta}_\perp -\gamma^{\delta}_{\perp}\frac{\slashed{p}_{2\, \perp}}{n_+(p-p_2)}\bigg)_{\gamma \beta}  \Bigg]u_{c,\beta}(p) \Bigg\}\Bigg\lvert_{p_1=p-p_2-k_+} \,  .\,
\end{eqnarray}
In this equation, only the term with the derivative 
$n_{-}\cdot\frac{\partial}{\partial p_{1}}$ gives a 
non-vanishing contribution to the derivative component $J_{1,2}$ 
of the collinear function defined in \eqref{eq:collFuncDecomp}, 
when it acts on the delta function.\footnote{In the factorization 
formula, once 
the derivative in  \eqref{eq:collFuncDecomp} is integrated by 
parts, it acts on the hard function, which is, however, constant 
at tree level. Hence the one-loop correction to $J_{1,2}$ 
contributes first at NNNLO together with the one-loop hard and 
soft function.} The remainder of the computation proceeds in the 
standard way, and we obtain a 
result valid to all orders in 
$\epsilon$. Expanding it here for illustration, we find 
\begin{eqnarray}
\label{8.2123S}
\langle g(k)^K|\mathcal{T}^{1g}_{\gamma f}(n_+q)|q(p)_e
\rangle_{{\rm{fig}}\ref{fig:dy2_colour2}} &=&   2\pi\,
\frac{g_s\alpha_s }{4\pi} \left(C_F -\frac{1}{2}C_A  \right) 
\frac{\mathbf{T}^K_{fe}}{(n_+p)}  
\bigg[ \frac{(n_+p)(n_-k)}{\mu^2}
\bigg]^{-\epsilon} 
\nn\\
&&\hspace{-3.7cm} \times \Bigg\{ \delta(n_+q -n_+p)\,
\bigg[ {2\delta_{\gamma\beta}\left( \frac{({n}_{+}k)}{(n_-k)} {n}^{\nu}_{-}
-n^{\nu}_{+}\right)}\nn \\
&& \hspace{-2cm}\nn
+  \,\delta_{\gamma\beta}{\left(\frac{k^2_{\perp}\,{n}^{\nu}_{-}}{(n_-k)^2}
-\frac{k^{\nu}_{\perp}}{( n_-k)}\right)}
\left(-\frac{2}{\epsilon ^2}-\frac{2}{\epsilon }+ 2+\frac{\pi^2}{6}\right)\, \nn\\
&& \hspace{-2.0cm}+\,\frac{\big[\gamma^{\nu}_{\perp}\,,\,\nn
\slashed{k}_{\perp}\big]_{\gamma\beta}}{(n_-k)}
\left(-\frac{1}{\epsilon ^2}+\frac{\pi^2}{12}
\right)  +\mathcal{O}(\epsilon)  \bigg] \\
&& \hspace{-3cm} +\,\frac{\partial}{\partial n_+q_{}}\,\delta(n_+q-n_+p)  \,\delta_{\gamma\beta} \left(\frac{({n}_{+}k)}{(n_-k)} {n}_-^\nu- {n}_+^\nu\right)\nn \\
&& \hspace{-2cm}\times \left( -\frac{2}{\epsilon^2}-\frac{2}{\epsilon}-4+\frac{\pi ^2}{6}  +\mathcal{O}(\epsilon) \right) \Bigg\}\,u_{c,\beta}(p)
\epsilon^{*\,}_{\nu }(k)\,.
\end{eqnarray}
The transversality and on-shell conditions  $k\cdot \epsilon^*=0$ 
and $k^2=0$, respectively, for the emitted gluon, have 
not yet been used in obtaining~\eqref{8.2123S}. 
\subsubsection{Amplitude calculation results}

The calculation of all diagrams in 
Figure~\ref{fig:CollinearFunctionAmplitudes}  gives the 
following result, after using the on-shell and transversality 
relations:  
\begin{eqnarray}
&& 
\langle g(k)^K|\mathcal{T}^{1g}_{\gamma f}(n_+q)|q(p)_e\rangle^{(1)} 
\nonumber\\
&& \hspace*{1cm} 
= \, 2\pi \,\frac{g_s\alpha_s}{4 \pi} \,\mathbf{T}^{K}_{fe}
\left[\frac{k_{\perp}^{\eta} }{(n_-k)}
-\frac{k_{\perp}^2 n_-^{\eta}}{(n_-k)^2} \right] 
\epsilon^{*}_{\eta  }(k)\, u_{c,\gamma}(p)\, 
\frac{\delta(n_+q -n_+p)}{n_+p}
\left(\frac{n_-k\, n_+p}{\mu ^2}\right)^{-\epsilon }
\nonumber\\[1ex]&&  
\hspace{1.5cm}\times    
\left( C_F\left(-\frac{4}{\epsilon}+3 
+8\epsilon+\epsilon^2  \right)-  C_A \left(-5+{8}{\epsilon}
+\epsilon ^2 \right)\right) 
\frac{e^{\epsilon \gamma_E} \,\Gamma[1+\epsilon ]
\Gamma [1-\epsilon]^2}{(-1+\epsilon)(1+\epsilon) 
\Gamma [2-2 \epsilon ]}\nn\\[1ex] 
&& \hspace{1.5cm}
+\,2\pi \,\frac{g_s\alpha_s}{4 \pi}  \,\mathbf{T}^{K}_{fe}
\left[ \frac{{k^{\mu}}_{\perp}
	{\epsilon}^{*\nu}_{\perp }(k)}{{n_-k}}-
\frac{{k^{\nu}}_{\perp}
	{\epsilon}^{*\mu}_{\perp }(k)}{{n_-k}} \right]
u_{c,\beta}(p) \,
\frac{\delta(n_+q -n_+p)}{n_+p}
\left(\frac{{n_-k}\, {n_+p}}{\mu ^2}\right)^{-\epsilon }
\nonumber\\
&&\hspace{1.5cm} \times\,\big[ \gamma^{\mu}_{\perp }
\gamma^{\nu}_{\perp }\big]_{\gamma\beta}
\left( {C_F}-{C_A} \right) 
\frac{e^{\epsilon \gamma_E} \, \Gamma[1+\epsilon ]
\Gamma[1-\epsilon ]^2 }{2\,\Gamma[2-2 \epsilon ]}
\,. \qquad   
\label{eq:oneloopt1amp}
\end{eqnarray}
These results constitute the left-hand side of the matching 
equation, that is, the extension of \eqref{eq:4.5} to one-loop
accuracy.  Remarkably, we find that 
the one-loop correction to the derivative delta-function term 
cancels exactly when all diagrams are added, which explains 
the absence of such term in the above equation.

\subsubsection{Collinear functions results at the 
one-loop order}

Comparing \eqref{eq:oneloopt1amp} to \eqref{eq:4.10} we obtain 
the one-loop correction to $J_1$ and $J_6$. 
We give the $d$-dimensional result and 
its expansion in $\epsilon=(4-d)/2$ in the following:
\begin{eqnarray}
\label{J1}
&& {J}^{K\,(1)}_{1,1;\gamma\beta, fe}
\left(n_+q, n_+p; \,\omega \right) = 
\frac{\alpha_s}{4 \pi} 
\delta_{\gamma\beta}\mathbf{T}_{fe}^{K} 
\, \frac{1}{(n_+p) } \left(\frac{n_+p\,\omega}
{\mu ^2}\right)^{-\epsilon }
\frac{e^{\epsilon\,\gamma_E}\,\Gamma[1+\epsilon ]
\Gamma [1-\epsilon]^2}
{(-1+\epsilon)(1+\epsilon) \Gamma [2-2 \epsilon ]}
\nn\\ 
&& \hspace{1.5cm}\times  
\left( C_F\left(-\frac{4}{\epsilon}+3 
+8\epsilon+\epsilon^2  \right)
-  C_A \left(-5+{8}{\epsilon}
+\epsilon ^2 \right)\right) \delta(n_+q -n_+p)\qquad
\\
&& \hspace{1cm} = \label{J1expanded}
\frac{ \alpha_s}{4 \pi} \,
\frac{1}{(n_+p)}  \, \delta_{\gamma\beta}
\mathbf{T}_{fe}^{K}
\bigg( C_F\left(\frac{4}{\epsilon }+5-4 \ln
\left(\frac{n_+p\,\omega}{\mu^2}\right) \right) 
-  5\, C_A \bigg)\,\delta(n_+q -n_+p)  
\nonumber\\
&& \hspace{1.5cm} +\,\mathcal{O}(\epsilon) \,,\qquad
\\[0.2cm]
&& {J}^{K\,(1)}_{1,2;\gamma\beta, fe}
\left(n_+q, n_+p; \,\omega \right) = 0\,,
\\[0.2cm]
&& {J}^{\mu\nu,\,K\,(1)}_{6;\gamma\beta,f e}
\left(n_+q, n_+p; \,\omega \right)= \label{J6}
\frac{\alpha_s}{4 \pi   } 
\frac{1}{( {n_+p})}
\,\left[ \gamma^{\mu}_{\perp }
\gamma^{\nu}_{\perp } \right]_{\gamma\beta}
\mathbf{T}^K_{fe} 
\,\left(\frac{ n_+p \,\omega }{\mu ^2}\right)^{-\epsilon }
\nn\\ && \hspace{1.5cm} \times  
\frac{e^{\epsilon\,\gamma_E} \, \Gamma[1+\epsilon]
\Gamma[1-\epsilon ]^2 }{2\,\Gamma[2-2 \epsilon ]}
\left(C_F-C_A\right)\delta(n_+q -n_+p)
\\ 
&& \hspace{1cm}= 
\frac{\alpha_s }{4 \pi   } \frac{ 1 }{2}
\frac{1}{( {n_+p})} \,\left[ \gamma^{\mu}_{\perp }
\gamma^{\nu}_{\perp } \right]_{\gamma\beta}
\mathbf{T}^K_{fe}
\, \left(C_F-C_A\right)\delta(n_+q -n_+p)
+\mathcal{O}(\epsilon) \, .    
\end{eqnarray}
It is noteworthy that there are no $1/\epsilon^2$ poles 
in the $\mathcal{O}(\alpha_s)$ collinear functions. This implies 
that there are no leading (double) logarithmic (LL) 
contributions from the collinear functions and 
confirms the finding of \cite{Beneke:2018gvs} from the 
consistency of LL resummation. The absence of the  $1/\epsilon^2$ pole  
results from a cancellation and after applying the 
equation-of-motion relation, as can be seen from the fact 
that individual diagrams do contain it, see \eqref{8.2123S}. 
Moreover, for the $C_A$ colour coefficient, even the single 
pole cancels. The above one-loop corrections to the collinear 
functions constitute the second main result of this work. 
As noted earlier, neither ${J}^{K\,(1)}_{1,2;\gamma\beta, fe}$  
nor  ${J}^{K\,(1)}_{6;\gamma\beta, fe}$ contribute to 
the NNLO DY cross section.

\subsubsection{Relation to the LBK ampltiude and the radiative jet function}

The study of an amplitude with a next-to-soft emission has a long 
history starting from the Low-Burnett-Kroll formula and its 
extension to soft gluon emission from jets \cite{DelDuca:1990gz}. 
The emergence of the next-to-soft LBK amplitude within SCET was 
discussed in \cite{Larkoski:2014bxa,Beneke:2017mmf}. The calculation 
of the collinear functions at the one-loop level presented above 
forms part of the generalization of the LBK formula to the one-loop 
order. The complete next-to-soft, one-loop amplitude is 
provided in App.~\ref{appendix:amplituderesults}, including terms 
that vanish at the cross-section level due to the interference 
with the complex-conjugated tree-level amplitude. The result 
does not display any suggestive structure, and indeed, to our knowledge 
there is no simple representation of the one-loop result 
in terms of the angular momentum operator that would generalize 
the well-known expression of the tree-level next-to-soft 
amplitude. 

Next-to-soft emission at the one-loop order in amplitudes with a colourless 
final state has been studied before within the diagrammatic approach  
\cite{Bonocore:2015esa,Bonocore:2016awd,DelDuca:2017twk}, where 
the concept of a ``radiative jet function''  \cite{DelDuca:1990gz}
is used to describe the soft emission from jets. Ultimately, 
the formalism presented here aims to capture the same physics,
however there are conceptual differences. 
The most important one is that the radiative 
jet function, as can be found in (2.12) of \cite{Bonocore:2016awd},
is not a single scale object unlike the collinear functions 
defined in \eqref{eq:genMatch}. This fact can be seen in the result for 
the one-loop radiative jet function given in (3.3) of \cite{Bonocore:2016awd}.
In addition to the collinear contributions, there exist subtraction terms
which account for the overlap of the radiative jet function with 
the soft function. No such complications arise here, which makes the
effective field theory construction more suitable for resummation using 
renormalization group techniques. (Nevertheless, NLP resummation near 
the Drell-Yan threshold using diagrammatic techniques has been achieved 
at LL accuracy \cite{Bahjat-Abbas:2019fqa} owing to the fact that 
the radiative jet or collinear functions do not contribute beyond 
tree level at this accuracy \cite{Beneke:2018gvs}.) 

Since the radiative jet function contains both collinear and soft 
contributions, in order to compare our collinear functions with 
results given in \cite{Bonocore:2016awd} it is necessary to 
multiply the collinear functions with their corresponding soft 
structures. At this point, it is most convenient to compare 
the radiative jet function in \cite{Bonocore:2015esa,Bonocore:2016awd}
with our results
for the soft emission amplitude at NLP calculated within SCET
and written in App.~\ref{appendix:amplituderesults}.
The relevant contributions are given
in   \eqref{eq:B4}, \eqref{eq:B5},  \eqref{eq:B6},
and  \eqref{eq:B7}. 
We compare these expressions (appropriately expanded in powers of $\epsilon$) with $J^{(1)}_{\mu,F}$ and $J^{(1)}_{\mu,A}$
given in \cite{Bonocore:2016awd}. We find agreement for all terms,\footnote{Noting the typo in (3.3) of
\cite{Bonocore:2016awd} 
where one must replace	$(-2p\cdot k)^{-\epsilon}\to (2p\cdot k)^{-\epsilon}$
and a overall minus sign error in one-loop results given in 
\cite{Bonocore:2015esa}.}
except for contributions \eqref{eq:B6} and  \eqref{eq:B7} proportional to $n_-^{\rho}/(n_-l)$.
Given that our calculation gives the full amplitude with the emission of a soft gluon,
we conclude that the radiative jet function in \cite{Bonocore:2015esa,Bonocore:2016awd} 
fails to reproduce the complete amplitude, 
although the missing terms do not contribute to the matrix element squared 
at NLP.  
It would be interesting to 
investigate further what is the underlying reason for this discrepancy,
which is beyond the scope of this work. 
We speculate that contributions similar to those from the 
$J^{A0,A1}$ and $J^{A0,B1}$ SCET currents are needed in the radiative jet
function formalism.

\section{Fixed-order results}
\label{sec:fixedresults}

There exists a number of NLP results for the DY process  
at NLO and NNLO in the 
strong coupling \cite{Hamberg:1990np,Laenen:2010uz,Bonocore:2014wua,Bonocore:2015esa,Bonocore:2016awd}, obtained from direct expansions of the QCD
diagrams. In this section we verify the correctness of the 
NLP factorization formula by comparing to these results and own 
results from the expansion-by-regions method \cite{Beneke:1997zp}.

\subsection{NLO}

Expanding the NLP factorization formula to NLO, one finds 
only one dynamical contribution to the cross section, 
since the soft function begins at $\mathcal{O}(\alpha_s)$. 
The one-loop soft function is given by\footnote{The expansion 
in $\epsilon$ was already presented in \cite{Beneke:2018gvs}.} 
\begin{eqnarray}\label{eq:softfunc}
  S^{(1)}_{1}\left(\Omega,\omega\right) = \frac{\alpha_sC_F}{2\pi}
  \frac{\mu^{2\epsilon}e^{\epsilon\gamma_E}}{\Gamma[1-\epsilon]}
  \,\frac{1}{\omega^{1+\epsilon}}\frac{1}{(\Omega-\omega)^{\epsilon}}
  \,\theta(\omega)\theta(\Omega-\omega) \,\,.\quad
\end{eqnarray}
Using this result in \eqref{eq:3.41} and performing the 
convolution integral over $\omega$ gives
\begin{eqnarray}\label{NLP-NLO-dyn}
   \Delta^{dyn\,(1)}_{{\rm{NLP}}}(z)&=& 
   \frac{\alpha_s}{4\pi}C_F\left(
   \frac{8}{\epsilon }-16 \ln (1-z)- \epsilon  
   \left(2\pi ^2-16 \ln^2(1-z)\right)+O(\epsilon ^2)
\right),
\end{eqnarray}
where we set $\mu=Q$.

In addition, at NLO we need to take into account the kinematic 
corrections $\Delta^{K1}_{{\rm{NLP}}}(\Omega)$, 
$\Delta^{K2}_{{\rm{NLP}}}(\Omega)$, and 
$\Delta^{K3}_{{\rm{NLP}}}(\Omega)$ in \eqref{3.29}, \eqref{3.30}, 
and \eqref{3.31}.  For the latter two, we can use  
\eqref{eq:softfunc} and  $H^{(0)}(Q^2)=1$, since no derivatives 
with respect to the coordinate $\vec{x}$ needs to be taken.
To compute $\Delta^{K1}_{{\rm{NLP}}}(\Omega)$ we use 
the result for the one-loop soft function with full $x$ dependence
from \cite{Beneke:2018gvs,Li:2011zp}. 
Upon  summing the three kinematic corrections
 we obtain
\begin{eqnarray} \label{NLP-NLO-kin}
\Delta^{kin\,(1)}_{{\rm{NLP}}}(z) = \frac{\alpha_s\,C_F}{4\pi}
\Big(8 -\epsilon \,16\ln(1-z)\Big) \,.
\end{eqnarray}

Results for the NLO NLP contribution to DY production have been 
presented in \cite{Laenen:2010uz} within a diagrammatic approach, 
in which power-suppressed soft radiation is described it terms 
of generalized next-to-soft Wilson lines. Our result 
\eqref{NLP-NLO-dyn} agrees with the corresponding expression 
Eq.~(6.17) of \cite{Laenen:2010uz}. The kinematic correction
\eqref{NLP-NLO-kin} is provided in Eq.~(6.13) of \cite{Laenen:2010uz} as
a correction to the LP matrix element. Agreement can be easily checked. After summing~\eqref{NLP-NLO-dyn} and 
\eqref{NLP-NLO-kin}, and applying the subtractions that arise 
from PDF renormalization, we also find agreement with the 
NLO NLP result reported in Eq.~(B.29) of~\cite{Hamberg:1990np}.

\subsection{NNLO}
\label{sec:NNLOresults}
In Sec.~\ref{sec3.3} the three possible dynamical NLP 
contributions to the cross section at NNLO have been discussed. 
These are collinear, hard, and soft contributions presented in 
\eqref{eq:3.43}, \eqref{eq:3.44} and \eqref{eq:3.34s}, 
respectively. In this section we explicitly compute and 
check the first two of these. The soft
contribution requires a full NLP NNLO soft 
function computation, which is beyond the scope of this work. 
However, we present the one-virtual, one-real soft contribution to the cross section here. Also, in  Appendix~\ref{appendix:amplituderesults}
we present complete results for the one-loop power-suppressed 
amplitude with one real soft emission, including 
the soft loop contribution. The latter forms part of the 
virtual-real contribution to the NNLO soft function. The missing
contribution comes from double real soft emission,
which we leave for future work. 

\subsubsection{Collinear contribution}  
   
This contribution comes from the one-loop collinear functions 
combined with the NLO soft function and tree-level hard 
function, see \eqref{eq:3.43}. We recall that the delta-function 
derivative term in the collinear function, spelled out 
in \eqref{eq:collFuncDecomp}, vanishes after partial integration, 
since the hard function at tree level is a constant. The 
one-loop collinear function that is required
is then given by \eqref{J1} with colour generator and Dirac-index Kronecker-symbol removed.   

For the purpose of deriving the NNLO fixed-order result, 
we keep must use the $d$-dimensional expression of the 
collinear function and perform the convolution with the 
$d$-dimensional soft function. 
Then expanding in $\epsilon$ and setting $\Omega=Q(1-z)$, 
$\mu=Q$ yields  
\begin{eqnarray}\label{sig9}
\Delta^{dyn\,(2)}_{{\rm{NLP-coll}}}(z)&=& 
\frac{  \alpha_s^2 }{(4\pi)^2} 
 \Bigg(C_F^2 \,\bigg(-\frac{16}{\epsilon ^2}
 +\frac{48 \ln (1-z)-20}{\epsilon } \nn \\ 
&& + 
 \left(-72 \ln ^2(1-z)+60 \ln (1-z)+8 \pi ^2-24\right) 
 +\mathcal{O}(\epsilon)  \bigg)\nonumber  \\
&& + \,{C_A}C_F
 \left( \frac{20}{\epsilon }-(60 \ln (1-z)-8)
 +\mathcal{O}(\epsilon)  \right)\Bigg)\,.
 \end{eqnarray} 
Notice that, as expected, there are no leading logarithms 
$\mathcal{O}(\alpha_s^2\ln^3(1-z))$ in the collinear contribution, 
since the highest power of 
the logarithm in the finite terms in the second line 
is NLL accuracy, $\ln^2(1-z)$. 

Results describing virtual collinear radiation
at one loop with emission of a soft gluon have 
been derived in \cite{Bonocore:2014wua} within the
expansion-by-regions approach~\cite{Beneke:1997zp}, 
and in \cite{Bonocore:2015esa,Bonocore:2016awd} within a 
diagrammatic approach, 
in which the effect of collinear loops is described in terms
of a ``radiative jet function''. The $C_F^2$ term in~\eqref{sig9} 
is in agreement with the corresponding contribution
in Eqs.~(13), (14) of \cite{Bonocore:2014wua} and Eq.~(4.22) 
of \cite{Bonocore:2015esa}, where the abelian contribution 
only is considered.\footnote{Notice also that these references drop
all contributions proportional to transcendental numbers, such as 
$\pi^2$ and $\zeta(3)$.} The $C_A C_F$ term in our result 
\eqref{sig9} is not provided separately in literature, but only 
in sum with the hard and soft contribution, that we  
consider in the following. 

\subsubsection{Hard contribution}

Next we check the contribution composed of the one-loop hard 
function, the tree-level collinear functions, and the 
one-loop soft function. In contrast to the collinear contribution, 
here the collinear function with the derivative contributes, 
since the hard matching coefficient is momentum-dependent 
beyond tree level. 

The relevant formula is now 
\eqref{eq:3.44}, which already made use of the
expressions for the collinear functions at tree level.
The one-loop soft function was given 
in \eqref{eq:softfunc}. The $d$-dimensional hard matching coefficient at the one-loop order can be found in Eq.~(2.23) of \cite{Gehrmann:2010ue}, 
\begin{eqnarray}\label{eq:WilsonC}
{C}^{A0,A0}(n_+p, n_-\bar{p}) &=& 
1+ \frac{\alpha_s}{4\pi}C_F
\left(\frac{-Q^2}{\mu^2}\right)^{\!-\epsilon}
\bigg(-\frac{2}{\epsilon^2}  -\frac{3}{\epsilon}-8
 +\frac{\pi^2}{6} \nn \\
&& +\,\epsilon\left(\frac{\pi^2}{4}+\frac{14\zeta(3)}{3}-16\right)+\mathcal{O}(\epsilon^2)\bigg) 
 +\mathcal{O}(\alpha^2) \,,
 \end{eqnarray}
where $Q^2=(n_+p)(n_-\bar{p})$. Taking care of the imaginary part, 
the one-loop hard function $H = |{C}^{A0,A0}|^2$ reads 
\bea
\label{hardfunction}
H(Q^2) &=& 1 + \frac{\alpha_s C_F}{4\pi}\bigg(-\frac{4}{\epsilon ^2} 
-\frac{1}{ \epsilon }\left(4 \ln \left(\frac{\mu^2}{Q^2}\right)+6\right)
\nn \\ 
&&\hspace{-1cm}- \left(2 \ln ^2\left(\frac{\mu^2}{Q^2}\right)
 +6 \ln \left(\frac{\mu^2}{Q^2}\right)-\frac{7\pi ^2}{3} +16\right)
 + \epsilon  \,\bigg(-\frac{2}{3} \ln ^3\left(\frac{\mu^2}{Q^2}\right)
-3 \ln ^2\left(\frac{\mu^2}{Q^2}\right)  \nonumber \\
&&\hspace{-1cm} 
 + \left(\frac{7}{3} \pi ^2-16\right) \ln \left(\frac{\mu^2}{Q^2}\right) 
 +\frac{28}{3} \zeta (3)+\frac{7}{2} \pi ^2-32\bigg) +\mathcal{O}(\epsilon ^2)
\bigg) + \mathcal{O}(\alpha^2_s)\,.\quad
\eea
Performing the $\omega$-integration in \eqref{eq:3.44}, 
setting $\mu=Q$, and expanding in $\epsilon$ leads to 
\begin{eqnarray}\label{sig13}
 \Delta^{dyn\,(2)}_{{\rm{NLP-hard}}}&=&  
 \frac{\alpha_s^2C^2_F}{(4\pi)^2}
    \bigg(-\frac{32}{\epsilon ^3}
 +\frac{64 \ln (1-z)-16}{\epsilon ^2} \nn\\
&& \hspace{0cm}+\,\frac{-64 \ln ^2(1-z)+32 \ln (1-z)
 +\frac{80}{3} \left(\pi ^2-3\right)}{\epsilon }
 \nonumber \\
&&\hspace{0cm}-\,\frac{8}{3} \,
 \Big(-16 \ln ^3(1-z)+12 \ln ^2(1-z)
 +20 \left(\pi ^2-3\right) \ln (1-z) \nn\\
&& \hspace{0cm}-\,56 \zeta (3)- 5 \pi ^2+48\Big)
 + \mathcal{O}(\epsilon)  \bigg)\,,   
 \end{eqnarray}
where $\zeta (3)$ is a Riemann zeta value.  In contrast to the 
NLP collinear contribution, LLs appear in this expression.
Resummation of the hard function is therefore necessary 
in order to sum LLs to all orders in $\alpha_s$, 
as was done in \cite{Beneke:2018gvs}.

In the literature, the  hard one-loop times  one real soft gluon 
 result has been considered before within the
expansion-by-regions method. The expression for the abelian 
$C_F^2$ term has been given in Eq.~(12) of 
\cite{Bonocore:2014wua}, and agrees with 
(\ref{sig13}).\footnote{We note the following typo in
\cite{Bonocore:2014wua}: in Eq.~(12) $\big[1+4\log(1-z)\big]/\epsilon^2$ 
should be $\big[-1+4\log(1-z)\big]/\epsilon^2$.} Within the 
diagrammatic approach~\cite{Bonocore:2015esa,
Bonocore:2016awd}, the hard contribution \eqref{sig13}
arises from dressing the non-radiative amplitude by a one real
soft gluon, according to 
the LBK theorem. For a discussion of the LBK theorem in 
the present approach, see \cite{Beneke:2017mmf}. 

\subsubsection{Soft contribution}
\label{sec:softNNLO}

The soft contribution provided here is the one-real, one-virtual piece of the full NNLO soft function as mentioned in the introduction of Sec \ref{sec:NNLOresults}. In \eqref{eq:3.34s} one can see that there are contributions to the NNLO soft function from different soft structures. However, as detailed in Appendix \ref{sec:softloopperp}, 
only one soft structure, $S_1$, and corresponding  tree-level collinear function actually contribute to this piece. Hence the simplified factorization formula is
\begin{eqnarray}\label{sig6s1r1v}
   \Delta^{dyn\,(2)1r1v}_{{\rm{NLP-soft}}}= {4}  Q\,H^{(0)}(Q^2)\,
   \int d\omega  J^{(0)}_{1,1}\left(x_a(n_+p_A);\omega\right)
   S^{(2)1r1v}_{1}(\Omega,\omega).  
 \end{eqnarray}
The result for one-real, one-virtual contribution to the 
two-loop soft function reads
 \begin{eqnarray}\label{4.1.2.10}
S^{1r1v}_{1}(\Omega,\omega)& =& 
-4\,\frac{ \alpha_s^2}{(4\pi)^2} C_FC_A
\left(-\frac{\omega^2(\Omega-\omega)^2}{\mu^4}
\right)^{\!-\epsilon}
      \frac{1}{\omega} \nn \\
&&\times
\,\frac{1}{\epsilon^2}  
   \frac{e^{2\epsilon\gamma_E}\,\Gamma[1-\epsilon]^2}{\Gamma[1-2\epsilon]}  
   \Gamma[1+\epsilon]^2\,\,\theta(\Omega-\omega)
   \theta(\omega).
\end{eqnarray}
Using \eqref{eq:j011} for the tree-level collinear function in \eqref{sig6s1r1v}, integrating 
over $\omega$ and expanding in $\epsilon$ yields
\begin{eqnarray} \label{sig6s1r1vB}
\Delta^{dyn\,(2)1r1v}_{{\rm{NLP-soft}}}&=& 
\frac{\alpha_s^2}{(4\pi)^2} 
\,C_FC_A\,\bigg(-\frac{8}{\epsilon ^3}+
\frac{32 \ln (1-z)}{\epsilon ^2}-
\frac{64 \ln ^2(1-z)}{\epsilon }+\frac{28\pi^2}{3 \epsilon }
\nonumber     \\
&&\hspace{0cm}+\frac{256}{3} \ln ^3(1-z)-\frac{112}{3} \pi ^2 \ln (1-z)+\frac{448 \zeta (3)}{3} +\mathcal{O}(\epsilon) \bigg)\,.
 \end{eqnarray}
In the literature the non-abelian $C_F C_A$ term of the 
one-real, one-virtual contribution has been provided as a sum of 
the collinear, soft and kinematic correction (see Eq.~(4.6) of 
\cite{Bonocore:2016awd}), thus \eqref{sig6s1r1vB}
cannot be  compared directly. We 
performed an independent calculation of the full one-real, one-virtual
correction within the expansion-by-regions method, and \eqref{sig6s1r1vB} 
agrees with the soft region, as it should be.

\subsubsection{Kinematic contribution}

The kinematic correction from the 
sum of terms \eqref{3.29}--\eqref{3.32} can also be obtained at 
NNLO by using the NNLO soft function with full $x$-dependence 
presented in \cite{Li:2011zp}. We find 
\bea
\Delta^{kin\,(2)}_{{\rm{NLP}}}(z) &=& 
\frac{\alpha_s^2}{(4\pi)^2}\, \Bigg[ C_F^2
\bigg(\frac{16}{\epsilon^2}-\frac{192 \ln(1-z)+96}{\epsilon}+512\ln^2(1-z)
\nonumber \\  
&& \hspace{1.0cm}
+\, 192\ln(1-z)- 40\pi^2 -256 \bigg)  + C_F C_A \,\bigg(
\frac{88}{3\epsilon}-\frac{352\ln(1-z)}{3} \nonumber\\
&&\hspace{1.0cm} 
-\, \frac{8\pi^2}{3}+ \frac{476}{9} \bigg) + C_F n_f \bigg( 
 -\frac{16}{3\epsilon}+\frac{64\ln(1-z)}{3}-\frac{56}{9} \bigg)\Bigg]\,.
\label{eq:NNLOkin}
\eea 
We note that there are no LLs due to kinematic corrections. 

The kinematic contribution has been calculated previously within
the expansion-by-regions or the diagrammatic approach as 
the NLP phase-space corrections to the LP matrix element, but the 
expression corresponding to (\ref{eq:NNLOkin}) has not been 
provided explicitly. (It is part of Eq.~(4.6) and~(5.2) 
of \cite{Bonocore:2016awd},
but it cannot be separated from the NLP matrix element.) 
We thus compare  (\ref{eq:NNLOkin}) with an own independent expansion-by-regions   
calculation, in which we take 
the matrix element at leading power (both one-real, one-virtual
and two-real diagrams), and integrate it against 
the NLP phase space, finding agreement.

\section{Ill-defined convolution}
\label{sec:convolution}

One of the primary uses of factorization formulas in SCET is to 
perform resummation using renormalization group equations. 
Soft-collinear factorization often involves convolutions 
$C\otimes F$ of hard functions with collinear factors, for 
example, in deep-inelastic scattering or in convolutions with 
PDFs for any hadronic scattering cross section, or 
$J\otimes S$ of jet with soft functions, for example in the 
description of radiation from final-state jets. Resummation 
relies on defining renormalized factors by subtracting their 
poles in dimensional regularization and deriving a renormalization 
group equation for the renormalized function, which usually 
also has a convolution form. 
Large logarithms are then summed by evolving one function 
to the characteristic scale of the other. Finally, the 
convolution of the two factors is done.

This procedure evidently requires that the final convolution 
integral of the renormalized factors is well defined. As we 
discuss now, this important requirement is not satisfied 
by the NLP factorization formula for the DY process. 

The issue is most clearly exposed when we focus on the functional 
form of the objects appearing in the one-loop collinear times 
one-loop soft NNLO term in factorization formula given in
\eqref{eq:3.43}. The one-loop collinear function 
$J^{(1)}_{1,1}$ is taken from \eqref{J1}
and the soft function from \eqref{eq:softfunc}. The 
convolution integral reads
\bea
\label{eq:divconvolution} 
 \int_0^\Omega d\omega \,\underbrace{\big(n_+p
 	\, \omega\big)^{-\epsilon}}_{{\rm{collinear\,piece}}}
 \,\underbrace{\frac{1}{\omega^{1+\epsilon}}
 	\frac{1}{(\Omega-\omega)^{\epsilon}}}_{{\rm{soft\,piece}}}
 \,.\quad
 \eea 
It is evident that the integral is well defined when keeping
the exact $\epsilon$ dependence in the integrand, as was 
done in the previous section in order to obtain and reproduce the 
fixed-order NNLO NLP results. However, as explained above, for 
resummation we would like to treat the parts originating in 
the collinear function, $(n_+p\, \omega)^{-\epsilon}$, 
and the soft function pieces, ${\omega^{-1-\epsilon}}\,
{(\Omega-\omega)^{-\epsilon}}$, independently. That is, we wish 
to expand each in $\epsilon$ and define renormalized functions. 
However,  it is  clear that there is a problem when this 
procedure is attempted in (\ref{eq:divconvolution}).  
Concretely, one encounters a divergent integral, or 
$\int d\omega \,\delta(\omega) \ln (\omega)$ and 
other ill-defined integrals after introducing the 
standard plus distribution for the $1/\omega^{1+\epsilon}$ 
factor.\footnote{In principle, one can move integer powers of $\omega$ from the collinear to the soft function by adjusting powers of $1/in_-\partial$. However, this does not solve the problem, since there will always be a factor of $\omega^{-n\epsilon}$ associated with the collinear function.}

In order to make the issue even more explicit, we take the 
$\epsilon$-expanded collinear function given in \eqref{J1expanded}
and also expand the one-loop soft function \eqref{eq:softfunc}
in $\epsilon$, 
\begin{eqnarray}\label{eq:softfuncExpanded}
  S^{(1)}_{1}\left(\Omega,\omega\right) 
  = \frac{\alpha_s C_F}{4\pi} \left( 2\,\delta(\omega)\,
  \theta(\Omega)\left(-\frac{1}{\epsilon}
  +\ln\left(\Omega^2/\mu^2\right) 
  \right)   +2\left[\frac{1}{\omega} \right]_+
  \theta(\omega)\theta(\Omega-\omega)\right).\quad
  \end{eqnarray}
The convolution of this expression with \eqref{J1expanded} 
according to \eqref{eq:3.43} (at $\mu=Q$ as in the section above) 
gives
\begin{eqnarray}
\label{sigColl_illdefined2}
\Delta^{dyn\,(2)}_{{\rm{NLP-coll}}}(z)&=& 
\frac{\alpha_s^2 }{(4\pi)^2} \,
\Bigg(C_F^2 \,\bigg(\!-\frac{32}{\epsilon^2}
-\frac{8}{\epsilon } 
\left[5-8 \ln (1-z)-4 \int d\omega\, \delta(\omega) \ln
 \left(\frac{ \omega }{Q}\right)\right]\bigg)\nn\\ 
&& +\, {C_A}C_F \,\frac{40 }{\epsilon } +
\mathcal{O}(\epsilon^0) \Bigg)\,\quad
\end{eqnarray} 
where only the pole terms in 
$\epsilon$ are shown. There are two issues with this result. First, 
one of the terms with $1/\epsilon$ pole is ill-defined 
as we encounter the integral $\int d\omega \,
\delta(\omega) \ln (\omega)$. Second, the coefficient of the  
$C_F^2/\epsilon^2$, $C_F C_A/\epsilon$ pole terms which are not 
divergent have changed with respect to the 
correct result from \eqref{sig9} obtained from expanding 
in $\epsilon$ {\em{after}} performing the convolution in $d$ 
dimensions. 

It is clear from the above that it is not possible to obtain the 
NLP logarithms of $(1-z)$ correctly from the standard 
renormalization procedure and four-dimensional convolutions. 
The leading logarithms in the $q\bar{q}$ $(gg)$  channels 
in DY (Higgs) production summed in 
\cite{Beneke:2018gvs,Beneke:2019mua} form an exception, 
since they require only tree-level collinear functions 
and since the loop corrections to the collinear functions do 
not contribute leading logarithms. The ill-defined convolution, 
however, hampers the extension of resummation to NLL. 
The convolution itself requires subtraction, and contributes 
to the logarithms, which can therefore not be obtained 
from the separate renormalization group equations for the 
renormalized collinear and soft functions. Nevertheless, the 
NLP formula derived in this paper factorizes the different 
momentum scales of the DY process consistently at the level 
of regularized matrix elements of the soft and collinear operators, 
and therefore can be justifiably called a factorization 
formula. It may be hoped that it provides the starting point 
for understanding how to renormalize $d$-dimensional 
convolutions, which appear to be a generic feature of 
NLP factorization.\footnote{See also \cite{Beneke:2019kgv}, where a 
different type of divergent convolution is discussed.}

\section{Summary}
\label{sec:summary}

In this work, we derived for the first time a factorization formula 
for DY production near threshold in the $q\bar{q}$-channel 
at general powers in the $(1-z)$ expansion. We then focused on  
the next-to-leading power, which entails several simplifications. 
The main result is the NLP factorization formula \eqref{eq:3.24}, 
which generalizes the LL-accurate formula 
in \cite{Beneke:2018gvs}. 

As one of the new key ingredients of the subleading-power 
factorization formula, we identify and discuss the emergence of 
collinear functions at the amplitude level. While the related 
concept of a ``radiative jet function''  \cite{DelDuca:1990gz} has 
been known to be relevant to power corrections at the DY 
threshold from diagrammatic studies \cite{Bonocore:2015esa,Bonocore:2016awd,DelDuca:2017twk}, the benefit of the present SCET 
treatment is an operator definition, which renders the 
function gauge-invariant by construction. More precisely, 
see (\ref{eq:genMatch}), the collinear functions are 
the perturbative matching coefficients, when threshold-collinear
fields are matched to $c$-PDF fields in the presence of 
external soft structures that describe the emission of one (or 
several) soft gluons. Due to the strict scale separation and 
systematic power expansion, the collinear functions are 
single-scale objects. They are extracted from partonic matrix 
elements since the threshold-collinear scale is assumed to 
be much larger than the scale of strong interactions, 
$Q(1-z)^{1/2}\gg \Lambda$.
 
The tree-level collinear function required for LL resummation in 
threshold DY (and Higgs) production has already been used in 
\cite{Beneke:2018gvs,Beneke:2019mua}. In this work, we computed 
the one-loop $\mathcal{O}(\alpha_s)$ corrections  \eqref{J1} and 
\eqref{J6} to the collinear functions, which can contribute to the 
DY cross section at NNLO, and to the one-loop one-gluon emission 
amplitude. These results confirm explicitly the observation made in
\cite{Beneke:2018gvs,Beneke:2019mua} that the DY collinear function 
cannot contain LLs. The one-loop calculation demonstrates the 
validity of the definition of these NLP objects and allows us 
to verify the correctness of the factorization formula at NNLO 
by comparing its expansion in powers of $\alpha_s$ with 
existing results obtained at this order with the expansion-by-regions 
method.  

However, our investigation also highlights that factorization at 
NLP is not yet understood at a similar level as at LP. The 
factorization formula separates the scales relevant to the 
DY threshold in the form of well-defined, dimensionally regulated 
collinear and soft functions, which have to be convoluted in 
the soft momentum variables $\omega_i$. The $\mathcal{O}(\alpha^2_s)$ 
calculation makes explicit what can already be seen from 
general scaling arguments that the convolutions exist only 
for the $d$-dimensional functions. When the expansion in 
$\epsilon$ is performed before the convolution, the latter is 
ill-defined and leads to a divergence. This implies that the 
formula is not yet in a form suitable for the resummation of 
large threshold logarithms beyond the 
LLs through the renormalization group equations for renormalized 
hard, collinear and soft functions. Nevertheless, it may be 
hoped that it provides the starting point for understanding how 
to renormalize $d$-dimensional convolutions, which would 
open the path to NLL resummations beyond LP.

\subsubsection*{Acknowledgments} 
We thank M.~Garny and J.~Wang for useful insights and
R.~Szafron for many fruitful discussions and technical advice. 
MB, AB and LV thank the Munich Institute for Particle and 
Astroparticle Physics (MIAPP) for hospitality during the 2017 
programme ``Automated, Resummed and Effective: Precision 
Computations for the LHC and Beyond'', where some of the 
concepts of this work have been developed.  LV is supported 
by Fellini - Fellowship for Innovation at INFN, funded by the 
European Union's Horizon 2020 research programme under the 
Marie Sk\l{}odowska-Curie Cofund Action, grant agreement no. 754496.
This work has been supported by the Bundesministerium f\"ur 
Bildung and Forschung (BMBF) grant no. 05H18WOCA1, by the Dutch National Organization for Scientific Research (NWO) and by the ERC Starting Grant REINVENT-714788.

\begin{appendix}

\section{Subleading SCET Lagrangian}
\label{sec:appendixA}

\subsection{Quark-gluon subleading SCET Lagrangian }
The quark-gluon interaction terms of subleading power  in the
soft-collinear SCET Lagrangian \cite{Beneke:2002ni} are given by 
\begin{eqnarray}
\mathcal{L}_{\xi}^{\left(1\right)}	
&=&\bar{\chi}_c ix_{\perp}^{\mu}
\left[in_{-}\partial\mathcal{B}_{\mu}^{+}\right]
\frac{\slashed n_{+}}{2}\chi_c ,
\nonumber\\
\mathcal{L}_{1\xi}^{\left(2\right)} 	
&=&\frac{1}{2}\bar{\chi}_c i\nm x\,\np^{\mu}
\left[in_{-}\partial\mathcal{B}_{\mu}^{+}\right]
\frac{\slashed n_{+}}{2}\chi_c ,
\nonumber\\
\mathcal{L}_{2\xi}^{\left(2\right)}	
&=&\frac{1}{2}\bar{\chi}_c x_{\perp}^{\mu}x_{\perp}^{\nu}
\left[i\partial_{\nu}in_{-}\partial\mathcal{B}_{\mu}^{+}\right]
\frac{\slashed n_{+}}{2}\chi_c ,
\nonumber\\
\mathcal{L}_{3\xi}^{\left(2\right)}
&=&\frac{1}{2}\bar{\chi}_c x_{\perp}^{\mu}x_{\perp}^{\nu}
\left[\mathcal{B}_{\nu}^{+},in_{-}\partial\mathcal{B}_{\mu}^{+}\right]
\frac{\slashed n_{+}}{2}\chi_c ,
\nonumber\\
\mathcal{L}_{4\xi}^{\left(2\right)}	
&=&\frac{1}{2}\bar{\chi}_c\left(i\slashed\partial_{\perp}+\sA_{c\perp}\right)
\frac{1}{i\np\partial}ix_{\perp}^{\mu}\gamma_{\perp}^{\nu}
\left[i\partial_{\nu}\mathcal{B}_{\mu}^{+}-i\partial_{\mu}\mathcal{B}_{\nu}^{+}
\right]\frac{\slashed n_{+}}{2}\chi_c + {\rm h.c.},
\nonumber\\
\mathcal{L}_{5\xi}^{\left(2\right)}	
&=&\frac{1}{2}\bar{\chi}_c\left(i\slashed\partial_{\perp}+\sA_{c\perp} \right)
\frac{1}{i\np\partial}ix_{\perp}^{\mu}\gamma_{\perp}^{\nu}
\left[\mathcal{B}_{\nu}^{+},\mathcal{B}_{\mu}^{+}\right]
\frac{\slashed n_{+}}{2}\chi_c+ {\rm h.c.},\nonumber\\
\mathcal{L}_{\xi q}^{\left(1\right)} 
&=&\bar{q}_{+}\sA_{c\perp}\chi_c+ {\rm h.c.},\nonumber\\
\mathcal{L}_{\xi q}^{(2)}&=& \bar q_+ 
\bigg[i n_- \partial + n_- {\cal A}_c + \Big(i {\slash \partial }_{\perp} + {\cal \slash A}_c \Big)
\frac{1}{i n_+ \partial} \Big(i {\slash \partial}_{\perp} + {\cal \slash A}_c \Big)\bigg] \frac{\slash n_+ }{2} \chi_{c} \nonumber \\
&&+\,\bar q_{+}  \Big(i \overleftarrow\partial^{\mu} + {\cal B}_+^{\mu} \Big) x_{\perp \mu} 
\Big(i {\slash \partial}_{\perp} + {\cal \slash A}_c \Big)\chi_{c} + {\rm h.c.} \,.
\label{eq:quarkint}
\end{eqnarray}

\subsection{YM subleading SCET Lagrangian}

The subleading-power gluon self-interaction terms of  the
soft-collinear Yang-Mills Lagrangian \cite{Beneke:2002ni} 
expressed in terms of the collinear and soft gauge-invariant 
fields are given by 
\label{appendix:YMSCETLagrangian}
\bea\label{Lym2} \nn
 {\cal L}_{\rm 1YM}^{(1)} &=& -\frac{1}{g_s^2}{\rm tr} \Big( 
 \big[ n_+ \partial \, {\cal A}^c_{\nu_{\perp}} \big] 
 \Big[
  x_{\perp}^{\rho} 
   \, i n_- \partial \, {\cal B}^+_{\rho} ,
   {\cal A}_c^{\nu_{\perp}} \Big] \Big),  \\ \nn 
  {\cal L}_{\rm 2YM}^{(1)} &=& -\frac{1}{g_s^2}{\rm tr} \Big( 
  \big[ n_+ \partial \, {\cal A}_c^{\nu_{\perp}} \big] 
  i n_- \partial \, {\cal B}^+_{\nu_{\perp}} \Big), \\ \nn
{\cal L}_{\rm 1YM}^{(2)} &=&
-\frac{1}{2g_s^2}{\rm tr} \Big( 
\big[ n_+ \partial \, {\cal A}^c_{\nu_{\perp}} \big] 
\Big[ n_-x\, i n_- \partial \, n_+{\cal B}^+, \, 
{\cal A}_c^{\nu_{\perp}} \Big] \Big),  \\ \nn
{\cal L}_{\rm 2YM}^{(2)} &=& 
- \frac{1}{2g^2_s}{\rm tr} \Big(
\big[ n_+ \partial \, {\cal A}^c_{\nu_{\perp}} \big]  \,
\Big[ x_{\perp}^{\rho} x_{\perp \omega} 
\big[ \partial^{\omega}, \, i n_- \partial \, {\cal B}^+_{\rho} \big],
\,  {\cal A}_c^{\nu_{\perp}} \Big] \Big),  \\ \nn
{\cal L}_{\rm 3YM}^{(2)} &=& 
- \frac{1}{2g^2_s}{\rm tr} \Big( 
\big[ n_+ \partial \, {\cal A}^c_{\nu_{\perp}} \big] \, 
\Big[ x_{\perp}^{\rho} x_{\perp \omega} 
\big[ {\cal B}_+^{\omega}, n_- \partial \, {\cal B}^+_{\rho} \big],
\,  {\cal A}_c^{\nu_{\perp}} \Big] \Big), \\ \nn
{\cal L}_{\rm 4YM}^{(2)} &=& -\frac{1}{2g_s^2} {\rm tr} 
\Big(\big[ n_+ \partial \, {\cal A}^c_{\nu_{\perp}} \big]   \,
\Big[x_{\perp}^{\rho} \big[i \partial_{\rho} {\cal B}^+_{\nu_{\perp}}
- i \partial_{\nu_{\perp}} {\cal B}^+_{\rho} \big], 
\, n_-{\cal A}_c \Big] \Big),  \\  \nn
{\cal L}_{\rm 5YM}^{(2)} &=& 
-\frac{1}{2g_s^2} {\rm tr} \Big( 
\big[ n_+ \partial \, {\cal A}^c_{\nu_{\perp}} \big]   \,
\Big[x_{\perp}^{\rho} \big[ {\cal B}^+_{\rho}, 
{\cal B}^+_{\nu_{\perp}} \big],\,  n_-{\cal A}_c \Big]  \Big),  \\  \nn
{\cal L}_{\rm 6YM}^{(2)} &=& -\frac{1}{g_s^2} {\rm tr} 
\Big(\big[i \partial^{\mu_{\perp}} {\cal A}_c^{\nu_{\perp}}
- i \partial^{\nu_{\perp}} {\cal A}_c^{\mu_{\perp}} \big]
\Big[ i x_{\perp}^{\rho} 
\big[i \partial_{\rho} {\cal B}^+_{\mu_{\perp}}
- i \partial_{\mu_{\perp}} {\cal B}^+_{\rho} \big], \,
{\cal A}^c_{\nu_{\perp}} \Big] \Big),  \\ \nn 
{\cal L}_{\rm 7YM}^{(2)} &=& -\frac{1}{g_s^2} {\rm tr} 
\Big(\big[ {\cal A}_c^{\mu_{\perp}} , {\cal A}_c^{\nu_{\perp}}\big]
\Big[ i x_{\perp}^{\rho} 
\big[i \partial_{\rho} {\cal B}^+_{\mu_{\perp}}
- i \partial_{\mu_{\perp}} {\cal B}^+_{\rho} \big], \,
{\cal A}^c_{\nu_{\perp}} \Big] \Big), \\ \nn
{\cal L}_{\rm 8YM}^{(2)} &=& -\frac{1}{g_s^2} {\rm tr} \Big( 
\big[i \partial^{\mu_{\perp}} {\cal A}_c^{\nu_{\perp}}
- i \partial^{\nu_{\perp}} {\cal A}_c^{\mu_{\perp}} \big]
\Big[ i x_{\perp}^{\rho} \big[{\cal B}^+_{\rho}, {\cal B}^+_{\mu_{\perp}} \big], 
\,{\cal A}^c_{\nu_{\perp}} \Big] \Big), \\ \nn
{\cal L}_{\rm 9YM}^{(2)} &=& -\frac{1}{g_s^2} {\rm tr} \Big( 
\big[ {\cal A}_c^{\mu_{\perp}} , {\cal A}_c^{\nu_{\perp}}\big]
\Big[ i x_{\perp}^{\rho} \big[{\cal B}^+_{\rho}, {\cal B}^+_{\mu_{\perp}} \big], 
\,  {\cal A}^c_{\nu_{\perp}} \Big] \Big), \\ \nn
{\cal L}_{\rm 10YM}^{(2)} &=& -\frac{1}{2g_s^2} {\rm tr} 
\Big(\big[ n_+ \partial \, n_-{\cal A}^c \big] \, 
n_- \partial \, n_+{\cal B}^+ \Big), \\ \nn
{\cal L}_{\rm 11YM}^{(2)} &=& \frac{1}{g_s^2} {\rm tr} \Big( 
\big(i \partial^{\mu_{\perp}} {\cal A}_c^{\nu_{\perp}}
- i \partial^{\nu_{\perp}} {\cal A}_c^{\mu_{\perp}} \big) \,
\big(i \partial_{\mu_{\perp}} {\cal B}^+_{\nu_{\perp}}
- i \partial_{\nu_{\perp}} {\cal B}^+_{\mu_{\perp}} \big) \Big),  \\ \nn 
{\cal L}_{\rm 12YM}^{(2)} &=& \frac{1}{g_s^2} {\rm tr} \Big( 
\big[ {\cal A}_c^{\mu_{\perp}} , {\cal A}_c^{\nu_{\perp}} \big] \, 
\big(i \partial_{\mu_{\perp}} {\cal B}^+_{\nu_{\perp}}
- i \partial_{\nu_{\perp}} {\cal B}^+_{\mu_{\perp}} \big) \Big), \\ \nn
{\cal L}_{\rm 13YM}^{(2)} &=& \frac{1}{g_s^2} {\rm tr} \Big( 
\big(i \partial^{\mu_{\perp}} {\cal A}_c^{\nu_{\perp}}
- i \partial^{\nu_{\perp}} {\cal A}_c^{\mu_{\perp}} \big) \,
\big[{\cal B}^+_{\mu_{\perp}}, {\cal B}^+_{\nu_{\perp}} \big] \Big), \\ \nn
{\cal L}_{\rm 14YM}^{(2)} &=& \frac{1}{g_s^2} {\rm tr} \Big( 
\big[ {\cal A}_c^{\mu_{\perp}} , {\cal A}_c^{\nu_{\perp}}\big] \,
\big[{\cal B}^+_{\mu_{\perp}}, {\cal B}^+_{\nu_{\perp}} \big]\Big), \\ \nn
{\cal L}_{\rm 15YM}^{(2)} &=& -\frac{1}{g_s^2} {\rm tr} \Big( 
\big[ n_+ \partial \, {\cal A}_c^{\nu_{\perp}} \big] x_{\perp\sigma}
\big[\partial^{\sigma} , n_-\partial B^+_{\nu_{\perp}} \big]\Big), \\ 
{\cal L}_{\rm 16YM}^{(2)} &=& \frac{1}{g_s^2} {\rm tr} \Big( 
\big[ n_+ \partial \, {\cal A}_c^{\nu_{\perp}} \big] x_{\perp\sigma}
\big[i B_+^{\sigma} , n_-\partial B^+_{\nu_{\perp}} \big]\Big). 
\eea

\section{One-loop single soft real emission 
amplitude}
\label{appendix:amplituderesults}

In the main body of the text we focused on the factorization 
formula at the cross-section level. As a by-product of the 
computation of the collinear functions, which are amplitude-level 
objects, we also obtained the power-suppressed one-loop one-soft 
emission DY amplitude, which we summarize here. The results below, 
computed directly in SCET, were shown to agree with in-house 
results obtained by applying the expansion-by-regions method  
to the same quantity. 
 
We consider the following operator, which is the right-hand side 
of~\eqref{eq:subleadingmatching} without the soft
current  $J_s$: 
\bea \label{eq:subleadingmatching2}
 \sum_{m_1, {m_2} }
 \int\, \{dt_k\} \, 
 \{d\bar{t}_{\bar{k}}\}  \,\widetilde{C}^{\,m_1,m_2}
 \left(\{t_k\} , \{\bar{t}_{\bar{k}}\}\right)
 J_\rho^{\,m_1,m_2}\left(\{t_k\} , \{\bar{t}_{\bar{k}}\}
 \right)
 \eea
where 
  \bea \label{eq:DYcurrent2}
  J_\rho^{\,m_1,m_2}\left(\{t_k\} , \{\bar{t}_{\bar{k}}\}
  \right) =  J^{m_1}_{\bar{c}}(\{\bar{t}_{\bar{k}}\})
\,\Gamma_{\rho}^{m_1,m_2}\,J^{m_2}_c\left(\{t_k\}
  \right)
\eea 
as in \eqref{eq:DYcurrent}. The variables appearing in this 
expression are defined in Sec.~\ref{sec:factGeneralPows}, and the 
sum is performed over the different power-suppressed currents in 
the $N$-jet SCET operator matched to the QCD current. 
  
Below we focus solely on the case in which the power 
suppression is in the collinear sector, thereby setting
 $m_1=A0$,  and allow for structures which give power-suppression 
up to $\mathcal{O}(\lambda^2)$ (NLP). Specifically,
we consider the time-ordered product of $J_\rho^{\,m_1,m_2}$ 
with subleading-power  Lagrangian insertions between an 
emitted soft gluon $\langle g(k)^K|$, and an incoming collinear quark 
and anticollinear antiquark, 
$|q(p)\, \bar{q}(l)\,\rangle $. This defines
the amplitude 
\bea \label{eq:ampdef}
\mathcal{M}^K_{\rho} = \langle g(k)^K|
 \sum_{ {m} }\int\, \{dt_k\} \, 
 d\bar{t} \,\widetilde{C}^{A0,m}
 \left(\{t_k\} , \bar{t} \,\right)
 \,J_\rho^{\,A0,m}\left(\{t_k\} , \bar{t}\,
 \right)|q(p)\, \bar{q}(l)\rangle\,,
\quad
\eea 
that we calculate at the one-loop order. Concretely, we consider 
only the time-ordered products of the collinear 
operator part $J^{m_2}_c$ in \eqref{eq:DYcurrent2} with 
subleading-power soft-collinear (not: soft-anticollinear) 
Lagrangian insertions. The complete result for the amplitude 
is obtained by subtracting from the contributions given below the 
corresponding ones with $n_+$ and $n_-$ interchanged. 

In the following sections we present the different contributions 
to this object. Partial results obtained when the virtual loop is 
collinear (soft) carry a subscript $c$ ($s$),
$\mathcal{M}_{c}$ ($\mathcal{M}_{s}$). The
NLO contributions from the one-loop hard matching coefficient 
are marked with $h$, $\mathcal{M}_{h}$.  
Moreover, we further split the results according
to the polarization of the off-shell DY photon $\gamma^*$ produced 
by the vector current, that is, we separate the 
amplitude into the terms proportional to 
$\gamma_{\perp\rho}$, $n_{+\rho}$, and $n_{-\rho}$. 
Notice that the $\gamma_{\perp\rho}$ structure 
appears due to the LP current in \eqref{eq:LPcurrent}, while 
$n_{\pm\rho}$ terms arise from the power-suppressed $A1$ and $B1$ 
currents in \eqref{JA1.1} and \eqref{JB1.1}, respectively. 


\subsection{Collinear loop: $\gamma_{\perp\rho}$}
\begin{figure}[t]
\begin{centering}
\includegraphics[width=0.90\textwidth]{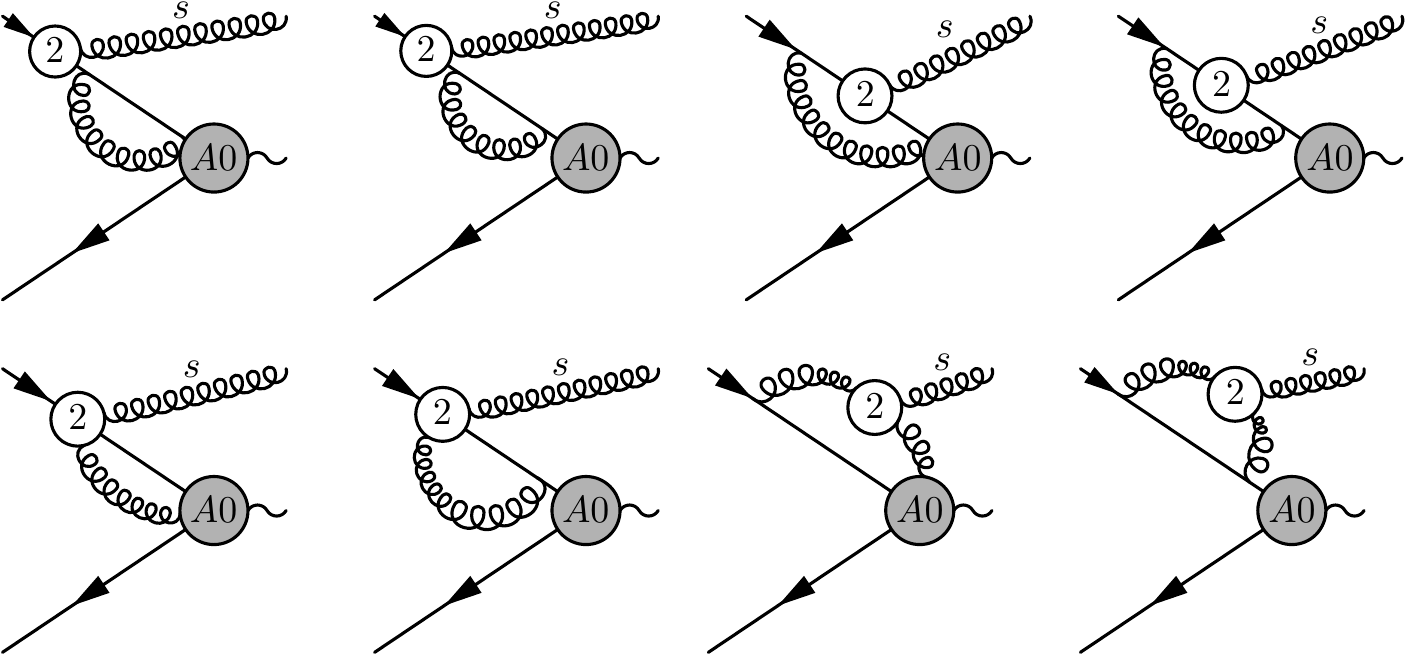}
\par\end{centering}
\caption{\label{fig:CollinearFunctionAmplitudes2} 
One-loop collinear diagrams with one soft gluon emission. Only the 
LP current, $A0$, is used here.  Power suppression is provided by 
the time-ordered product insertion of $\mathcal{L}^{(2)}$ Lagrangian 
terms. The collinear gluon in the loop attaches either to the 
collinear quark or the collinear Wilson line of the $\chi_c$ field, 
which is part of the $A0$ current. Note the difference
in the drawing of the diagrams in those in 
Figure~\ref{fig:CollinearFunctionAmplitudes}: here 
we included the anticollinear leg and hard current. }
\end{figure}

We begin with the results for the set of diagrams in which the 
virtual loop has collinear momentum scaling and the virtual photon 
created by the vector current has a transverse $\rho$ index.
In \eqref{eq:ampdef} this means taking the LP current, and index $m$
spans over time-ordered product insertions of the 
$\mathcal{L}^{(2)}$ Lagrangian. The  
equations below are in fact related to the results presented in 
\eqref{eq:oneloopt1amp} and come from calculation of the 
diagrams in Figure~\ref{fig:CollinearFunctionAmplitudes2}. 
 
We separate the resulting expression into the amplitude 
with colour factor $C_F$ and $C_A$. The former 
receives contributions from the diagrams in the 
top line of Figure \ref{fig:CollinearFunctionAmplitudes2}, 
the latter from those in the bottom line and the 
non-abelian part of the last two diagrams in the top line. 
We find  
\begin{eqnarray}\label{eq:B4}
\mathcal{M}^{\gamma_{\perp}\rho\, K}_{c,C_F}&=&
\bar{v}_{\bar{ c}}(l) \gamma^{\rho}_{\perp} 
\frac{ig_s\alpha_s}{(4\pi)}
\bigg[ \frac{(n_+p)(n_-k)}{\mu^2} \bigg]^{-\epsilon} 
\frac{C_F\mathbf{T}^K }{(n_+p)(n_-k)}\frac{e^{\epsilon \gamma_E}\Gamma[1+\epsilon]
	\Gamma[1-\epsilon]^2}{(1+\epsilon)(1-\epsilon)
	\Gamma[2-2\epsilon]} \nn\\ 
&&\times 
\,\bigg\{ (n_+k)n_{-\nu}  \left(\frac{3}{\epsilon }-4
-7 \epsilon  \right) + [\slashed{k}_{\perp},
\gamma_{\perp\nu}]\, \frac{1}{2}
\left(1-\epsilon ^2 \right)
\\
&&\nn +\,k_{\perp\nu} \left(\frac{2}{\epsilon }-5
-6 \epsilon + \epsilon ^2 \right) 
+(n_-k)n_{+\nu}  \left(-\frac{1}{\epsilon }-1
+  \epsilon + \epsilon ^2 \right)
 \bigg\}\, u_c(p)\epsilon^{*\nu}(k)\,,
\quad\\[0.2cm]
\label{eq:B5}
\mathcal{M}^{\gamma_{\perp}\rho\, K}_{c,C_A}&=&
\bar{v}_{\bar{ c}}(l) \gamma^{\rho}_{\perp} 
\frac{ig_s\alpha_s}{(4\pi)}
\bigg[ \frac{(n_+p)(n_-k)}{\mu^2} \bigg]^{-\epsilon} 
\frac{C_A\mathbf{T}^K }{(n_+p)(n_-k)}\frac{ e^{\epsilon \gamma_E}\Gamma[1+\epsilon]
	\Gamma[1-\epsilon]^2}{(1+\epsilon)(1-\epsilon)
	\Gamma[2-2\epsilon]} \nn\\
&&\hspace{-1.4cm}  \times 
\bigg\{ (n_+k)n_{-\nu}\,\frac{1}{2} \, \bigg(-\frac{1}{\, \epsilon ^2}
-\frac{1}{ \epsilon }-2 +11 \epsilon + \epsilon ^2\bigg) 
+ [\slashed{k}_{\perp},\gamma_{\perp\nu}]\,\frac{1}{2} 
\left( -1+\epsilon ^2 \right)
\\ 
&& \hspace{-1.4cm}\nn
+\,k_{\perp\nu} \left(-\frac{1}{\epsilon ^2}-
\frac{1}{\epsilon }+3+3 \epsilon  \right) 
+(n_-k)n_{+\nu}\,\frac{1}{2} \,\bigg(-\frac{1}{\epsilon ^2}-
\frac{1}{ \epsilon } +8
-5 \epsilon   - \epsilon ^2\bigg)
\bigg\} \,u_c(p)\epsilon^{*\nu}(k)\,.
\end{eqnarray}
In this appendix, we use the on-shell condition $k^2=0$ to rewrite $k^2_{\perp}=-(n_-k)(n_+k)$, but we do not impose the transversality relation \eqref{eq:onshelltransversality}.
Notice that in \eqref{eq:B5} there are still $1/\epsilon^2$
poles. These only cancel once soft structures 
are combined as described in the main text.  

\begin{figure}[t]
\begin{centering}
\includegraphics[width=0.9\textwidth]{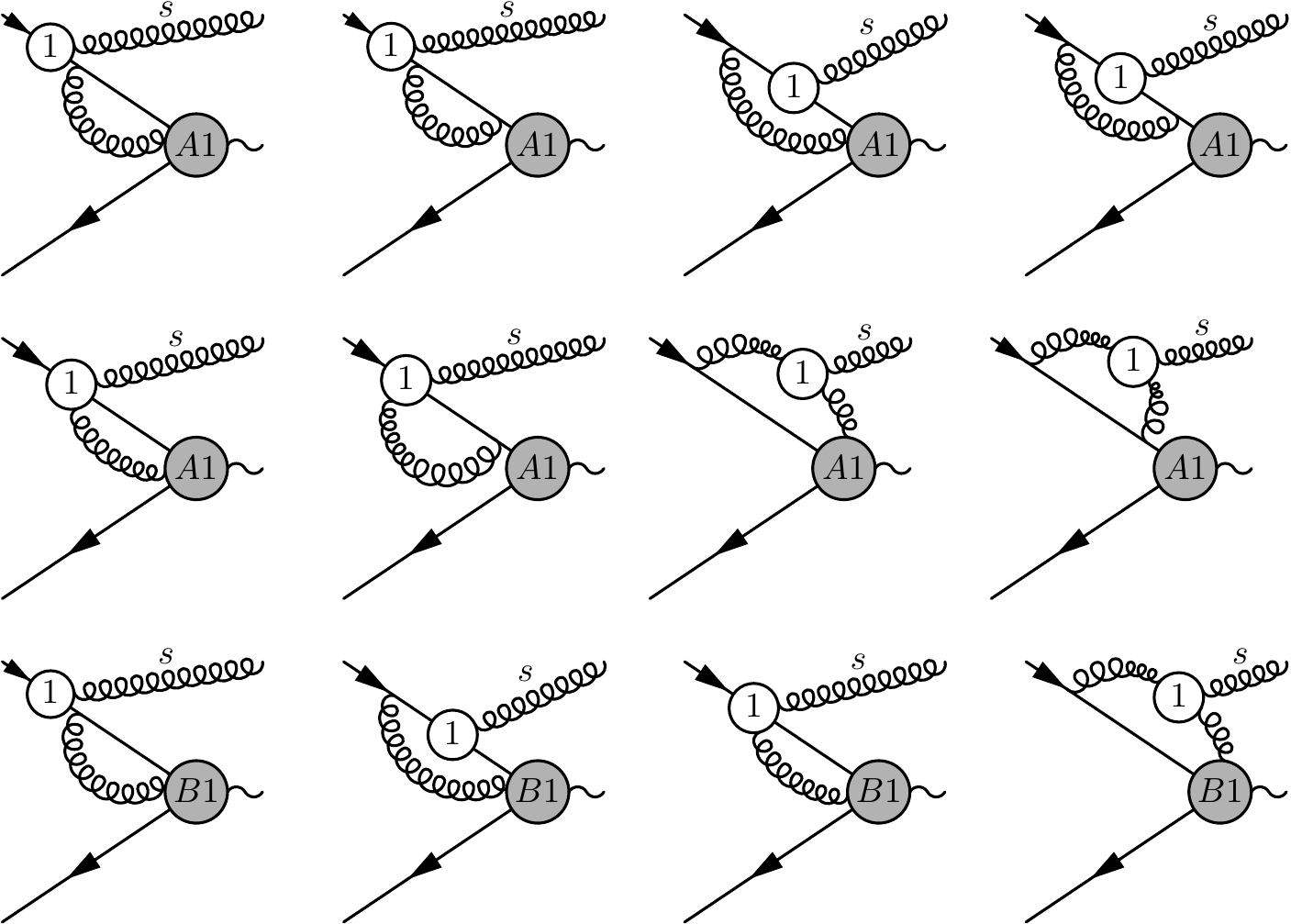}
\par\end{centering}
\caption{\label{fig:CollinearFunctionAmplitudes3} 
Collinear one-loop diagrams with one soft gluon emission. The 
$\mathcal{O}(\lambda^1)$ power-suppressed currents $A1$ and $B1$ 
defined in \eqref{JA1.1} and \eqref{JB1.1}, respectively, are 
used here. The collinear virtual gluon must attach to the 
$B1$ current, because of the additional $\mathcal{A}_{c\perp}$ 
gluon field 
present in this subleading current.}
\end{figure}

\subsection{Collinear loop: $n^{\rho}_{-}$ and $n^{\rho}_{+}$ }

These contributions are due to  time-ordered products of the 
power-suppressed hard currents defined in \eqref{JA1.1} and 
\eqref{JB1.1} with  $\mathcal{L}^{(1)}$ Lagrangian insertions. The 
corresponding diagrams are shown in 
Figure \ref{fig:CollinearFunctionAmplitudes3}. Separating the 
two colour structures, we find
\begin{eqnarray}\label{eq:B6}
\mathcal{M}^{n_{\pm}\rho\, K}_{c,C_F}&=&
\bar{v}_{\bar{ c}}(l)\left(\frac{n^{\rho}_{-}}{n_-l}
-\frac{n^{\rho}_{+}}{n_+p}\right)
\frac{ig_s\alpha_s}{(4\pi)}
\bigg[\frac{(n_+p)(n_-k)}{\mu^2} \bigg]^{-\epsilon}
C_F\mathbf{T}^K 
\frac{ e^{\epsilon\gamma_E}\Gamma[1+\epsilon]
	\Gamma[1-\epsilon]^2}{(1+\epsilon)(1-\epsilon)
	\Gamma[2-2\epsilon]}\nn\\&&\times 
\left(\gamma_{\perp\nu}-
\frac{\slashed{k}_{\perp}{n}_{-\nu}}{({n}_{-}k)}
\right) \left(1+2 \epsilon +  \epsilon ^2 \right) u_c(p)
\epsilon^{*\nu}(k)\,,\\[0.2cm]\label{eq:B7}
\mathcal{M}^{n_{\pm}\rho\, K}_{c,C_A}&=&
\bar{v}_{\bar{ c}}(l)\left(\frac{n^{\rho}_{-}}{n_-l}
-\frac{n^{\rho}_{+}}{n_+p}\right)
\frac{ig_s\alpha_s }{(4\pi)}
\bigg[\frac{(n_+p)(n_-k)}{\mu^2} \bigg]^{-\epsilon}
C_A\mathbf{T}^K    
\frac{e^{\epsilon\gamma_E}\Gamma[1+\epsilon]
	\Gamma[1-\epsilon]^2}{(1+\epsilon)(1-\epsilon)
	\Gamma[2-2\epsilon]}\nn\\&&\times
\left(\gamma_{\perp\nu}-
\frac{\slashed{k}_{\perp}{n}_{-\nu}}{({n}_{-}k)}
\right)\left(\frac{1}{\epsilon }-2\, 
\epsilon -    \epsilon ^2  \right)u_c(p)\epsilon^{*\nu}(k)\,.
\end{eqnarray}

\subsection{Soft loop: $\gamma_{\perp\rho}$}
\label{sec:softloopperp}

\begin{figure}[t]
\begin{centering}
\includegraphics[width=0.23\textwidth]{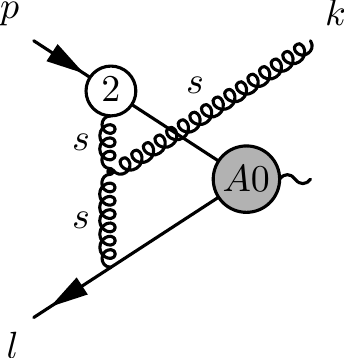}
\par\end{centering}
\caption{\label{fig:dy6lamb2_1} 
The only diagram relevant to the one virtual, one-real contribution 
to the two-loop soft function. Here the power suppression 
is placed on the collinear leg as indicated by the  
$\mathcal{O}(\lambda^2)$ vertex.}
\end{figure}

In this section we present the result for the soft one-virtual, 
one-real soft gluon amplitude proportional to $\gamma_{\perp\rho}$. 
Only one SCET diagram, shown in 
Figure~\ref{fig:dy6lamb2_1}, is needed to
reproduce the corresponding virtual-real contribution
from the expansion-by-regions method. 
Hence only non-abelian contributions arise here
and we find 
\begin{eqnarray}\label{a10.5163}
\mathcal{M}^{\gamma_{\perp}\rho\, K}_{s,C_A}
&=&\bar{v}_{\bar{ c}}(l) \,  \gamma^{\rho}_{\perp} 
\,   \frac{ig_s\alpha_s}{(4\pi)}
\left(\frac{-(n_-k)(n_+k)}{\mu^2}\right)^{-\epsilon}
\frac{{C_A}\mathbf{T}^K}{(n_+p)(n_-k)}
\frac{e^{\epsilon\gamma_E} \Gamma[1+\epsilon]^2
	\Gamma[1-\epsilon]^3}{\Gamma[2-2\epsilon]}
\nonumber\\&&\times  \Big(n_+k\,n_{-\nu}\, 
+ k_{\perp\nu}   \,+\frac{1}{2}[\slashed{k}_{\perp},\gamma_{\perp\nu}]
\Big)  \,   \left( \frac{1}{\epsilon^2}-
\frac{2}{\epsilon}\right) \, u_c(p)  \epsilon^{*\nu}(k)\,.
\end{eqnarray}  

\begin{figure}[t]
\begin{centering}
\includegraphics[width=0.75\textwidth]{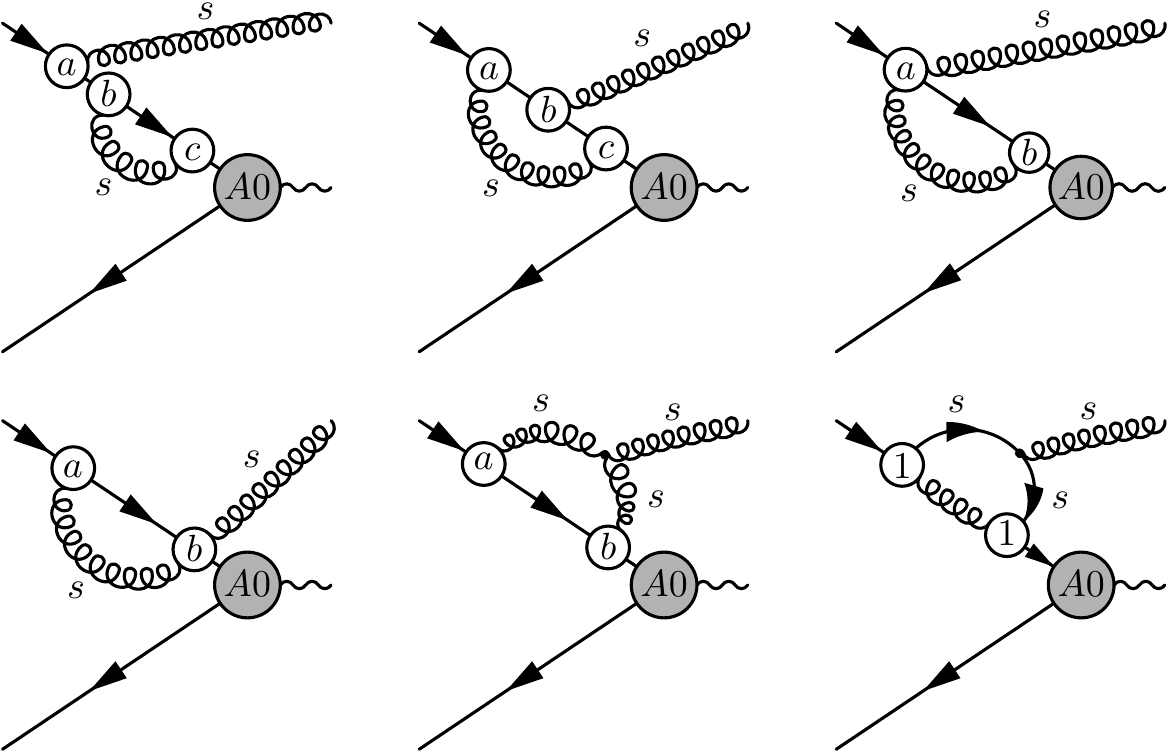}
\par\end{centering}
\caption{\label{fig:SoftAmplitudesAll_1} 
Diagrams with one soft emitted gluon and one soft loop. Since all the 
diagrams here include the LP $J^{A0,A0}$ current, the 
$\mathcal{O}(\lambda^2)$ power suppression must be provided by 
Lagrangian insertions. This means using all possible insertions such 
that $a+b\,(+\,c)=2$ at the indicated vertices. Out of the 20 
possibilities, many vanish immediately due to 
contractions which yield $n_{\pm}^2=n_{\pm}\cdot \gamma_{\perp}=0$ or 
propagators which give zero due to the vanishing external transverse 
momentum. The 
remaining integrals, where the integrand does not immediately 
vanish, are either scaleless or vanish by Cauchy's theorem, because 
all propagator poles lie in one half of the complex momentum plane.
}
\end{figure}

\begin{figure}[t]
\begin{centering}
\includegraphics[width=0.95\textwidth]{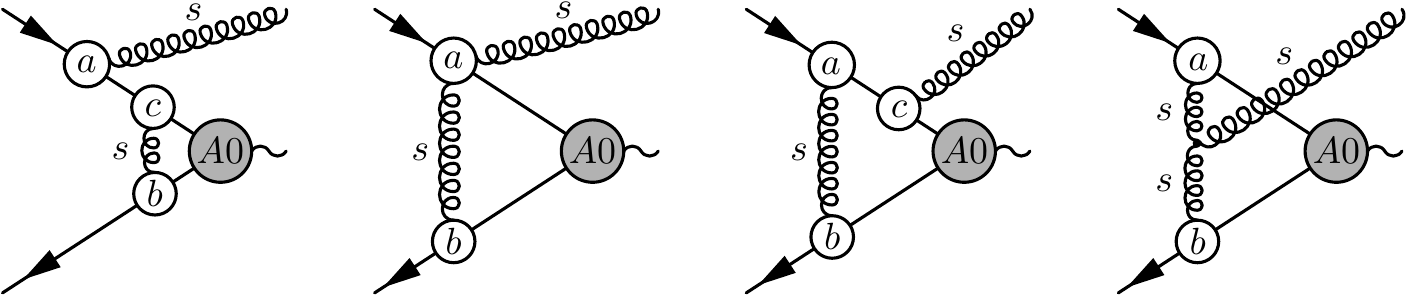}
\par\end{centering}
\caption{\label{fig:SoftAmplitudesAll_2} 
Soft one-loop diagrams with one emitted soft gluon. As in the 
previous figure, only the LP current is present in these diagrams, 
however now the virtual soft gluon connects the collinear and 
anticollinear legs. Lagrangian insertions must again be chosen 
such that  $a+b\,(+\,c)=2$. Note that all diagrams
with $b=1$ vanish, since a single leg 
cannot carry a $\mathcal{O}(\lambda)$ suppression as explained 
in Sec.~\ref{sec:NLPfact}. Only the last diagram
with $a=2$ or $b=2$ gives a non-vanishing result. The others 
are either scaleless or vanish after momentum conservation is 
imposed.}
\end{figure}

Details on the vanishing of numerous other a priori possible 
diagrams are provided in Figures 
\ref{fig:SoftAmplitudesAll_1} and \ref{fig:SoftAmplitudesAll_2}. 
Note that the latter figure also includes diagrams that 
represent insertions of both, the 
collinear (on the upper leg) and anticollinear (on the lower leg) 
subleading soft-collinear interactions, when $a=b=1$. However, 
as all these terms vanish, there is a unique separation of 
contributions from collinear Lagrangian insertions and from 
anticollinear Lagrangian insertions. In (\ref{a10.5163}) 
have we have given the $a=2$, $b=0$ contribution from the 
last diagram in Figure~\ref{fig:SoftAmplitudesAll_2}, while 
the $a=0,b=2$ anticollinear 
one is obtained by exchanging $n_+\leftrightarrow n_-$. 
We further note that the absence of a contribution 
of the second diagram in Figure~\ref{fig:SoftAmplitudesAll_2}, 
containing a power-suppressed two-soft gluon vertex, implies 
the statement made in Sec.~\ref{sec:softNNLO} that only 
the single soft-gluon structures with their corresponding 
soft functions $S_1$, $S_6$ contribute at NNLO, of which only 
$S_1$ is relevant at cross-section level as explained in the 
main text. 

\subsection{Soft loop:  $n^{\rho}_{+}$ }
\label{sec:softloopplus}
The relevant diagram is again the topology of 
Figure \ref{fig:dy6lamb2_1}. However, since one power
of $\lambda$ is used up by the power-suppressed 
current, at the soft-collinear vertex we now 
insert the $\mathcal{L}^{(1)}$ term from the 
SCET Lagrangian. 

The $J^{A0,B1}$ current cannot give a contribution 
here since it produces a collinear gluon, that cannot 
be contracted to form a soft loop. 

The diagrams shown in Figures \ref{fig:SoftAmplitudesAll_1}
and \ref{fig:SoftAmplitudesAll_2} are also present here. 
The only change is that the LP hard current is 
replaced by $J^{A0,A1}$ and the sum of $a+b\,(+\,c)=1$
only. Once again only the last diagram in Figure
 \ref{fig:SoftAmplitudesAll_2} does not vanish,  and we find 
\begin{eqnarray}
\mathcal{M}^{n_{+}\rho\, K}_{s,C_A}&=&
\bar{v}_{\bar{ c}}(l) \, 
n^{\rho}_{+}    
\frac{ig_s\alpha_s}{(4\pi)} 
\left( \frac{-(n_-k)(n_+k)}{\mu^2}\right)^{-\epsilon}
\frac{C_A\mathbf{T}^K}{(n_+p)}
\frac{e^{\epsilon\gamma_E}\Gamma[1+\epsilon]^2\,  
	\Gamma[1-\epsilon]^3}{\Gamma[2-2\epsilon]} 
 \\&&\hspace{-1cm}\nonumber\times\Bigg[
{\gamma_{\perp\nu}}\frac{1}{\epsilon^2}
+\frac{k_{\perp\nu} \slashed{k}_{\perp}}{ (n_-k)(n_+k)}
\frac{1}{\epsilon^2}+ \left(\frac{ n_{+  \nu }}{(n_+k)}
- \frac{ n_{-\nu} }{(n_-k)} \right) 
\slashed{k}_{\perp}   \left(\frac{1}{2\epsilon^2}
-\frac{1}{\epsilon}\right) \Bigg]   \, u_c(p)\epsilon^{*\nu}(k)\,.
\quad\end{eqnarray} 
There is no term proportional to $n_-^\rho$. 

\subsection{Hard loop: $\gamma_{\perp\rho}$}

As discussed in the main text, there exists also  a contribution
to the NLO NLP amplitude from the one-loop 
hard matching coefficient ${C}^{\,A0,A0}$  given in~\eqref{eq:WilsonC}. We obtain  
\begin{eqnarray}
\mathcal{M}^{\gamma_{\perp}\rho\, K}_{h,C_F}&=&
\bar{v}_{\bar{ c}}(l) \,  \gamma^{\rho}_{\perp} 
\,   \frac{ig_s \alpha_s}{(4\pi)}
\left(\frac{-(n_-l)(n_+p)}{\mu^2}\right)^{-\epsilon}
\,\frac{C_F\mathbf{T}^K}{(n_+p)(n_-k)}
\\&&\hspace{0.0cm}\times
\Bigg( (n_+k)n_{-\nu}
\bigg( \frac{2}{\epsilon ^2}+\frac{1}{\epsilon }
+5-\frac{\pi^2}{6} \bigg) 
+\big[ \slashed{k}_{\perp},\gamma_{\perp\nu}
\big]
\bigg(\frac{1}{\epsilon ^2}
+\frac{3}{2 \epsilon }-\frac{\pi ^2}{12}+4
\bigg)\nonumber\\&&\nonumber
\hspace{0.5cm}+\,k_{\perp\nu}\bigg(\frac{2}{\epsilon ^2}
+\frac{3}{\epsilon }-\frac{\pi ^2}{6}+8\bigg) 
+(n_-k)n_{+\nu}
\bigg(\frac{2}{\epsilon }+3
\bigg) +\mathcal{O}( \epsilon) 
\Bigg)\,u_c(p) \epsilon^{*\nu}(k)\,,
\end{eqnarray}
and $\mathcal{M}^{\gamma_{\perp}\rho\, K}_{h,C_A}=0$. 

\subsection{Hard loop:  $n^{\rho}_{+}$ }

This contribution  comes from the one-loop correction 
to the matching coefficient ${C}^{\,A0,A1}$ of the 
$J^{A0,A1}$ current together with an insertion of the 
$\mathcal{O}(\lambda)$ piece of quark SCET Lagrangian.  
${C}^{\,A0,A1}$ is related to ${C}^{\,A0,A0}$ 
by reparametrization invariance~\cite{Marcantonini:2008qn}. 
With the definition (\ref{JA1.1}) the relation reads 
${C}^{\,A0,A1} = -1/(n_+p)\, {C}^{\,A0,A0}$. 
We then find 
\bea\label{13.24}
\mathcal{M}^{n_{+}\rho\, K}_{h,C_F}&=& \bar{v}_{\bar{c}}(l)\,n^{\rho}_{+}
\frac{ig_s\alpha_s}{4\pi} 
\,\left(\frac{-(n_-l)(n_+p)}{\mu^2}\right)^{\!-\epsilon}
\frac{C_F\mathbf{T}^K}{(n_+p)(n_-k)}
\,\big(\slashed{k}_{\perp}
n_{-\nu}-(n_-k)\gamma_{\perp\nu}\big)
\nn\\&& \times\left(-\frac{2}{\epsilon^2}-\frac{3}{\epsilon}-8
+\frac{\pi^2}{6}+\mathcal{O}(\epsilon)\right)\,u_c(p)\epsilon^{*\nu}(k)\,.
\eea
There is no term proportional to $n_-^\rho$. 

\bibliography{NLP}

\providecommand{\href}[2]{#2}\begingroup\raggedright\begin{thebibliography}{10}

\bibitem{Sterman:1986aj}
G.~Sterman, \emph{{Summation of Large Corrections to Short Distance Hadronic
  Cross-Sections}}, {\emph{Nucl. Phys.} {\bfseries B281} (1987) 310}.

\bibitem{Catani:1989ne}
S.~Catani and L.~Trentadue, \emph{{Resummation of the QCD Perturbative Series
  for Hard Processes}},
  \href{https://doi.org/10.1016/0550-3213(89)90273-3}{\emph{Nucl. Phys.}
  {\bfseries B327} (1989) 323--352}.

\bibitem{Idilbi:2005ky}
A.~Idilbi and X.-d. Ji, \emph{{Threshold resummation for Drell-Yan process in
  soft-collinear effective theory}},
  \href{https://doi.org/10.1103/PhysRevD.72.054016}{\emph{Phys. Rev.}
  {\bfseries D72} (2005) 054016},
  [\href{https://arxiv.org/abs/hep-ph/0501006}{{\ttfamily hep-ph/0501006}}].

\bibitem{Idilbi:2006dg}
A.~Idilbi, X.-d. Ji and F.~Yuan, \emph{{Resummation of threshold logarithms in
  effective field theory for DIS, Drell-Yan and Higgs production}},
  \href{https://doi.org/10.1016/j.nuclphysb.2006.07.002}{\emph{Nucl. Phys.}
  {\bfseries B753} (2006) 42--68},
  [\href{https://arxiv.org/abs/hep-ph/0605068}{{\ttfamily hep-ph/0605068}}].

\bibitem{Becher:2007ty}
T.~Becher, M.~Neubert and G.~Xu, \emph{{Dynamical Threshold Enhancement and
  Resummation in Drell- Yan Production}},
  \href{https://doi.org/10.1088/1126-6708/2008/07/030}{\emph{JHEP} {\bfseries
  07} (2008) 030}, [\href{https://arxiv.org/abs/0710.0680}{{\ttfamily
  0710.0680}}].

\bibitem{Moch:2005ky}
S.~Moch and A.~Vogt, \emph{{Higher-order soft corrections to lepton pair and
  Higgs boson production}},
  \href{https://doi.org/10.1016/j.physletb.2005.09.061}{\emph{Phys. Lett.}
  {\bfseries B631} (2005) 48--57},
  [\href{https://arxiv.org/abs/hep-ph/0508265}{{\ttfamily hep-ph/0508265}}].

\bibitem{Bonocore:2014wua}
D.~Bonocore, E.~Laenen, L.~Magnea, L.~Vernazza and C.~D. White, \emph{{The
  method of regions and next-to-soft corrections in Drell-Yan production}},
  \href{https://doi.org/10.1016/j.physletb.2015.02.008}{\emph{Phys. Lett.}
  {\bfseries B742} (2015) 375--382},
  [\href{https://arxiv.org/abs/1410.6406}{{\ttfamily 1410.6406}}].

\bibitem{Bahjat-Abbas:2018hpv}
N.~Bahjat-Abbas, J.~Sinninghe~Damsté, L.~Vernazza and C.~D. White, \emph{{On
  next-to-leading power threshold corrections in Drell-Yan production at
  N$^3$LO}}, \href{https://doi.org/10.1007/JHEP10(2018)144}{\emph{JHEP}
  {\bfseries 10} (2018) 144},
  [\href{https://arxiv.org/abs/1807.09246}{{\ttfamily 1807.09246}}].

\bibitem{Laenen:2008ux}
E.~Laenen, L.~Magnea and G.~Stavenga, \emph{{On next-to-eikonal corrections to
  threshold resummation for the Drell-Yan and DIS cross sections}},
  \href{https://doi.org/10.1016/j.physletb.2008.09.037}{\emph{Phys. Lett.}
  {\bfseries B669} (2008) 173--179},
  [\href{https://arxiv.org/abs/0807.4412}{{\ttfamily 0807.4412}}].

\bibitem{Laenen:2008gt}
E.~Laenen, G.~Stavenga and C.~D. White, \emph{{Path integral approach to
  eikonal and next-to-eikonal exponentiation}},
  \href{https://doi.org/10.1088/1126-6708/2009/03/054}{\emph{JHEP} {\bfseries
  03} (2009) 054}, [\href{https://arxiv.org/abs/0811.2067}{{\ttfamily
  0811.2067}}].

\bibitem{Laenen:2010uz}
E.~Laenen, L.~Magnea, G.~Stavenga and C.~D. White, \emph{{Next-to-eikonal
  corrections to soft gluon radiation: a diagrammatic approach}},
  \href{https://doi.org/10.1007/JHEP01(2011)141}{\emph{JHEP} {\bfseries 01}
  (2011) 141}, [\href{https://arxiv.org/abs/1010.1860}{{\ttfamily 1010.1860}}].

\bibitem{Bonocore:2015esa}
D.~Bonocore, E.~Laenen, L.~Magnea, S.~Melville, L.~Vernazza and C.~D. White,
  \emph{{A factorization approach to next-to-leading-power threshold
  logarithms}}, \href{https://doi.org/10.1007/JHEP06(2015)008}{\emph{JHEP}
  {\bfseries 06} (2015) 008},
  [\href{https://arxiv.org/abs/1503.05156}{{\ttfamily 1503.05156}}].

\bibitem{Bonocore:2016awd}
D.~Bonocore, E.~Laenen, L.~Magnea, L.~Vernazza and C.~D. White,
  \emph{{Non-abelian factorisation for next-to-leading-power threshold
  logarithms}}, \href{https://doi.org/10.1007/JHEP12(2016)121}{\emph{JHEP}
  {\bfseries 12} (2016) 121},
  [\href{https://arxiv.org/abs/1610.06842}{{\ttfamily 1610.06842}}].

\bibitem{Beneke:2018gvs}
M.~Beneke, A.~Broggio, M.~Garny, S.~Jaskiewicz, R.~Szafron, L.~Vernazza et~al.,
  \emph{{Leading-logarithmic threshold resummation of the Drell-Yan process at
  next-to-leading power}},
  \href{https://doi.org/10.1007/JHEP03(2019)043}{\emph{JHEP} {\bfseries 03}
  (2019) 043}, [\href{https://arxiv.org/abs/1809.10631}{{\ttfamily
  1809.10631}}].

\bibitem{Beneke:2019mua}
M.~Beneke, M.~Garny, S.~Jaskiewicz, R.~Szafron, L.~Vernazza and J.~Wang,
  \emph{{Leading-logarithmic threshold resummation of Higgs production in gluon
  fusion at next-to-leading power}},
  \href{https://doi.org/10.1007/JHEP01(2020)094}{\emph{JHEP} {\bfseries 01}
  (2020) 094}, [\href{https://arxiv.org/abs/1910.12685}{{\ttfamily
  1910.12685}}].

\bibitem{Bahjat-Abbas:2019fqa}
N.~Bahjat-Abbas, D.~Bonocore, J.~Sinninghe~Damsté, E.~Laenen, L.~Magnea,
  L.~Vernazza et~al., \emph{{Diagrammatic resummation of leading-logarithmic
  threshold effects at next-to-leading power}},
  \href{https://doi.org/10.1007/JHEP11(2019)002}{\emph{JHEP} {\bfseries 11}
  (2019) 002}, [\href{https://arxiv.org/abs/1905.13710}{{\ttfamily
  1905.13710}}].

\bibitem{Boughezal:2016zws}
R.~Boughezal, X.~Liu and F.~Petriello, \emph{{Power Corrections in the
  N-jettiness Subtraction Scheme}},
  \href{https://doi.org/10.1007/JHEP03(2017)160}{\emph{JHEP} {\bfseries 03}
  (2017) 160}, [\href{https://arxiv.org/abs/1612.02911}{{\ttfamily
  1612.02911}}].

\bibitem{Moult:2016fqy}
I.~Moult, L.~Rothen, I.~W. Stewart, F.~J. Tackmann and H.~X. Zhu,
  \emph{{Subleading Power Corrections for N-Jettiness Subtractions}},
  \href{https://doi.org/10.1103/PhysRevD.95.074023}{\emph{Phys. Rev.}
  {\bfseries D95} (2017) 074023},
  [\href{https://arxiv.org/abs/1612.00450}{{\ttfamily 1612.00450}}].

\bibitem{Moult:2017jsg}
I.~Moult, L.~Rothen, I.~W. Stewart, F.~J. Tackmann and H.~X. Zhu, \emph{{N
  -jettiness subtractions for $gg\to H$ at subleading power}},
  \href{https://doi.org/10.1103/PhysRevD.97.014013}{\emph{Phys. Rev.}
  {\bfseries D97} (2018) 014013},
  [\href{https://arxiv.org/abs/1710.03227}{{\ttfamily 1710.03227}}].

\bibitem{Ebert:2018lzn}
M.~A. Ebert, I.~Moult, I.~W. Stewart, F.~J. Tackmann, G.~Vita and H.~X. Zhu,
  \emph{{Power Corrections for N-Jettiness Subtractions at ${\cal
  O}(\alpha_s)$}}, \href{https://doi.org/10.1007/JHEP12(2018)084}{\emph{JHEP}
  {\bfseries 12} (2018) 084},
  [\href{https://arxiv.org/abs/1807.10764}{{\ttfamily 1807.10764}}].

\bibitem{Boughezal:2018mvf}
R.~Boughezal, A.~Isgr\'o and F.~Petriello, \emph{{Next-to-leading-logarithmic
  power corrections for $N$-jettiness subtraction in color-singlet
  production}}, \href{https://doi.org/10.1103/PhysRevD.97.076006}{\emph{Phys.
  Rev.} {\bfseries D97} (2018) 076006},
  [\href{https://arxiv.org/abs/1802.00456}{{\ttfamily 1802.00456}}].

\bibitem{Boughezal:2019ggi}
R.~Boughezal, A.~Isgr\'o and F.~Petriello, \emph{{Next-to-leading power
  corrections to $V+1$ jet production in $N$-jettiness subtraction}},
  \href{https://doi.org/10.1103/PhysRevD.101.016005}{\emph{Phys. Rev.}
  {\bfseries D101} (2020) 016005},
  [\href{https://arxiv.org/abs/1907.12213}{{\ttfamily 1907.12213}}].

\bibitem{Ebert:2018gsn}
M.~A. Ebert, I.~Moult, I.~W. Stewart, F.~J. Tackmann, G.~Vita and H.~X. Zhu,
  \emph{{Subleading power rapidity divergences and power corrections for
  q$_{T}$}}, \href{https://doi.org/10.1007/JHEP04(2019)123}{\emph{JHEP}
  {\bfseries 04} (2019) 123},
  [\href{https://arxiv.org/abs/1812.08189}{{\ttfamily 1812.08189}}].

\bibitem{Cieri:2019tfv}
L.~Cieri, C.~Oleari and M.~Rocco, \emph{{Higher-order power corrections in a
  transverse-momentum cut for colour-singlet production at NLO}},
  \href{https://doi.org/10.1140/epjc/s10052-019-7361-8}{\emph{Eur. Phys. J.}
  {\bfseries C79} (2019) 852},
  [\href{https://arxiv.org/abs/1906.09044}{{\ttfamily 1906.09044}}].

\bibitem{Moult:2018jjd}
I.~Moult, I.~W. Stewart, G.~Vita and H.~X. Zhu, \emph{{First Subleading Power
  Resummation for Event Shapes}},
  \href{https://doi.org/10.1007/JHEP08(2018)013}{\emph{JHEP} {\bfseries 08}
  (2018) 013}, [\href{https://arxiv.org/abs/1804.04665}{{\ttfamily
  1804.04665}}].

\bibitem{DelDuca:1990gz}
V.~Del~Duca, \emph{{High-energy Bremsstrahlung Theorems for Soft Photons}},
  \href{https://doi.org/10.1016/0550-3213(90)90392-Q}{\emph{Nucl. Phys.}
  {\bfseries B345} (1990) 369--388}.

\bibitem{DelDuca:2017twk}
V.~Del~Duca, E.~Laenen, L.~Magnea, L.~Vernazza and C.~D. White,
  \emph{{Universality of next-to-leading power threshold effects for colourless
  final states in hadronic collisions}},
  \href{https://doi.org/10.1007/JHEP11(2017)057}{\emph{JHEP} {\bfseries 11}
  (2017) 057}, [\href{https://arxiv.org/abs/1706.04018}{{\ttfamily
  1706.04018}}].

\bibitem{Beneke:2002ph}
M.~Beneke, A.~P. Chapovsky, M.~Diehl and T.~Feldmann, \emph{{Soft collinear
  effective theory and heavy to light currents beyond leading power}},
  \href{https://doi.org/10.1016/S0550-3213(02)00687-9}{\emph{Nucl. Phys.}
  {\bfseries B643} (2002) 431--476},
  [\href{https://arxiv.org/abs/hep-ph/0206152}{{\ttfamily hep-ph/0206152}}].

\bibitem{Beneke:2002ni}
M.~Beneke and T.~Feldmann, \emph{{Multipole expanded soft collinear effective
  theory with non-abelian gauge symmetry}},
  \href{https://doi.org/10.1016/S0370-2693(02)03204-5}{\emph{Phys. Lett.}
  {\bfseries B553} (2003) 267--276},
  [\href{https://arxiv.org/abs/hep-ph/0211358}{{\ttfamily hep-ph/0211358}}].

\bibitem{Bauer:2000yr}
C.~W. Bauer, S.~Fleming, D.~Pirjol and I.~W. Stewart, \emph{{An Effective field
  theory for collinear and soft gluons: Heavy to light decays}},
  \href{https://doi.org/10.1103/PhysRevD.63.114020}{\emph{Phys. Rev.}
  {\bfseries D63} (2001) 114020},
  [\href{https://arxiv.org/abs/hep-ph/0011336}{{\ttfamily hep-ph/0011336}}].

\bibitem{Bauer:2001yt}
C.~W. Bauer, D.~Pirjol and I.~W. Stewart, \emph{{Soft collinear factorization
  in effective field theory}},
  \href{https://doi.org/10.1103/PhysRevD.65.054022}{\emph{Phys. Rev.}
  {\bfseries D65} (2002) 054022},
  [\href{https://arxiv.org/abs/hep-ph/0109045}{{\ttfamily hep-ph/0109045}}].

\bibitem{Korchemsky:1993uz}
G.~P. Korchemsky and G.~Marchesini, \emph{{Resummation of large infrared
  corrections using Wilson loops}},
  \href{https://doi.org/10.1016/0370-2693(93)90015-A}{\emph{Phys. Lett.}
  {\bfseries B313} (1993) 433--440}.

\bibitem{Beneke:2004in}
M.~Beneke, F.~Campanario, T.~Mannel and B.~D. Pecjak, \emph{{Power corrections
  to $\bar{B} \to X_u \ell \bar{\nu} \,(X_s \gamma)$ decay spectra in the
  `shape-function' region}},
  \href{https://doi.org/10.1088/1126-6708/2005/06/071}{\emph{JHEP} {\bfseries
  06} (2005) 071}, [\href{https://arxiv.org/abs/hep-ph/0411395}{{\ttfamily
  hep-ph/0411395}}].

\bibitem{Beneke:2017ztn}
M.~Beneke, M.~Garny, R.~Szafron and J.~Wang, \emph{{Anomalous dimension of
  subleading-power N-jet operators}},
  \href{https://doi.org/10.1007/JHEP03(2018)001}{\emph{JHEP} {\bfseries 03}
  (2018) 001}, [\href{https://arxiv.org/abs/1712.04416}{{\ttfamily
  1712.04416}}].

\bibitem{Beneke:2018rbh}
M.~Beneke, M.~Garny, R.~Szafron and J.~Wang, \emph{{Anomalous dimension of
  subleading-power $N$-jet operators. Part II}},
  \href{https://doi.org/10.1007/JHEP11(2018)112}{\emph{JHEP} {\bfseries 11}
  (2018) 112}, [\href{https://arxiv.org/abs/1808.04742}{{\ttfamily
  1808.04742}}].

\bibitem{Beneke:2019kgv}
M.~Beneke, M.~Garny, R.~Szafron and J.~Wang, \emph{{Violation of the
  Kluberg-Stern-Zuber theorem in SCET}},
  \href{https://doi.org/10.1007/JHEP09(2019)101}{\emph{JHEP} {\bfseries 09}
  (2019) 101}, [\href{https://arxiv.org/abs/1907.05463}{{\ttfamily
  1907.05463}}].

\bibitem{Matsuura:1987wt}
T.~Matsuura and W.~L. van Neerven, \emph{{Second Order Logarithmic Corrections
  to the {Drell-Yan} Cross-section}},
  \href{https://doi.org/10.1007/BF01624369}{\emph{Z. Phys.} {\bfseries C38}
  (1988) 623}.

\bibitem{Bosch:2004cb}
S.~W. Bosch, M.~Neubert and G.~Paz, \emph{{Subleading shape functions in
  inclusive B decays}},
  \href{https://doi.org/10.1088/1126-6708/2004/11/073}{\emph{JHEP} {\bfseries
  11} (2004) 073}, [\href{https://arxiv.org/abs/hep-ph/0409115}{{\ttfamily
  hep-ph/0409115}}].

\bibitem{Lee:2004ja}
K.~S.~M. Lee and I.~W. Stewart, \emph{{Factorization for power corrections to
  $B \to X_s \gamma$ and $B \to X_u \ell \bar{\nu}$}},
  \href{https://doi.org/10.1016/j.nuclphysb.2005.05.004}{\emph{Nucl. Phys.}
  {\bfseries B721} (2005) 325--406},
  [\href{https://arxiv.org/abs/hep-ph/0409045}{{\ttfamily hep-ph/0409045}}].

\bibitem{Larkoski:2014bxa}
A.~J. Larkoski, D.~Neill and I.~W. Stewart, \emph{{Soft Theorems from Effective
  Field Theory}}, \href{https://doi.org/10.1007/JHEP06(2015)077}{\emph{JHEP}
  {\bfseries 06} (2015) 077},
  [\href{https://arxiv.org/abs/1412.3108}{{\ttfamily 1412.3108}}].

\bibitem{Beneke:2017mmf}
M.~Beneke, M.~Garny, R.~Szafron and J.~Wang, \emph{{Subleading-power $N$-jet
  operators and the LBK amplitude in SCET}},
  \href{https://doi.org/10.22323/1.290.0048}{\emph{PoS} {\bfseries RADCOR2017}
  (2017) 048}, [\href{https://arxiv.org/abs/1712.07462}{{\ttfamily
  1712.07462}}].

\bibitem{Hamberg:1990np}
R.~Hamberg, W.~L. van Neerven and T.~Matsuura, \emph{{A complete calculation of
  the order $\alpha_s^{2}$ correction to the Drell-Yan $K$ factor}},
  \href{https://doi.org/10.1016/S0550-3213(02)00814-3,
  10.1016/0550-3213(91)90064-5}{\emph{Nucl. Phys.} {\bfseries B359} (1991)
  343--405}.

\bibitem{Beneke:1997zp}
M.~Beneke and V.~A. Smirnov, \emph{{Asymptotic expansion of Feynman integrals
  near threshold}},
  \href{https://doi.org/10.1016/S0550-3213(98)00138-2}{\emph{Nucl. Phys.}
  {\bfseries B522} (1998) 321--344},
  [\href{https://arxiv.org/abs/hep-ph/9711391}{{\ttfamily hep-ph/9711391}}].

\bibitem{Li:2011zp}
Y.~Li, S.~Mantry and F.~Petriello, \emph{{An Exclusive Soft Function for
  Drell-Yan at Next-to-Next-to-Leading Order}},
  \href{https://doi.org/10.1103/PhysRevD.84.094014}{\emph{Phys. Rev.}
  {\bfseries D84} (2011) 094014},
  [\href{https://arxiv.org/abs/1105.5171}{{\ttfamily 1105.5171}}].

\bibitem{Gehrmann:2010ue}
T.~Gehrmann, E.~W.~N. Glover, T.~Huber, N.~Ikizlerli and C.~Studerus,
  \emph{{Calculation of the quark and gluon form factors to three loops in
  QCD}}, \href{https://doi.org/10.1007/JHEP06(2010)094}{\emph{JHEP} {\bfseries
  06} (2010) 094}, [\href{https://arxiv.org/abs/1004.3653}{{\ttfamily
  1004.3653}}].

\bibitem{Marcantonini:2008qn}
C.~Marcantonini and I.~W. Stewart, \emph{{Reparameterization Invariant
  Collinear Operators}},
  \href{https://doi.org/10.1103/PhysRevD.79.065028}{\emph{Phys. Rev.}
  {\bfseries D79} (2009) 065028},
  [\href{https://arxiv.org/abs/0809.1093}{{\ttfamily 0809.1093}}].

\end{thebibliography}\endgroup

\end{appendix}

\end{document}